\documentclass[twocolumn]{aastex631}
\hypersetup{linkcolor=red,citecolor=blue,filecolor=cyan,urlcolor=magenta}

\newcommand{\mgii}{Mg\,{\sc ii}}
\newcommand{\mgi}{Mg\,{\sc i}}
\newcommand{\aliii}{Al\,{\sc iii}}
\newcommand{\alii}{Al\,{\sc ii}}
\newcommand{\siii}{Si\,{\sc ii}}

\newcommand{\ovi}{O\,{\sc vi}}
\newcommand{\civ}{C\;{\sc iv}}
\newcommand{\cii}{C\;{\sc ii}}
\newcommand{\feii}{Fe\;{\sc ii}}
\newcommand{\oi}{O\;{\sc i}}
\newcommand{\siiv}{Si\;{\sc iv}}
\newcommand{\lya}{Ly-$\alpha$}
\newcommand{\kms}{$\rm{km s^{-1}}$}
\newcommand{\msunyr}{$\rm{M_{sun}\; yr^{-1}}$}

\begin{document}

\title{EIGER IV: The cool 10$^4$K circumgalactic environment of high-$z$ galaxies reveals remarkably efficient IGM enrichment}

\correspondingauthor{Rongmon Bordoloi}
\email{rbordol@ncsu.edu}\tabletypesize{\footnotesize}

\author[0000-0002-3120-7173]{Rongmon Bordoloi}
\affiliation{Department of Physics, North Carolina State University, Raleigh, 27695, North Carolina, USA}

\author[0000-0003-3769-9559]{Robert A.~Simcoe}
\affiliation{MIT Kavli Institute for Astrophysics and Space Research, 77 Massachusetts Ave., Cambridge, MA 02139, USA}

\author[0000-0003-2871-127X]{Jorryt Matthee}
\affiliation{Department of Physics, ETH Z{\"u}rich, Wolfgang-Pauli-Strasse 27, Z{\"u}rich, 8093, Switzerland}

\author[0000-0001-9044-1747]{Daichi Kashino}
\affiliation{Institute for Advanced Research, Nagoya University, Nagoya 464-8601, Japan}
\affiliation{Department of Physics, Graduate School of Science, Nagoya University, Nagoya 464-8602, Japan}

\author[0000-0003-0417-385X]{Ruari Mackenzie}
\affiliation{Department of Physics, ETH Z{\"u}rich, Wolfgang-Pauli-Strasse 27, Z{\"u}rich, 8093, Switzerland}

\author[0000-0002-6423-3597]{Simon J.~Lilly}
\affiliation{Department of Physics, ETH Z{\"u}rich, Wolfgang-Pauli-Strasse 27, Z{\"u}rich, 8093, Switzerland}

\author[0000-0003-2895-6218]{Anna-Christina Eilers}
\affiliation{MIT Kavli Institute for Astrophysics and Space Research, 77 Massachusetts Ave., Cambridge, MA 02139, USA}

\author[0000-0003-4491-4122]{Bin Liu}
\affiliation{Department of Physics, North Carolina State University, Raleigh, 27695, North Carolina, USA}

\author[0000-0000-0000-0000]{David DePalma} 
\affiliation{MIT Kavli Institute for Astrophysics and Space Research, 77 Massachusetts Ave., Cambridge, MA 02139, USA}

\author[0000-0002-5367-8021]{Minghao Yue}
\affiliation{MIT Kavli Institute for Astrophysics and Space Research, 77 Massachusetts Ave., Cambridge, MA 02139, USA}

\author[0000-0003-3997-5705]{Rohan~P.~Naidu}
\altaffiliation{NASA Hubble Fellow}
\affiliation{MIT Kavli Institute for Astrophysics and Space Research, 77 Massachusetts Ave., Cambridge, MA 02139, USA}

\begin{abstract}
We report new observations of the cool diffuse gas around 29, $2.3<z<6.3$ galaxies, using  deep \textit{JWST}/NIRCam slitless grism spectroscopy around the sightline to the quasar J0100+2802. The galaxies span a stellar mass range of $7.1 \leq \log M_{*}/M_{sun} \leq 10.7$, and star-formation rates of $-0.1 < \log$ SFR/{\msunyr} $<2.3$. We find galaxies for seven {\mgii} absorption systems within 300 kpc of the quasar sightline. The {\mgii} radial absorption profile falls off sharply with radii, with most of the absorption extending out to 2-3$R_{200}$ of the host galaxies. Six out of seven {\mgii} absorption systems are detected around galaxies with $\log M_{*}/M_{sun} >$9. {\mgii} absorption kinematics are shifted from the systemic redshift of host galaxies with a median absolute velocity $\approx$135 {\kms} and standard deviation $\approx$ 85 {\kms}. The high kinematic offset and large radial separation ($R> 1.3 R_{200}$), suggest that five out of the seven {\mgii} absorption systems are gravitationally not bound to the galaxies. In contrast, most cool circumgalactic media at $z<1$ are gravitationally bound. The high incidence of unbound {\mgii} gas in this work suggests that towards the end of reionization, galaxy halos are in a state of remarkable disequilibrium, and are highly efficient in enriching the intergalactic medium. Two strongest {\mgii} absorption systems are detected at $z\sim$ 4.22 and 4.5, the former associated with a merging galaxy system and the latter associated with three kinematically close galaxies. Both these galaxies reside in local galaxy over-densities, indicating the presence of cool {\mgii} absorption in two ``proto-groups" at $z>4$.

\end{abstract}


\section{Introduction} \label{sec:intro}
 
The commissioning of the $JWST$ has ushered in a new era for spectroscopy of galaxies and intergalactic matter near the Epoch of Reionization (EoR) \citep{RigbyJWST2022}. Strong rest frame optical lines (e.g., H-$\alpha$, [OIII]) are finally observable at $z \gtrsim 3.5$; which combined with JWST's efficient spectroscopic modes have enabled large scale spectroscopic surveys of galaxies at EoR \citep{Kashino2023,Matthee2023b,Wang2023,Oesch2023}. 
By performing carefully constructed spectroscopic experiments, where the galaxy fields also have bright high-$z$ quasars, one can extend the study of galaxies to characterize their gaseous halos \citep{Kashino2023}. Deep ground based near infra-red (NIR) spectroscopy of these quasars often reveal intervening metal absorption line systems associated with galaxies along the line of sight: a signature of diffuse baryonic reservoir of gas around galaxies \citep[e.g.,][]{Cooper2019}. These cosmic ecosystems fuel the growth of stellar mass in galaxies and serve as reservoirs of gas recycling. At $z<2$, such cosmic ecosystems have been successfully characterized as the circumgalactic medium (CGM) around galaxies \citep{Tumlinson2017}. Over the last two decades, large galaxy and quasar surveys have enabled detailed characterization of the CGM, establishing it as an ubiquitous reservoir of diffuse gas around galaxies \citep{Chen2001,Chen2010,Bordoloi2011,Tumlinson2013,Nielsen2013,Zhu2014,Huang2016,Burchett2016,Johnson2017}.

In the last 7 Gyrs of the history of the Universe ($z<1$), comparison of absorption line systems observed in background spectra of bright background quasars or galaxies with their host galaxy populations have revolutionized our understanding of the late-time CGM and its role in galaxy formation (see, \citealt{Tumlinson2017} for a detailed review). These studies have revealed that both highly ionized metals (traced by {\ovi}, {\civ}) and low ionized metals (traced by {\mgii}) show strong trends with increasing galaxy star-formation rates, and stellar masses 
\citep{Chen2001,Prochaska2011,Chen2010,Tumlinson2011,Bordoloi2011,Nielsen2013,Liang2014,Lan2018}. Cool circumgalactic gas traced by {\mgii} shows strong dependence on morphology and orientation of outflows vs inflows \citep{Bordoloi2011,Bordoloi2014c,Bordoloi2014b,Kacprzak2012,Bouche2012,Rubin2014,Martin2019,Lundgren2021}.  Diffuse warm-hot intra-group medium is also detected around groups of galaxies \citep{Bordoloi2011,Johnson2015,Chen2020,Dutta2021,McCabe2021}. CGM gas appears bimodal in metallicity, with a portion tracing enriched outflows ($\sim$20-100\% solar) and a distinct metal-poor component ($\sim$ 5\% solar) that may resemble the long-sought “cold accretion” entering galaxies from the IGM \citep{Lehner2013,Wotta2019}. Most crucially mass and metal census of the CGM gas at $z \sim 0.2$ suggests that the CGM may host a large share of galactic baryons, with the CGM mass content outweighing the total stellar mass content of galaxies \citep{Stocke2013,Werk2014,Prochaska2017}, and hosts a massive reservoir of galactic metals, with galaxies having ejected at least as much metal mass as they have retained \citep{Tumlinson2011,Bordoloi2014b,Peeples2014}. At these redshifts, most of the CGM gas is bound to the dark matter halo of its host galaxy, suggesting that most of the gas will be recycled back to the ISM \citep{Tumlinson2013,Bordoloi2014b,Ford2014,Huang2016,Bordoloi2018}.  

At $z\sim 2$, the CGM contain a large metal reservoir, but more of this gas is kinematically consistent with not being bound to the host galaxy's dark matter halo than their $z<1$ counterparts \citep{Rudie2019}. Observations by lensed QSOs and spatially extended lensed arcs reveal that individual CGM gas clouds are small and show large variations even within a single halo \citep{Rauch1999,Rubin2018,Lopez2018,Bordoloi2022}. A consensus has emerged that the CGM is a ubiquitous feature of galaxies from $z \sim 3$ to $z \sim 0$ \citep{Rudie2012,Tumlinson2017,Peroux2020,Lehner2022}.

Within the first Gyr of cosmic history, the rest-frame UV transitions that are instrumental in characterizing the low-$z$ CGM, are redshifted into the near infrared (NIR). Moreover, any transition blueward of {\lya} will not be observable owing to the intervening neutral IGM. Therefore, heavy element absorption systems in high-$z$ quasar spectra provide our only access to the chemical enrichment and ionization taking place in this environment. NIR spectroscopy of QSOs at $z > 6$ (e.g., \citealt{Becker2001,Simcoe2012,Banados2018,Cooper2019,Yang2020,Davies2023})  has pushed these investigations within the first Gyr of the Big Bang. These studies find that the number density evolution of strong {\mgii} absorption systems trace the cosmic star-formation history of the Universe whereas the weaker systems show no evolution out to $z \sim 6$  \citep{Matejek2012}. The incidence of detection of low ionization species (e.g., \cii, \oi) remain high even at high-$z$, whereas the incidence of detection of high ionized species (e.g., \civ, \siiv) drop off sharply beyond $z \sim 5.7$ \citep{Cooper2019,Becker2019}. This might be owing to change in UV background at the early Universe or reflect some fundamental change in galaxy properties. Therefore, identifying the galaxies associated with these absorbers at high-$z$ might give crucial insight into the latter stages of EoR.

Prior to \textit{JWST}, it has been prohibitive to conduct detailed galaxy surveys to identify the host galaxies of these absorbers. But initial studies of detection of host galaxies suggest a strong correlation between the absorption lines detected at high-$z$ with host galaxies.

Recently two works reported the presence of galaxy overdensities near a strong {\civ} absorption system at $z \sim $ 5.72, four {\lya} emitting galaxies (LAE) \citep{Diaz2021} and two [CII]158 $\mu$m emitting galaxies \citep{Kashino2023b}. Additionally, another host galaxy associated with a $z=5.9$ OI absorption system using [CII]158 $\mu$m line was reported \citep{Wu2021}. Cross correlation of LAE galaxies from deep MUSE observations also suggest a link between strong {\civ} absorbers and bright LAE galaxies out to $z\sim 4$ \citep{Galbiati2023}.

These promising early results suggest that a systematic multi-wavelength galaxy survey is warranted to study the CGM host galaxies at $z > 4$, where detailed galaxy properties can be studied.

In this work we present the first CGM measurements traced by {\mgii} absorption around 29 ($2<z<6$) galaxies in the EIGER survey (\textit{Emission-line Galaxies and Intergalactic Gas in the Epoch of Reioniation}; \citealt{Kashino2023}). We focus on the first observations around the hyper-luminous $z = 6.33$ quasar J0100+2802. 

This paper is organized as follows. In section 2 we describe the observations and summarize the survey strategy. In section 3 we describe the measurements of galaxy properties and the CGM absorber properties. In section 4 we describe the results. In section 5 we present the summary and discussion of the results. Throughout this paper we follow a flat $\Lambda$CDM cosmology with $H_{0} =$ 67.7 $\rm{km\, s^{-1} Mpc^{-1}}$, $\Omega_{M}$ = 0.31, and $\Omega_{\Lambda}$ = 0.69 \citep{Planck2020}. All magnitudes are listed in the AB system.  Unless stated otherwise, all distances are quoted in units of physical kpc.

\begin{figure*}
\includegraphics[width=1\columnwidth]{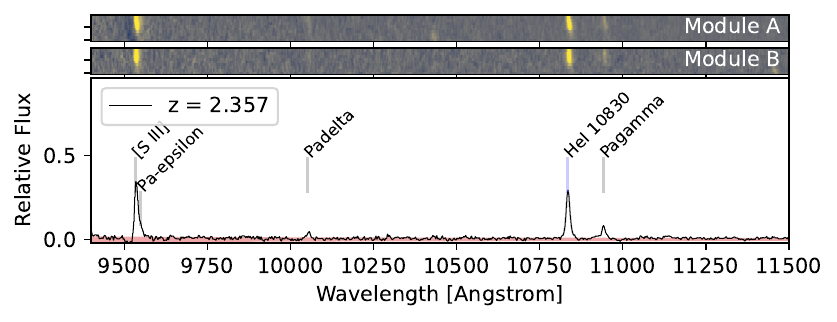}
\includegraphics[width=1\columnwidth]{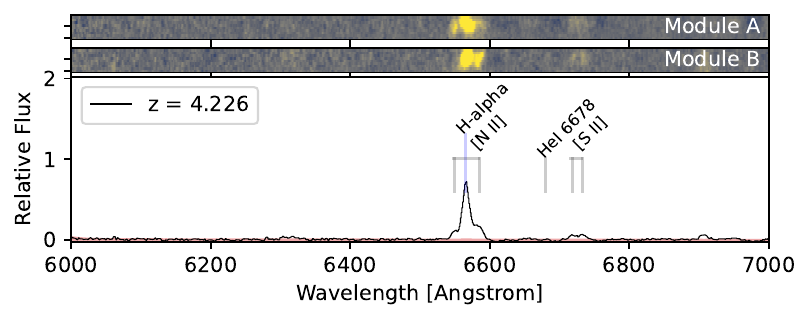}
\begin{center}
    \includegraphics[width=1\columnwidth]{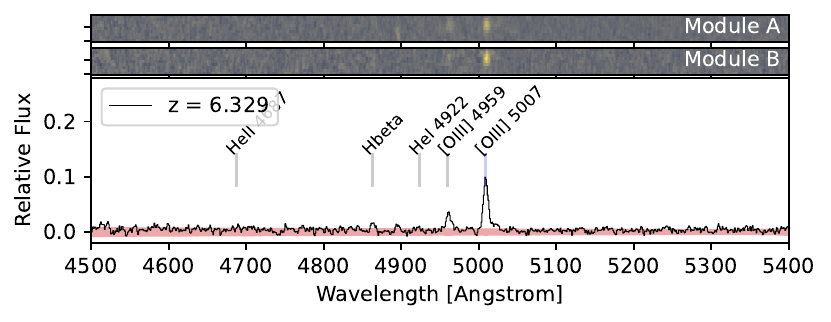}
\end{center}
\caption{JWST NIRCam/F356W grism spectra of three galaxies at $z \sim 2.3$ (top left), $z \sim 4.2$ (top right), and $z \sim 6.3$ (bottom). The 2D emission line spectra from each module is presented for each galaxy. The 1D optimally extracted spectra is shown in the main panel for each galaxy with prominent emission lines marked out. The red shaded region shows the 1-$\sigma$ uncertainty on the 1D spectra.
\label{fig:galaxy_spectra}}
\end{figure*}

\section{Observations}

\subsection{The EIGER Survey}
The EIGER survey is a 126.5 hour \textit{JWST} GTO (PID:1243, PI: S. J. Lilly) program, that performs NIRCam wide field slitless spectroscopy (WFSS) around six extra-galactic fields, each centered on a hyper-luminous ($6 \lesssim z \lesssim 7$) quasar. We refer the reader to \citet{Kashino2023} for a detailed description of survey design rationale and data reduction methods. Below we briefly summarize different aspects of the observations.

\subsection{NIRCam observations of J0100+2802}
In this work, we focus on metal absorption in the vicinity of galaxies detected in the $z=6.33$ quasar field of J0100+2802. \textit{JWST}/NIRCam WFSS spectroscopy is performed with the F356W filter using the reverse grisms (GRISMR), which both disperse the spectra horizontally on the NIRCam sensors but with opposite parity. The spectral resolution of the observations is $R \sim 1500$. Simultaneously with spectroscopy, F115W, F200W imaging of the field is performed. Direct and out-of-field imaging in the F356W filter is performed to cover the full spectroscopic field of view \citep{Kashino2023}. 

A four pointing mosaic strategy is employed, that ensures that the central $40\arcsec \times 40\arcsec$ has the maximum depth of $\sim$ 35 kilo-seconds. The total spectroscopic field of view around J0100+2802 is $\sim$ 25.9 arcmin$^2$, and the total exposure time ranges from 8-35 kilo-seconds. In this work we focus only on the central $\approx$ 4.6 arcmin$^2$ of the field, which is covered by both the NIRCam modules A and B (with reversed
dispersion directions). This enables us to accurately identify single emission line objects and measure their redshifts \citep{Matthee2023}.

NIRCam imaging data is reduced as described in \cite{Kashino2023} using \texttt{jwst} pipeline \citep[v1.8.2;][]{bushouse_howard_2022_7325378}. Additional post-processing steps are performed to mask strong cosmic ray hits following \cite{Merlin2022}. Astrometry is calibrated by aligning known stars from the Gaia Data Release 2 catalog \citep{Gaia2018}. Several known artefacts (e.g., stray-light features, 1/f noise, residual sky background) are subtracted \citep{Kashino2023} before obtaining a final co-added image with pixel sizes of 0.03 $\arcsec$/pixel.

These final co-added images are used to perform aperture-matched photometry with \texttt{SExtractor} in dual mode. The F356W image is used as the detection image. All images are convolved to match the point spread function of the F356W image. Kron aperture magnitudes are measured and photometric uncertainties are estimated by measuring random blank sky variations for apertures of different sizes, scaled to the local variance propagated by the pipeline (see, \citealt{Kashino2023}).

NIRCam WFSS data reduction is performed using a combination of \texttt{jwst} pipeline \citep[v1.7.0;][]{bushouse_howard_2022_7071140} and custom in-house tools as described in detail in \cite{Kashino2023} and \cite{Matthee2023}. To summarize, each individual exposure is processed with the \texttt{Detector1} step in the \texttt{jwst} pipeline and assigned a WCS with the \texttt{Spec2} step. The frames are flat fielded and additional $1/f$ noise and sky background variation are removed by subtracting the median flux of each column to create the science frames. A continuum map is created by using a running median filter along the dispersion direction. The median filter kernel size is adaptive and has a hole in the center to ensure that it does not over subtract the emission lines. This continuum map is subtracted from each science frame to create an emission line map for each exposure. We stress that the continuum subtraction process does not rely on the source position or any trace model. We refer the reader to \citep{Kashino2023} for detailed description of this process.

For each object detected in F356W imaging, a 2D spectrum is extracted based on \texttt{grismconf}\footnote{\url{https://github.com/npirzkal/GRISMCONF}} with the V4 trace models\footnote{\url{https://github.com/npirzkal/GRISM_NIRCAM}}. 
We perform additional pixel-level correction to the trace models to optimize the extraction based on our own empirical calibration using spectra of bright stars. Individual exposures are divided by the relevant sensitivity curve, rectified for small curvature, and re-sampled onto a common observed wavelength grid ($3\; \mu m \leq \lambda \leq 4\; \mu m$, with pixel size of 9.75 {\AA}). For each individual module, these exposures are co-added with sigma clipping to produce the final 2D spectrum of a galaxy. For a given object position, we extract one independent spectra from each module. This results in two independent spectra obtained from the two NIRCam modules for each source. Since these two spectra have reverse dispersion directions, only emission lines that are truly coming from the object of interest will appear at the same observed wavelength in both the spectra. All other lines will shift in wavelength and/or disappear completely. This is the primary diagnostic to remove unrelated emission lines that are coming from other objects.  Figure \ref{fig:galaxy_spectra}, top panels show the continuum subtracted 2D emission line spectra of three galaxies at $z\sim$2.3, 4.2 and 6.3, respectively. Common emission lines are independently detected on both modules, further verifying their robustness.
 
\begin{figure*}[t]
\centering
\includegraphics[width=\columnwidth]{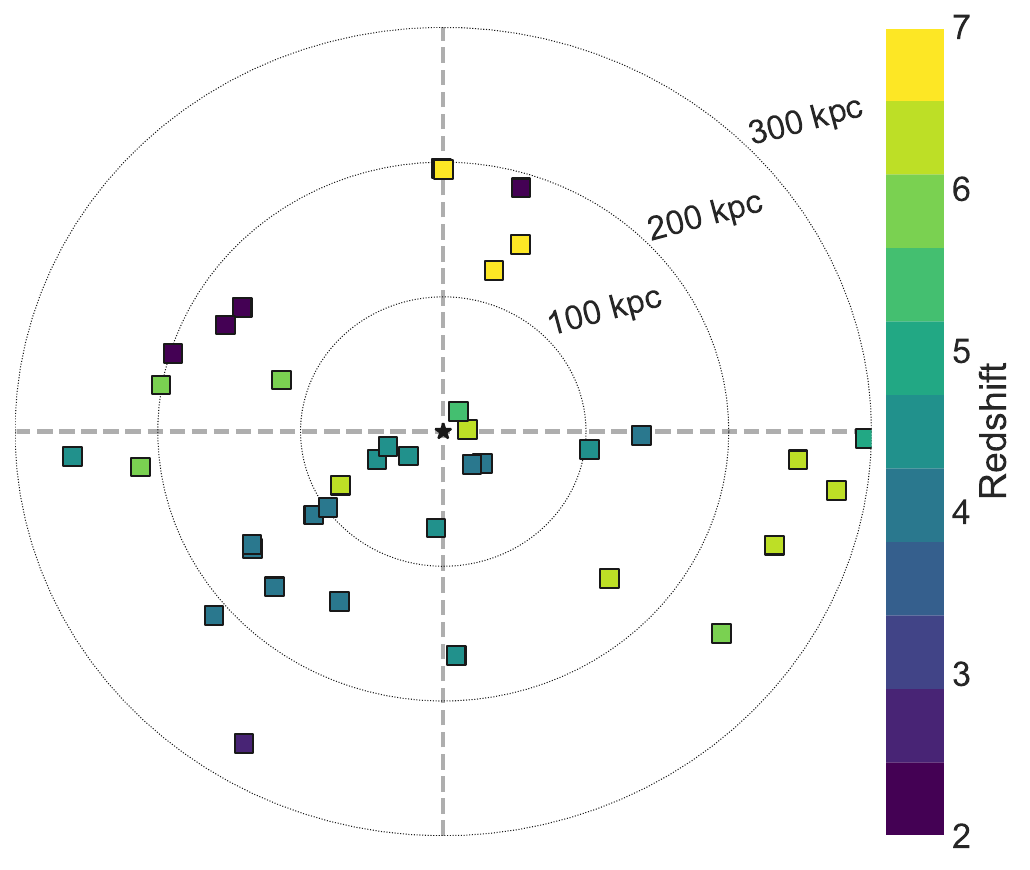}
\includegraphics[width=\columnwidth]{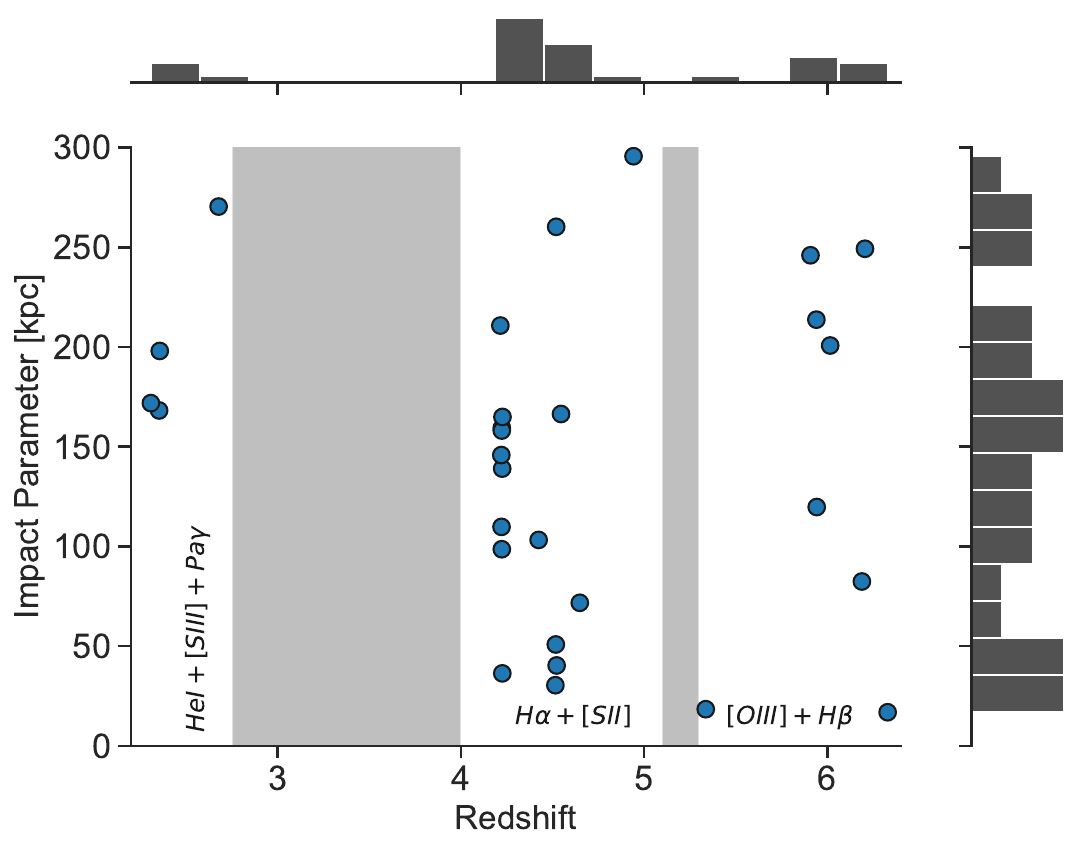}
\caption{ Redshift and impact parameter distribution of the galaxies within 300 kpc of the quasar J0100+2802. \textit{Left Panel:} The distribution of all spectroscopically confirmed galaxies within 300 kpc from the quasar sightline. Each square represents a galaxy color coded as a function of its redshift. The quasar J0100+2802 is in the center of the figure (black star). \textit{Right Panel:} Redshift and impact parameter distribution of the 29 galaxies analyzed in this work. Prominent emission lines used for redshift measurement are shown. These galaxies are chosen to be at $z<6.33$ and at impact parameters $<$ 300 kpc. The gray boxes denote redshift ranges where no strong galaxy emission lines shift into the observed frame of NIRCam 3.5$\mu$m WFSS spectra. \label{fig:galaxy_prop}}
\end{figure*}

\subsection{HST observations of J0100+2802}
We obtain HST/ACS imaging of the J0100+2802 field in F850LP, F775W, and F606W filters respectively (HST PID: 15085, 13605). The total exposure time for these observations is 24,450 seconds. We query MAST for the individual flc.fits exposures, which are corrected for charge transfer inefficiency but have not been re-sampled. We align the individual exposures to the NIRCam F356W mosaic, and drizzle the images to the common pixel grid of the NIRCam mosaics using \texttt{DrizzlePac} \citep{Mack2022}. The routine masks cosmic rays, performs median blotting and matches the sky of each exposure before calculating the median co-added image. To create PSF-matched images for precise multi-band photometry we calculate matching kernels, using PSFs derived from \texttt{TinyTim} \citep{Krist2011} and re-sample to the 0.03$\arcsec$ pixel scale.

\subsection{Ground Based spectroscopy of Quasar J0100+2802}

Deep ground based optical and NIR spectroscopic observations are performed on the $z= 6.33$ quasar J0100+2802 using both the Magellan/FIRE and the VLT/X-shooter instruments. The target has been observed for a total of 16.8 hours with 5.8 hours of Magellan/FIRE (PI: Simcoe) and 11 hours of VLT/X-shooter (program ID: 096.A-0095; PI: Pettini) observations, respectively. Additionally, high-resolution ($R\approx 50,000$), Keck/HIRES observations are performed on J0100+2802, to cover the optical ($0.86 \mu m \leq \lambda \leq 9.9\mu m$) part of the spectrum \citep{Cooper2019}. The total integration time for HIRES observations are 3.8 and 3 hours, respectively in two different grating setups.
 
These observations are self consistently reduced with the \texttt{PypeIt} data reduction pipeline \citep{PypeIt2020}, and a final co-added, flux calibrated spectra is produced for both the instruments. We refer the reader to \cite{Eilers2023} for a detailed description of the data reduction method. The final FIRE/XShooter spectra have a median signal to noise ratios (SNR) $\sim$ 138/126 per resolution element across the full wavelength range. The 16$^{th}$ and 84$^{th}$ percentile SNR per resolution element for FIRE/XShooter spectra are 75/32 and 204/214, respectively. The variation of SNR per resolution element as a function of observed frame wavelength is presented in Table \ref{tab:SNR}.

\section{Methods} 
Here we describe how emission line galaxy properties are measured and how absorption lines are identified and analyzed.

\begin{figure}
\centering
\includegraphics[width=\columnwidth]{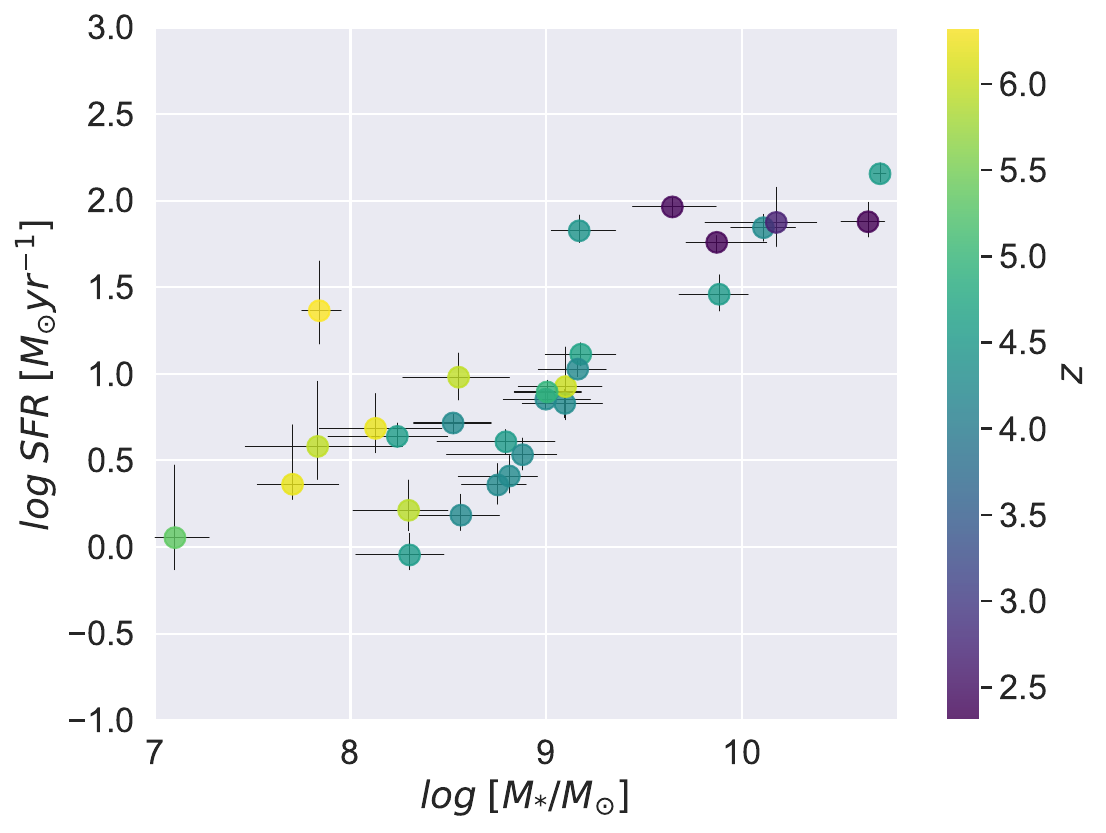}
\caption{Stellar mass and star-formation rate estimates of galaxies used in this work. Each circle is color coded to reflect the spectroscopic redshift of the galaxy. The error-bars represent the 16$^{th}$ and 84$^{th}$ percentile uncertainty in stellar masses and star-formation rates obtained from SED fits.
\label{fig:galaxy_sfr}}
\end{figure}

\begin{deluxetable*}{ccccc}
\tablecolumns{4}
\tablewidth{0pt}
\tablecaption{Signal to noise ratio per resolution element \label{tab:SNR}}
\tablehead{
\colhead{Instrument}&
\colhead{$9.5  \leq \lambda [\mu m] \leq 13.5$\tablenotemark{a}}&
\colhead{$14.1  \leq \lambda [\mu m] \leq 17.8$\tablenotemark{a}}&
\colhead{$19.4  \leq \lambda [\mu m] \leq 24.96$\tablenotemark{a}}&
}
\startdata
FIRE & 102, 143, 189 & 83, 180,244 & 59, 106, 183 \\
XShooter & 80, 150, 188 & 80, 196, 245 & 20, 37, 65 \\
\enddata
\tablenotetext{a}{The 16$^{th}$, 50$^{th}$ and 84$^{th}$ percentile SNR per resolution element are reported.}
\end{deluxetable*}

\subsection{Galaxy Redshifts}

We create a candidate emission line galaxy list using the following two criteria. We search for emission line objects (with SNR per module $>$ 7) in the central $\approx$ 4.6 arcmin$^2$ of the EIGER footprint, which is covered by both the NIRCam grism modules (A and B). This choice ensures that the total SNR of each emission line object is $>$ 10. We further restrict our search to objects whose spectrum contains a verified emission line that would plausibly be at a redshift within $\Delta z \approx$ 0.02 of the identified intervening metal absorption lines. For each identified object, we extract 2D galaxy spectra as described in Section 2.2. Additionally, we look for higher redshift galaxies near the quasar identified with [OIII]/H-$\beta$ emission lines \citep{Matthee2023}. This yields a sample of 127 objects within $\approx$ 108 arcsec of the J0100+2802 quasar. Monte Carlo simulations of injecting and recovering  synthetic emission lines in the central $\approx$ 4.6 arcmin$^2$ of the J0100+28 field, yield a spectroscopic completeness of 50\% for emission line flux of 1.6$\rm{\times 10^{-18}\; erg\; s^{-1}\; cm^{-2}}$, and 90\% for emission line flux of 2.8$\rm{\times 10^{-18}\; erg\; s^{-1}\; cm^{-2}}$, respectively (Mackenzie et al. in prep). We quantify the mass completeness of the spectroscopic sample presented in this work by using mock observations from the \texttt{JAGUAR} (JWST extra-galactic mock) catalog \citep{Williams2018}. We compute the spectroscopic mass completeness as a function of stellar mass bins in different redshift slices. In each redshift slice, the mass completeness of an individual stellar mass bin is defined as the ratio between the number of galaxies above the F356W flux detection threshold and the total number of galaxies in that bin. At $\langle z \rangle \sim$ 6 we find that the spectroscopic sample presented in this work is 50\% mass complete at $\log M_{*}/M_{\odot} \sim $ 8.5, and at $\langle z \rangle \sim$4.2 the galaxy sample is 50\% mass complete at $\log M_{*}/M_{\odot} \sim  $ 8.13, respectively.

Each spectrum is individually inspected using a custom python API \citep[\texttt{zgui}, ][]{rbcodes}, which is used to independently extract an 1D spectrum for each module. We identify individual emission lines and fit a Gaussian profile to measure the redshift of the galaxy. Photo-$z$ posterior distribution function for each object is also inspected to identify any foreground contaminating object. For the same object, it is crucial to inspect the spectra from both modules to identify and mask out any contaminating feature from other galaxies. Since the two spectra are extracted from the two NIRCam modules with reversed dispersion directions, only ``real" lines associated with the extracted object will appear consistently at the same wavelength in the two modules. Each object is visually inspected by at least two individuals, and only objects with secure redshifts are considered for analysis.   This creates a total sample of 87 galaxies with $0.4 < z < 6.8$. Figure \ref{fig:galaxy_prop}, left panel shows the spatial distribution of these galaxies within 300 physical kpc of the J0100+2802 quasar sightline. Each individual galaxy is color coded to reflect its spectroscopic redshift. Independent redshift measurements from both modules yield a typical redshift accuracy of $\approx$ 140 {\kms} for each galaxy. Since we focus only on the CGM host of the {\mgii} absorbers along the J0100+2802 sightline, we further select only galaxies within 300 kpc from the J0100+2802 quasar sightline and at a redshift lower than the quasar ($2.3 < z < 6.33$). The lower redshift limit is chosen to match the lowest redshift {\mgii} absorber detected along this line of sight. This yields a final sample of 29 galaxies. Figure \ref{fig:galaxy_prop}, right panel shows the redshift and impact parameter distribution of this final sample of galaxies with a mean redshift of $\langle z \rangle \;= 4.5478 \pm 0.201 $. Throughout the rest of the paper, we will only focus on this sample of galaxies. 

The gray shaded regions (Figure \ref{fig:galaxy_prop}, right panel) mark the redshift ranges where no strong galaxy rest frame optical emission lines shift into the observed wavelength range of EIGER NIRCam/grism spectroscopy. Our survey is most sensitive to the three redshift windows, $2.3<z<2.7$ (using HeI+[SIII]+Pa-$\gamma$ lines), $4<z<5.1$ (using H-$\alpha$+[SII] lines), and $5.3<z<7$ (using [OIII]+H-$\beta$ lines). Future, ground based spectroscopic follow-ups or additional \textit{JWST} spectroscopy with different gratings will enable us to cover these missing redshift ranges.

We note that since our galaxy search is explicitly within $\Delta z \approx$ 0.02 ($\Delta v \approx \pm6000$ {\kms}) of the identified absorption lines, the galaxy sample is not selected blindly without any knowledge of the absorption systems. This ``galaxy centric"  \citep[e.g.,][]{Bordoloi2011,Tumlinson2013} approach is essential to characterize the covering fraction of the CGM gas or to measure the total metal mass budget of the CGM \citep{Tumlinson2011, Bordoloi2014a}. Owing to challenges of identifying single emission line galaxies, and in mitigating contamination from other sources, we restrict this work to focus only on galaxies within $\Delta z \approx$ 0.02 of identified absorption line systems. This search window is large enough that we can still detect galaxies not associated with these absorption systems. However, this work does not search for all galaxies outside the selection window and does not attempt to quantifying the {\mgii} absorption covering fraction and metal mass budget around the EIGER galaxies. A complete ``galaxy centric" analysis of the CGM of the full EIGER survey will be presented in a future work.

\subsection{Emission line measurement and SED fitting}
We follow the procedure introduced in \cite{Matthee2023} to measure the emission line flux of the strongest lines in a galaxy spectrum (e.g., H-$\alpha$, He-$10830$, [OIII]) from the grism emission line observations, which is summarized as follows. We start with the 2D emission line spectra (Top panels, Figure \ref{fig:galaxy_spectra}), and select a spectral region within $\pm 50${\AA} of the emission line of interest in the rest frame. We collapse this emission line in the spectral direction and fit the spatial profile with single or multiple Gaussian profiles. This spatial profile is used to optimally extract an 1D continuum filtered spectrum for the galaxy. We follow \cite{Matthee2023} and re-scale the noise levels of the 2D emission line spectrum, by evaluating the standard deviation of empty sky pixels and setting it equal to the mean noise level of our 1D spectrum. This procedure is performed independently on each module. Figure \ref{fig:galaxy_spectra} bottom panels show the three representative optimally extracted 1D spectra of galaxies at $z \approx$ 2.3, 4.2 and 6.3, respectively. The strong emission line features in each spectrum are marked.

We use these optimally extracted 1D spectra to fit the emission lines of interest. We fit the emission lines with Gaussian profiles (between 1 and 3, depending on complexity) and measure their total line flux.

These line fluxes are used to perform spectral energy distribution (SED) fits, along with the photometric data at the spectroscopic redshift of the galaxy. We use broad-band photometry from three HST (F6060W, F776W, F850LP) and three \textit{JWST} (F115W, F200W, F356W) images (see Section 2.2, 2.3). We use the SED fitting code \texttt{prospector} \citep{Johnson2021} to perform fits to these six photometric measurements along with F356W grism line fluxes. Following \cite{Matthee2023}, we assume a 5\% error on the spectro-photometric calibration of the observations. \texttt{Prospector} fits model the total stellar mass, gas-phase and stellar metallicity, the star-formation history of the galaxy, dust attenuation and the ionization parameter. We assume a \cite{Chabrier2003} initial mass function (IMF) and use the MIST isochrone models \citep{Dotter2016,Choi2016}. We use a delayed-$\tau$ star-formation history model, and apply a dust attenuation correction following \cite{Calzetti2000}.

Figure \ref{fig:galaxy_sfr}, shows the stellar mass and star-formation rate (SFR) estimates for these galaxies from the SED fits. Each circle represents a galaxy and is color coded as a function of their spectroscopic redshifts. The estimated stellar masses span four decades in range ($7.1 \leq \log M_{*}/{M_{\odot}} \leq 10.6$, and almost two dex in star-formation rate estimates.  Most of the high-$z$ ($z>5$) galaxies are of lower stellar mass ($\log M_{*}/{M_{\odot}} <9.2$).

\begin{figure*}
\centering
\includegraphics[width=2\columnwidth]{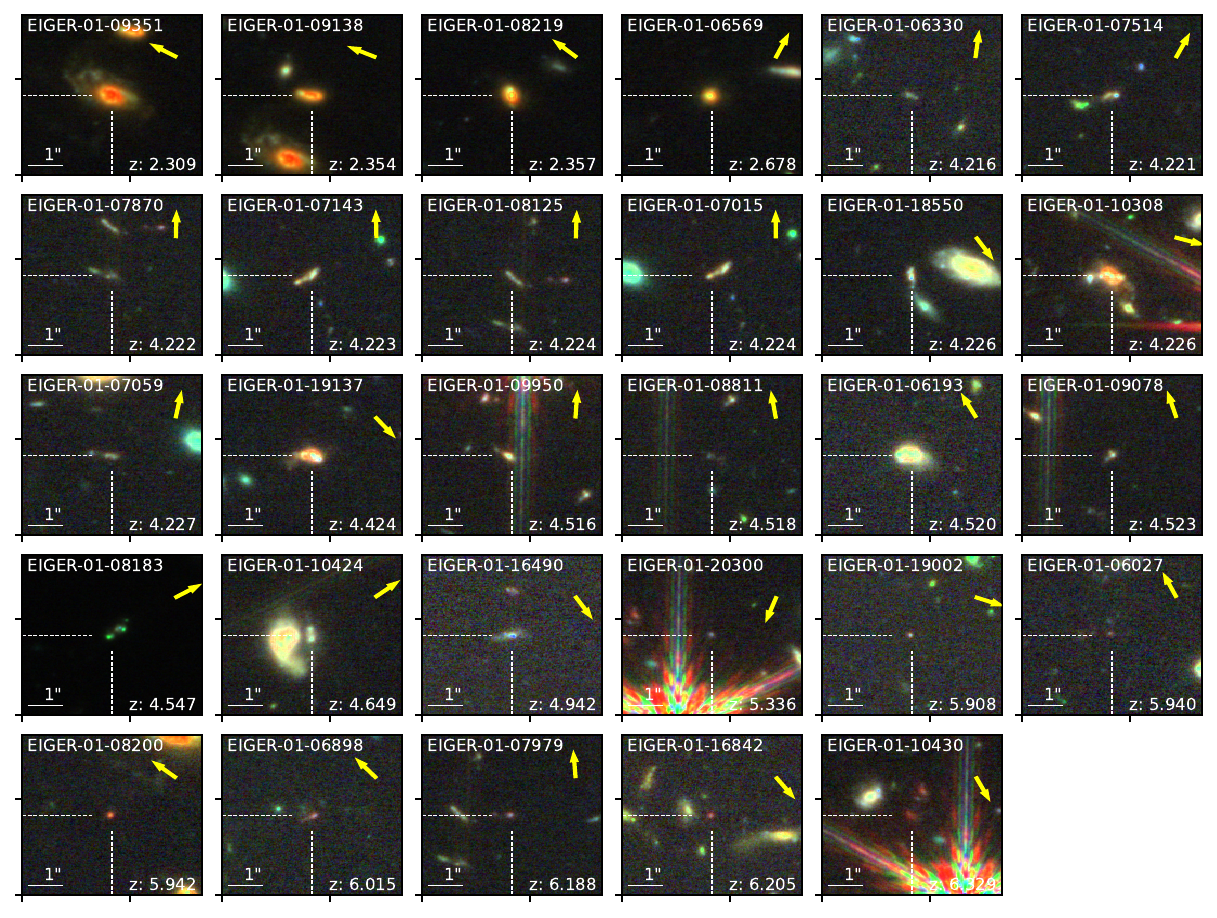}
\caption{False-color \textit{JWST}/NIRCam F115W/F200W/F356W stamps of galaxies within 300 kpc of the background J0100+2802 quasar. Each galaxy has a confirmed spectroscopic redshift. The yellow arrow is a position vector directed towards the quasar line of sight from the host galaxy. The location of the position vector is arbitrary in each stamp.
\label{fig:galaxy_stamps}}
\end{figure*}

We estimate the halo mass of the galaxies using the abundance matching relation from \cite{Behroozi2019}.  In this work, we quantify the virial radius of a galaxy as $R_{200}$, the radius at which the halo mass density is 200 times the critical density of the Universe. We write it as 
\begin{equation}
    R_{200}^{3}\;=\;=\frac{M_{halo} G}{100H^{2}(z)},
\end{equation}

where $M_{halo}$ is the halo mass, $G$ is the universal constant of gravitation and $H(z)$ is the Hubble parameter at the redshift of interest. The uncertainty on stellar mass and abundance matching relations are propagated through to halo mass and virial radii estimates. The $R_{200}$ measurements are within 5\% of $R_{vir}$ estimates derived from \citealt{Bryan1998}, well within the uncertainty of the $R_{200}$ estimates. The galaxy properties are presented in Table \ref{tab:galaxy_info}. We create false color \textit{JWST}/NIRCam (F115W, F200W, F356W) rgb images of each galaxy presented in this work. Each image is a $5\arcsec \times 5\arcsec$ stamp and is presented in Figure \ref{fig:galaxy_stamps}.

\subsection{Absorption line measurements}
We visually inspect the spectra of the quasar J0100 obtained from FIRE, X-shooter and HIRES instruments and search for intervening absorption line systems. We use a python-based API from the \texttt{rbcodes} package \citep{rbcodes} to identify and tabulate these systems. Along the J0100+2802 quasar line of sight, 22 unique intervening absorption line systems are identified within $2.3 < z < 6.33$. These lines include strong absorption systems traced by {\alii}, {\aliii}, {\cii}, {\civ}, {\feii}, {\mgii}, {\mgi}, {\oi}, {\siii}, {\siiv} transitions. Among these, there are 16 unique {\mgii} absorption systems between $2.3 < z < 6.14$. We compute the rest frame {\mgii} equivalent width detection limit (3-$\sigma$ detection threshold) per two resolution elements for both the XShooter and FIRE spectrum of J0100+2802, respectively. The mean 3-$\sigma$ detection threshold for XShooter and FIRE spectra are 11 m{\AA} and 18 m{\AA}, respectively. The 50\% equivalent width completeness limit, computed as the median 3$\sigma$ detection threshold per two resolution elements within different redshift bins are presented in Table \ref{tab:EW_limit}.

\begin{deluxetable}{ccc}
\tablecolumns{3}
\tablewidth{0pt}
\tablecaption{{\mgii} 50\% equivalent width completeness limit [m\AA] \label{tab:EW_limit}}
\tablehead{
\colhead{Redshift bins}&
\colhead{FIRE}&
\colhead{XShooter}
}
\startdata
2.3--3.8 & 14 & 8 \\
3.8--5.5& 10 & 5 \\
5.8--6.4 & 14 & 13 \\
\enddata
\vspace{-0.5cm}
\end{deluxetable}

In this paper we will focus primarily on the hosts of {\mgii} absorption line systems within the redshift windows of $2.3<z<2.7$, $4<z<5.1$, and $5.3<z<6.3$ (see Figure \ref{fig:galaxy_prop}). This redshift range corresponds to the observer frame wavelength range with optical galaxy emission lines covered by EIGER NIRCam/grism spectroscopy. We use a semi-automated framework to measure the absorption line strengths and kinematics associated with each identified foreground galaxy as follows: we first shift the reduced final quasar spectrum to the rest-frame of the foreground galaxy, using the spectroscopic redshift of the galaxy as described in the previous section. We focus on the common atomic absorption lines at predictable observed frame wavelength ranges. We quantify a detected absorption system to be associated with a host galaxy if it is within 300 physical kpc of the J0100+2802 quasar sightline and within $\pm$400 {\kms} of the systemic redshift of the galaxy. Our emission line search criterion (galaxy emission line SNR $>7$ per module), can detect galaxies at $\log \; M_{*}/M_{\odot} \approx $ 7.1 (Figure \ref{fig:galaxy_sfr}). However, it is possible that some faint emission line galaxies are missed in this search. We adopt a conservative approach and only focus on the reliably detected emission line galaxies in this work. The search for fainter (lower SNR) emission line galaxies would be carried out in a future work incorporating all the six quasar fields of the EIGER survey (Bordoloi et al. in prep).    

We extract short slices of quasar spectra around $\pm$600 {\kms} of the systemic redshift of the galaxy, for each line of interest. These lines include {\mgii}, {\feii}, C IV, Si IV etc. We continuum normalize each slice using a multi-ordered Legendre polynomial and measure the rest frame equivalent width and apparent optical depth (AOD) column density of each transition. We visually inspect each transition to confirm its presence, and set the velocity range for AOD column density integration. We use the identified absorption line list to minimize contamination from other intervening absorption line systems, and when setting the velocity integration range. We attempt to identify every detection feature within $\pm$600 {\kms} of the lines of interest. Most such features are not associated with the {\mgii} host galaxy and are positively identified to be associated with other intervening absorption line systems. We pay particular attention to the identified {\mgii} absorption doublet and ensure that the AOD ratios between the doublet range from 2:1 to 1:1 and require that the {\mgii} absorption profiles are aligned within 200 {\kms} from other detected metal absorption lines in that system. 

Additionally, we fit Voigt profiles to both the {\mgii} absorption doublet, to quantify the kinematic component structure and to estimate column densities for severely blended lines. This is a crucial step, as the Voigt profile fitting improves on the AOD column density measurement using information about line shape, and location to constrain the fit. Further, for several saturated {\mgii} absorption features line saturation is taken into account as the line spread function is accounted for in these fits. 

\begin{figure*}
\centering
\includegraphics[width=\columnwidth,height=1.5\columnwidth]{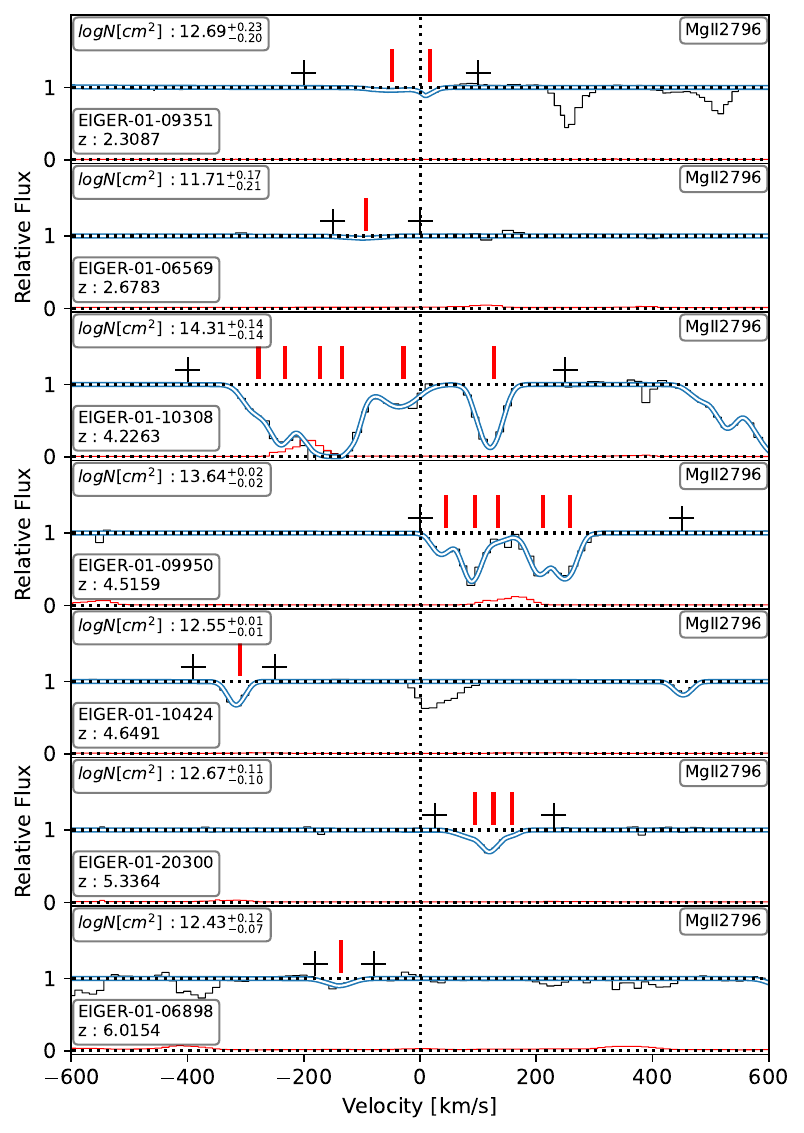}
\includegraphics[width=\columnwidth,height=1.5\columnwidth]{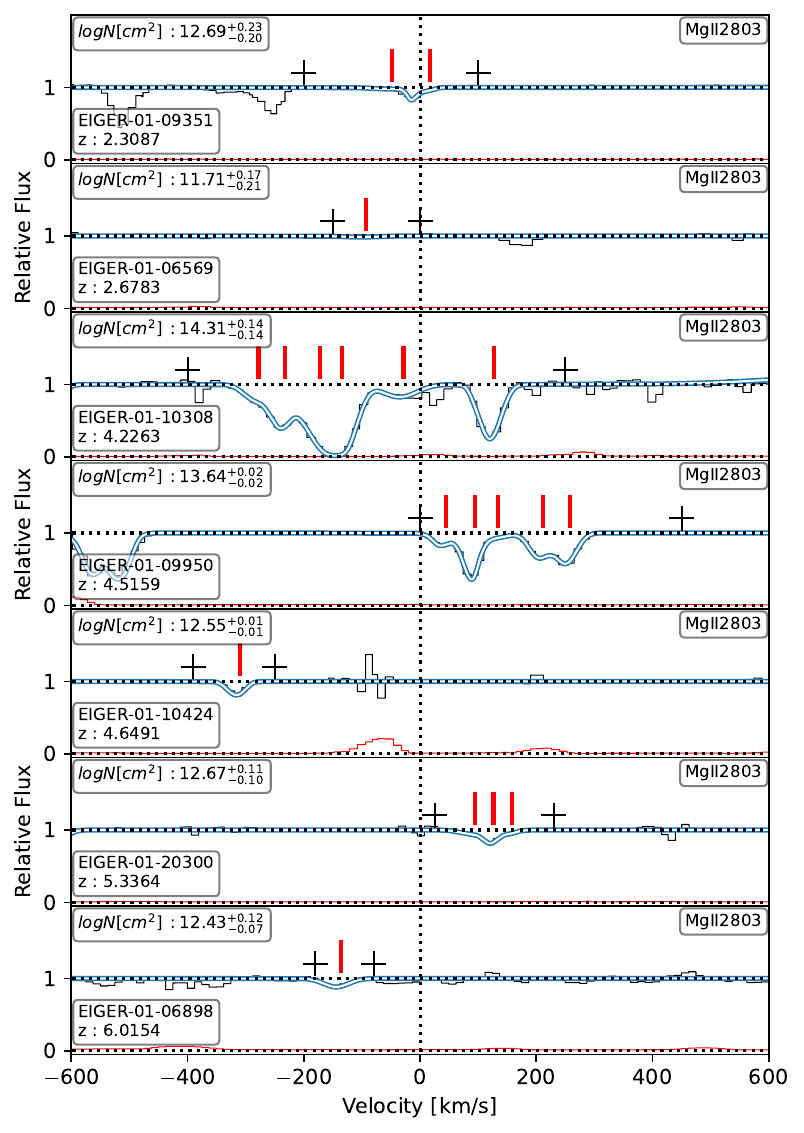}
\caption{Normalized quasar spectra of J0100+28002 showing the detected {\mgii} 2796 (left panel) and 2803 (right panel) absorption profiles along with their corresponding Voigt profile fits (blue lines). The vertical red ticks mark the individual Voigt profile components. For each system, the corresponding galaxy ID, redshift, and the integrated total column density are shown. The plus signs bracket the velocity range over which equivalent width is computed.
\label{fig:absorption_profile}}
\end{figure*}

We use a python-based Bayesian Markov chain Monte Carlo (MCMC) Voigt profile fitting toolbox \texttt{rbvfit} \citep{rongmon_bordoloi_2023_10403232} to perform simultaneous fits to the {\mgii} absorption doublet. This approach fits column density ($N$), Doppler $b$ parameter and velocity offset $v$ for each component simultaneously for each {\mgii} doublet. We assume flat priors on each of these parameters with reasonable physical bounds. The number of components and the initial guess of the velocity offsets are obtained via visual inspection of the data. The fitting procedure generates posterior distributions for the model parameters. We chose the median of each distribution as the best fit model parameters and the 16$^{th}$ and 84$^{th}$ percentile as the upper and lower bounds on the best fit parameters. One advantage of using a Bayesian MCMC approach over a frequentist $\chi$-squared minimization approach is that our approach yields marginalized posterior distribution of fitted parameters. This results in accurate column density estimates, even if simultaneously the Doppler $b$ parameters are not well constrained in moderate resolution spectrum used in this work. The {\mgii} absorption systems presented in this work, and the best fit Voigt profile parameters are reported in Table \ref{tab:vfit_info}, and Figure \ref{fig:absorption_profile}.

\begin{figure*}
\centering
\includegraphics[width=\columnwidth]{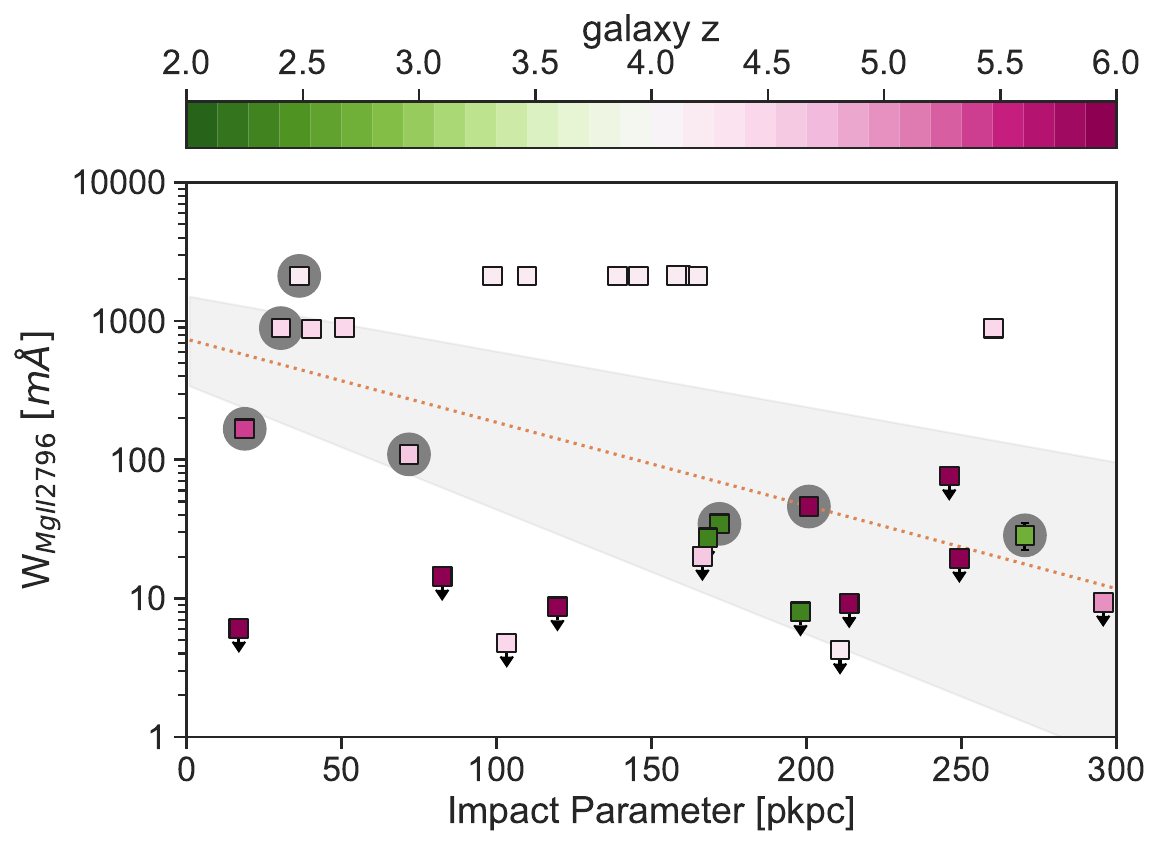}
\includegraphics[width=\columnwidth]{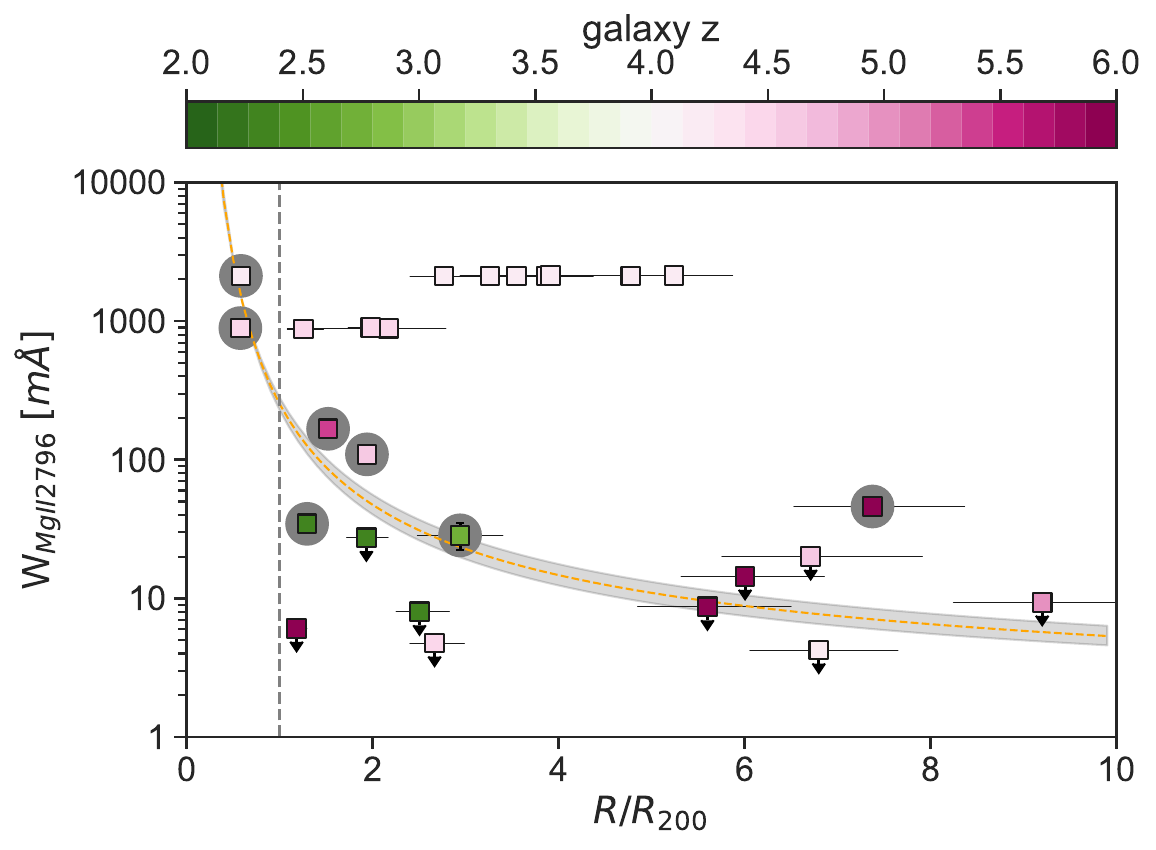}
\caption{{\mgii} radial absorption profile as a function of impact parameter (left panel), and normalized virial radius (right panel), respectively. On both panels, the squares represent detection, and the squares with downward arrows represent the 2-$\sigma$ upper limits of non-detection. Each square is color coded as a function of redshift of the galaxy. The gray circles show the closest galaxy associated with a {\mgii} absorber at a distinct redshift. The horizontal error bars show the uncertainty on the normalized virial radii. Gray shaded regions (left  panel) indicate the best fit {\mgii} absorption radial profile for $z\approx 2$ galaxies \citep{Dutta2021}. Galaxies associated with {\mgii} absorbers are seen out to 300 kpc but most of the associated galaxies are well beyond the inferred virial radii of the galaxies. Vertical dashed line (right panel) marks $R/R_{200}=$1. Dashed orange line (right panel) with shaded region, show the best fit power law to the EIGER {\mgii} radial profile normalized to the inferred virial radii of the galaxies.
\label{fig:radial_profile}}
\end{figure*}

\begin{figure*}
\centering
\includegraphics[width=\columnwidth]{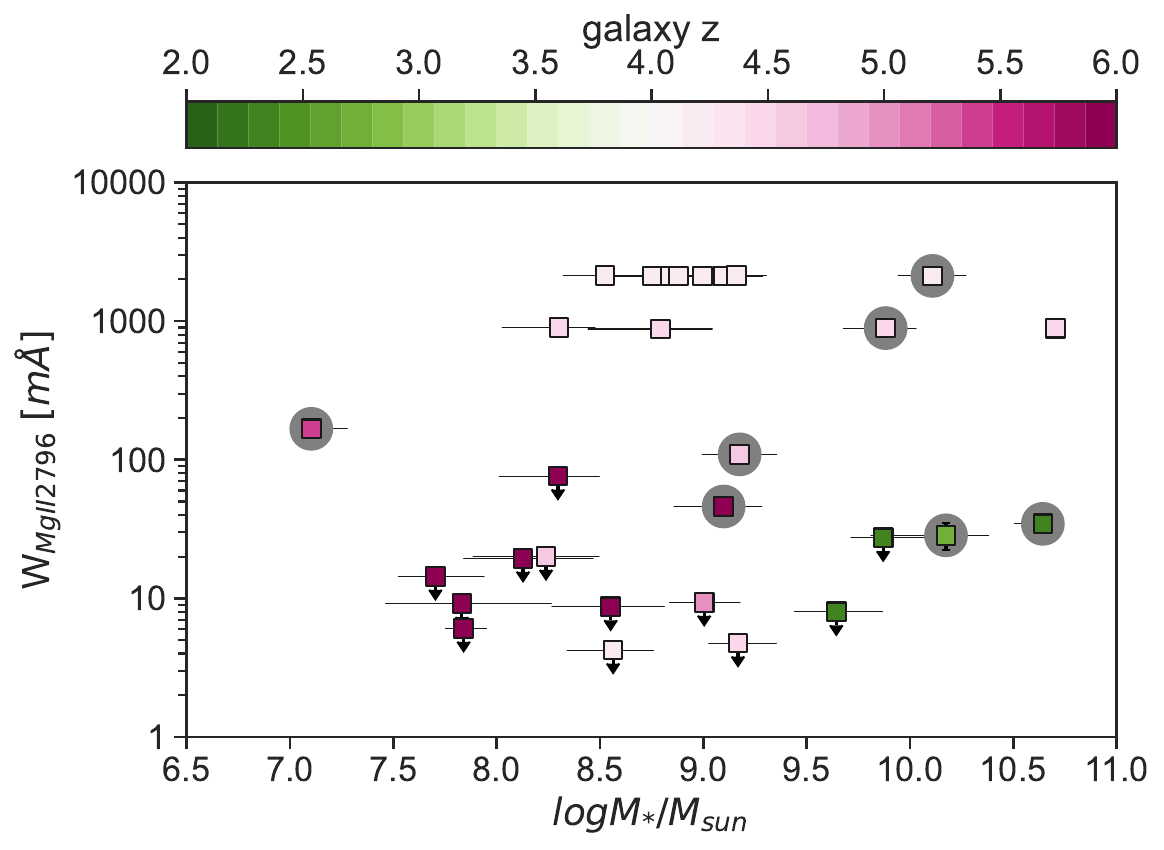}
\includegraphics[width=\columnwidth]{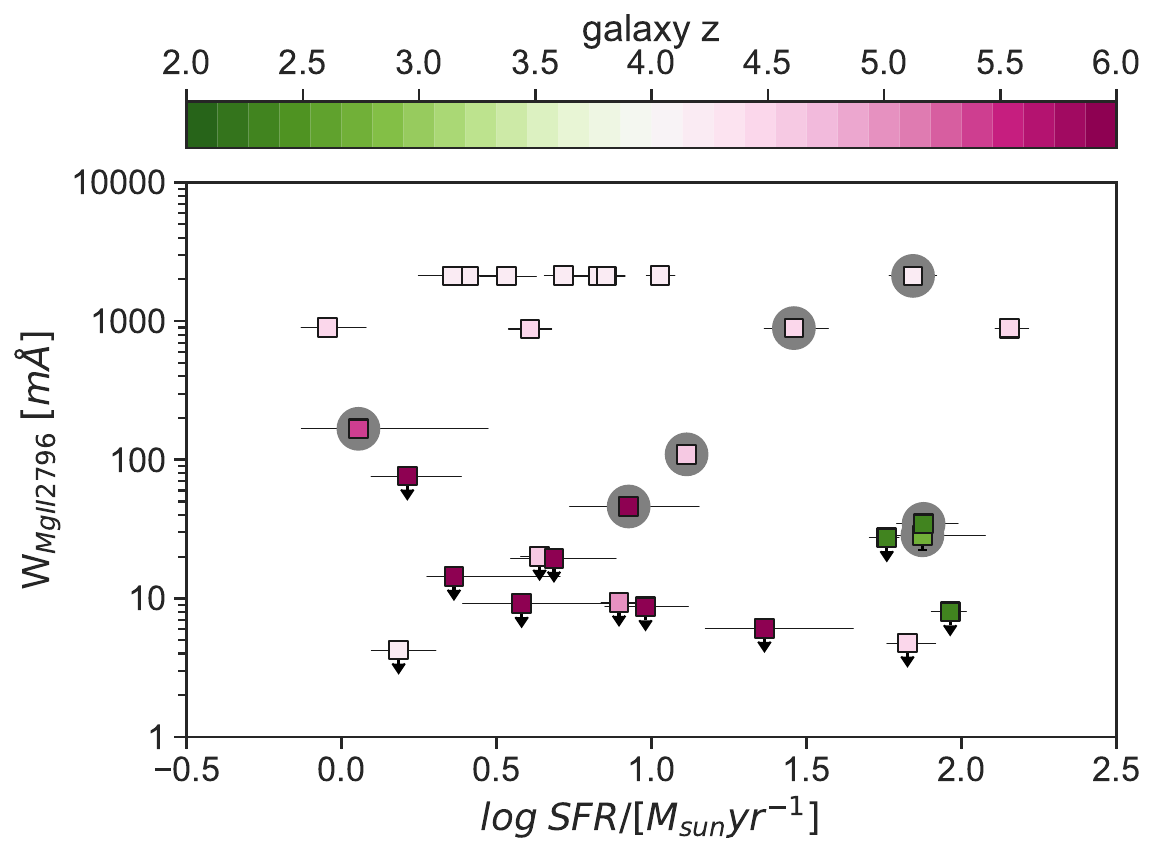}
\includegraphics[width=\columnwidth]{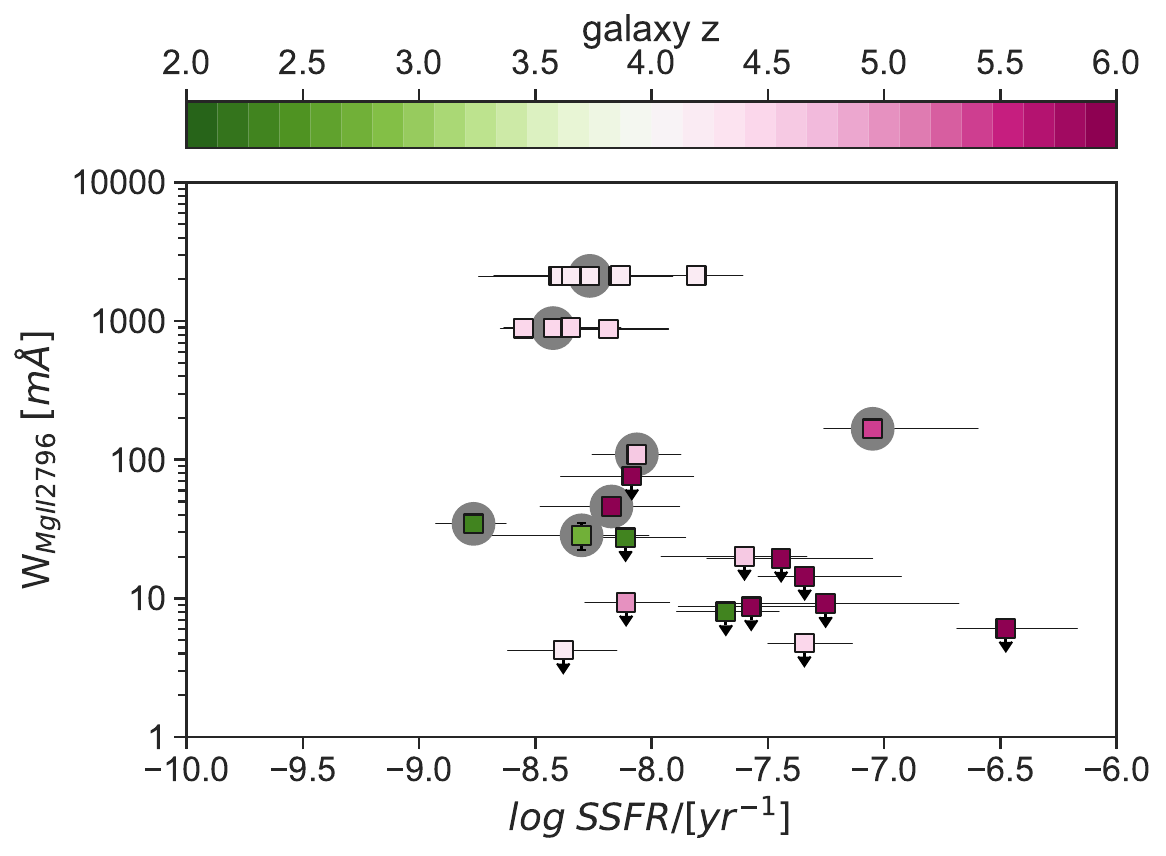}
\caption{Variation of {\mgii} absorption strength with galaxy stellar mass (top left panel), star-formation rate (top right panel) and specific star-formation rate (bottom panel) respectively. At each discrete absorber redshift, the closest galaxy is marked with the gray circle. Each galaxy is color coded to show its redshift. Horizontal error bars show the uncertainty on stellar mass, SFR and sSFR  respectively. The strongest {\mgii} absorption systems are detected near the most massive and most vigorously star-forming galaxies.
\label{fig:CGM_galaxy_property}}
\end{figure*}

\begin{figure*}
\centering
\includegraphics[width=\columnwidth]{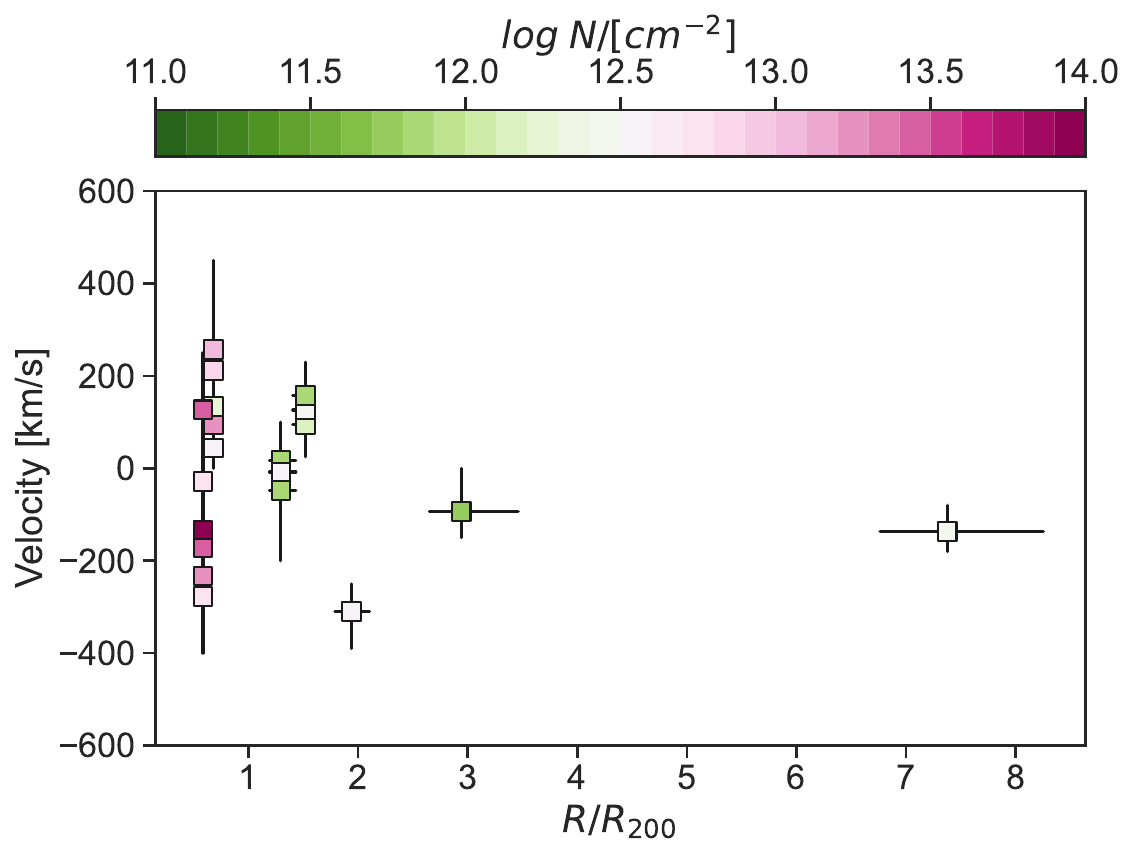}
\includegraphics[width=0.9\columnwidth]{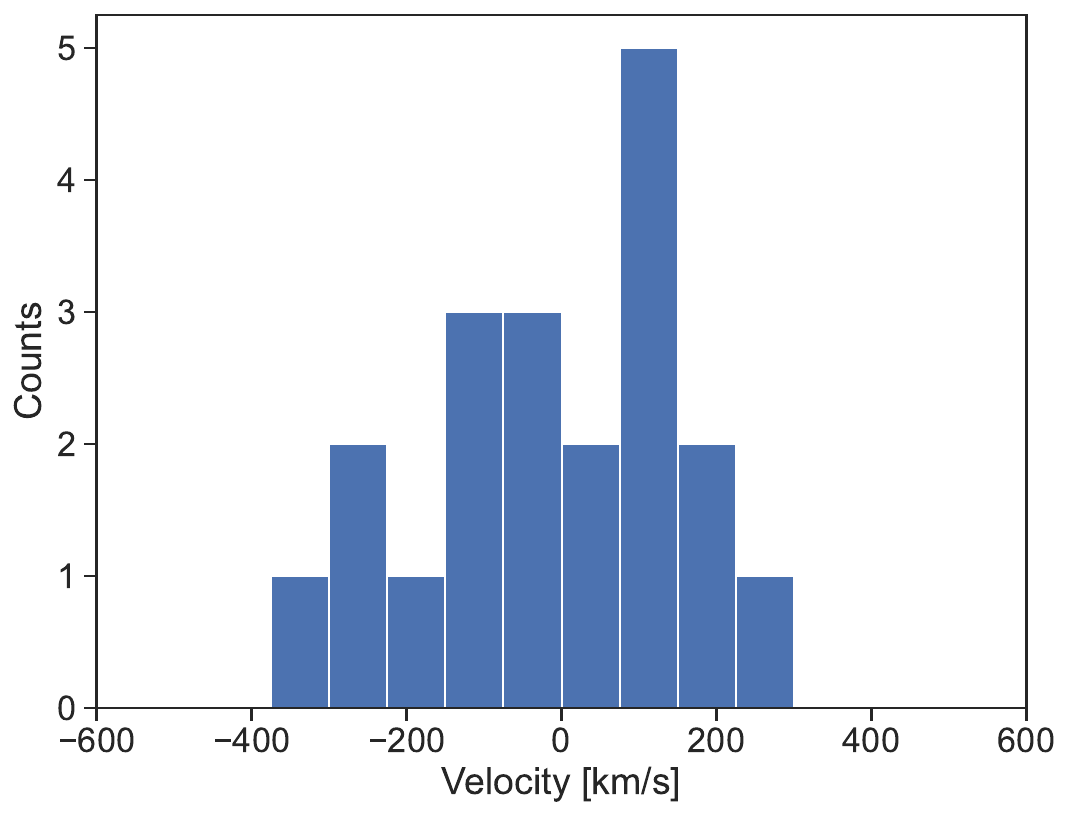} 
\caption{ \textit{Left Panel:} The {\mgii} absorption component velocity centroids with respect to the galaxies’ systemic redshifts as a function of impact parameter normalized to the virial radius of the galaxy. The vertical range bars indicate the maximum projected kinematic extent of {\mgii} absorption for each system. The horizontal bars show the uncertainty in $R/R_{200}$ estimates. Each component is color coded to show the Voigt profile fitted column density. {\mgii} absorption associated with EIGER-01-09950 is offset by 0.1$R/R_{200}$ along x-axis for visual clarity.
\textit{Right panel:} The distribution of
individual {\mgii} absorption components in each Voigt fit.  \label{fig:galaxy_kinematics}}
\end{figure*}

\begin{figure}
\centering
\includegraphics[width=\columnwidth]{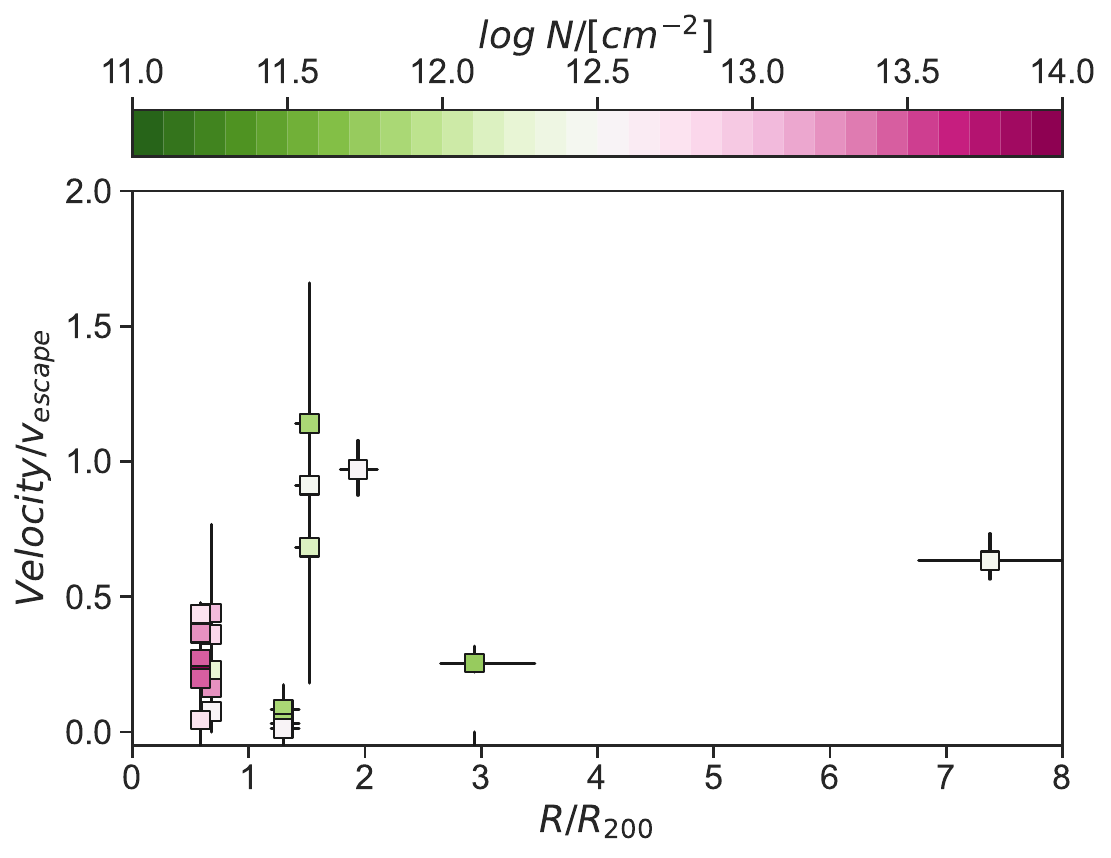}
\caption{Individual {\mgii} absorption component velocity normalized with the escape velocity of their associated galaxy at that $R$, as a function of $R/R_{200}$ of that galaxy. Components are color coded to show their column densities. The components that are at $R/R_{200} <1$ and with velocity/$v_{escape}\; <$1 are from two individual galaxies at very close impact parameters. For all other five galaxies, the {\mgii} absorption is consistent with not being bound to the host dark matter halos. {\mgii} absorption associated with EIGER-01-09950 is offset by 0.1$R/R_{200}$ along x-axis for visual clarity.
\label{fig:galaxy_kinematics_escape}}
\end{figure}

\begin{figure*}
\centering
\includegraphics[width=0.85\columnwidth]{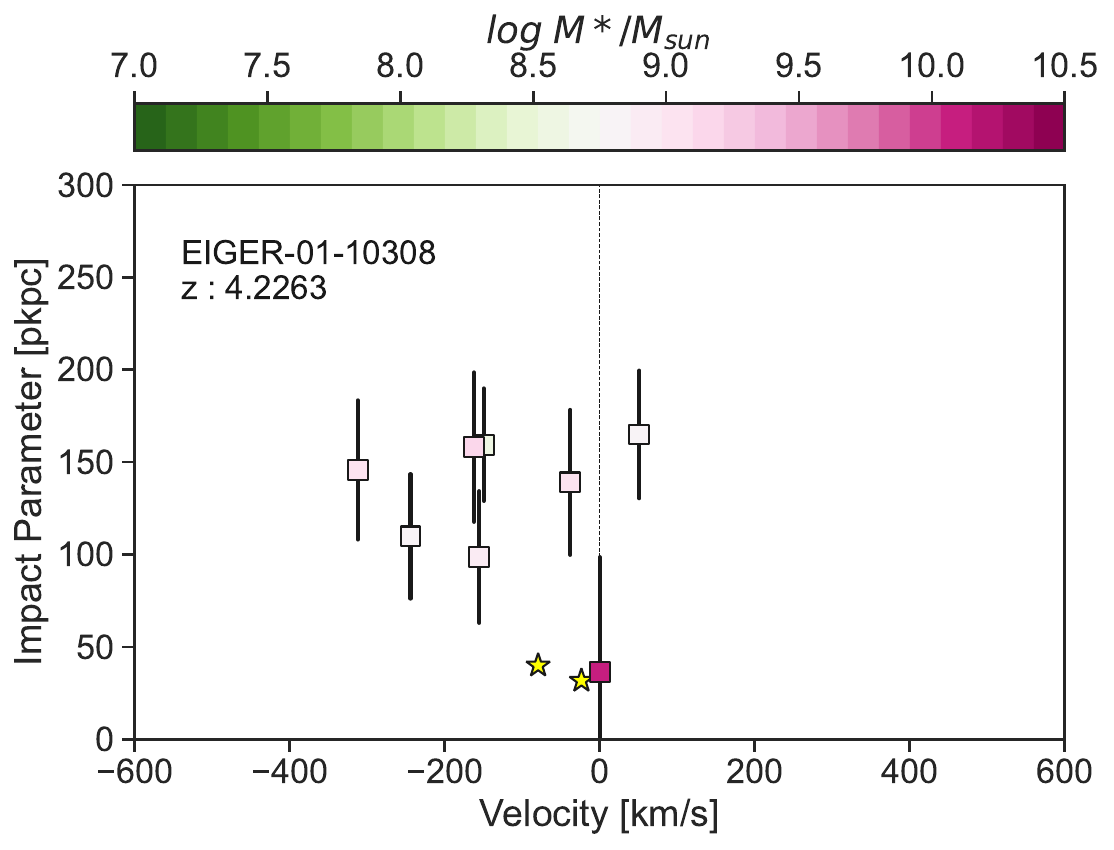}
\includegraphics[width=0.85\columnwidth]{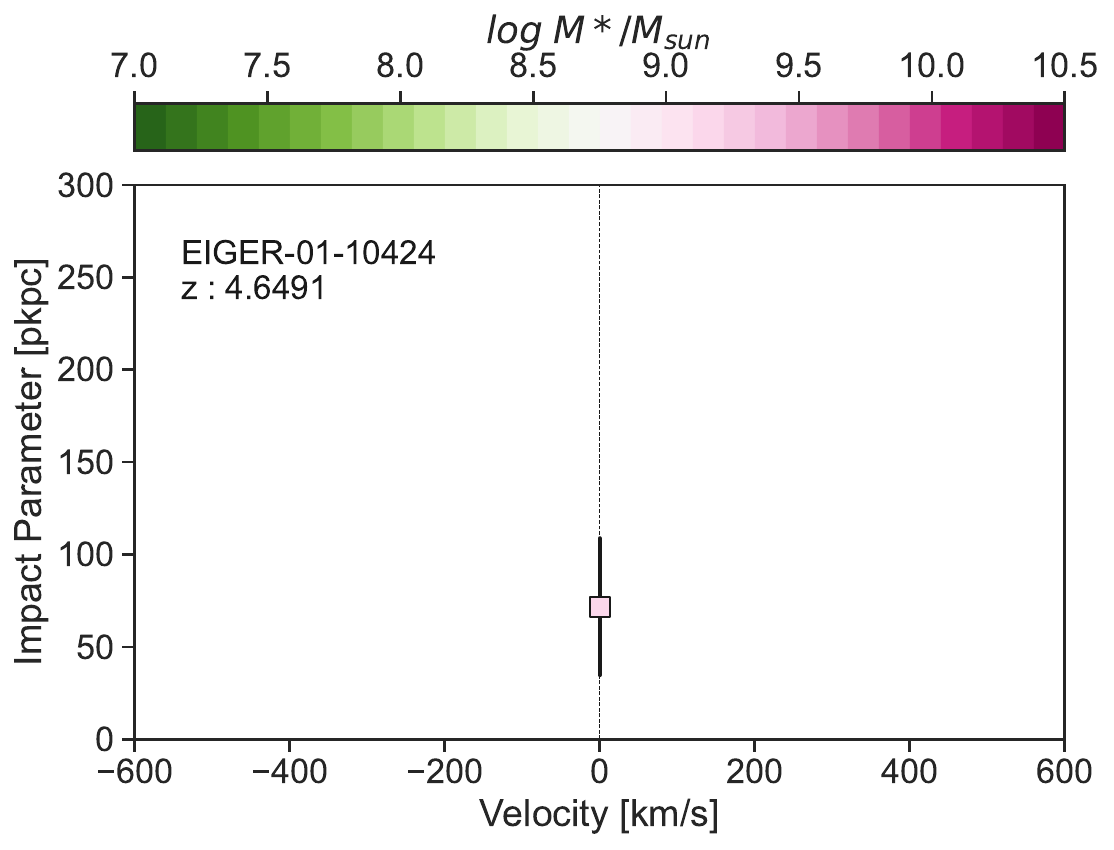}
\includegraphics[width=0.85\columnwidth]{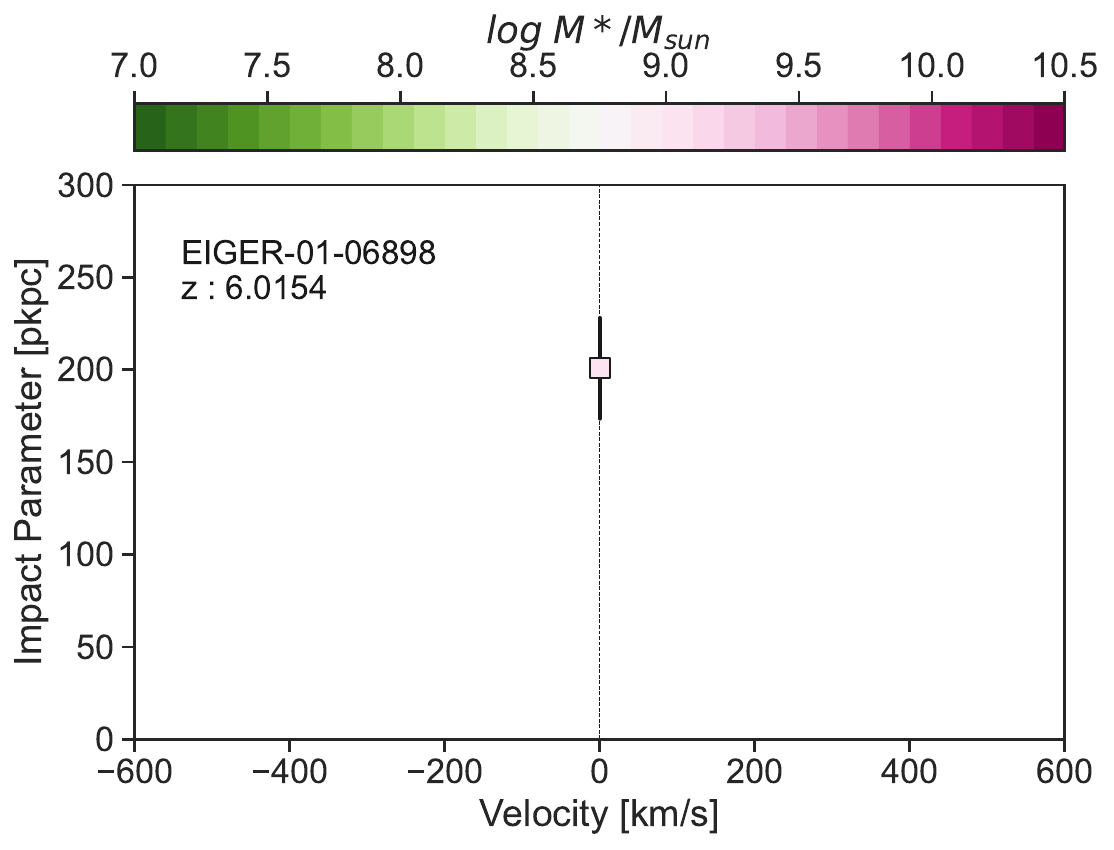}
\includegraphics[width=0.85\columnwidth]{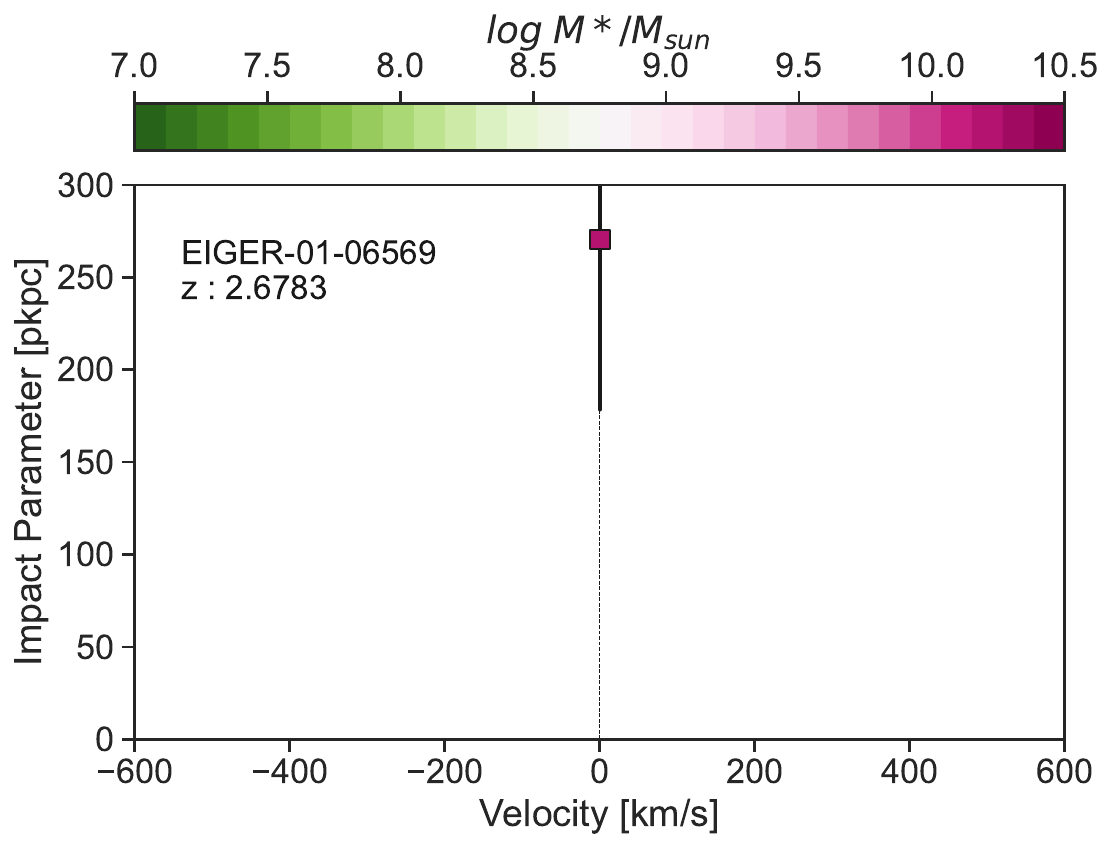}
\includegraphics[width=0.85\columnwidth]{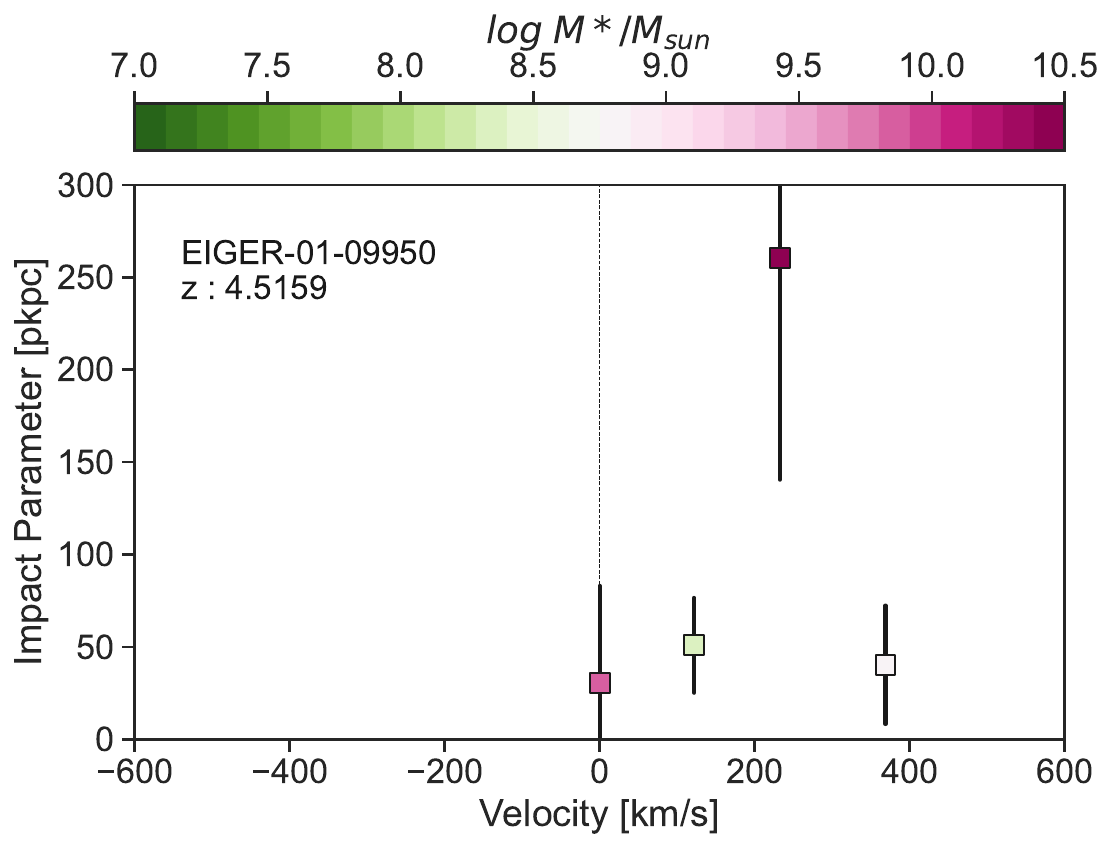}
\includegraphics[width=0.85\columnwidth]{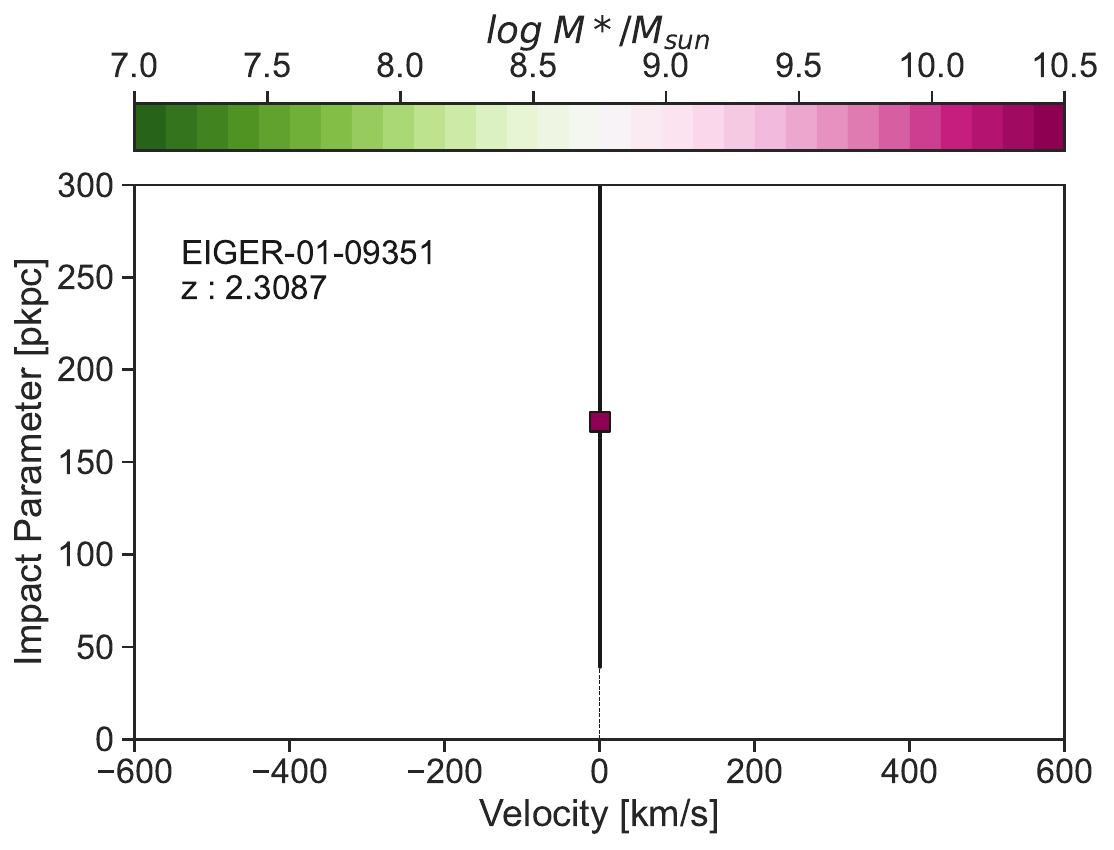}
\includegraphics[width=0.85\columnwidth]{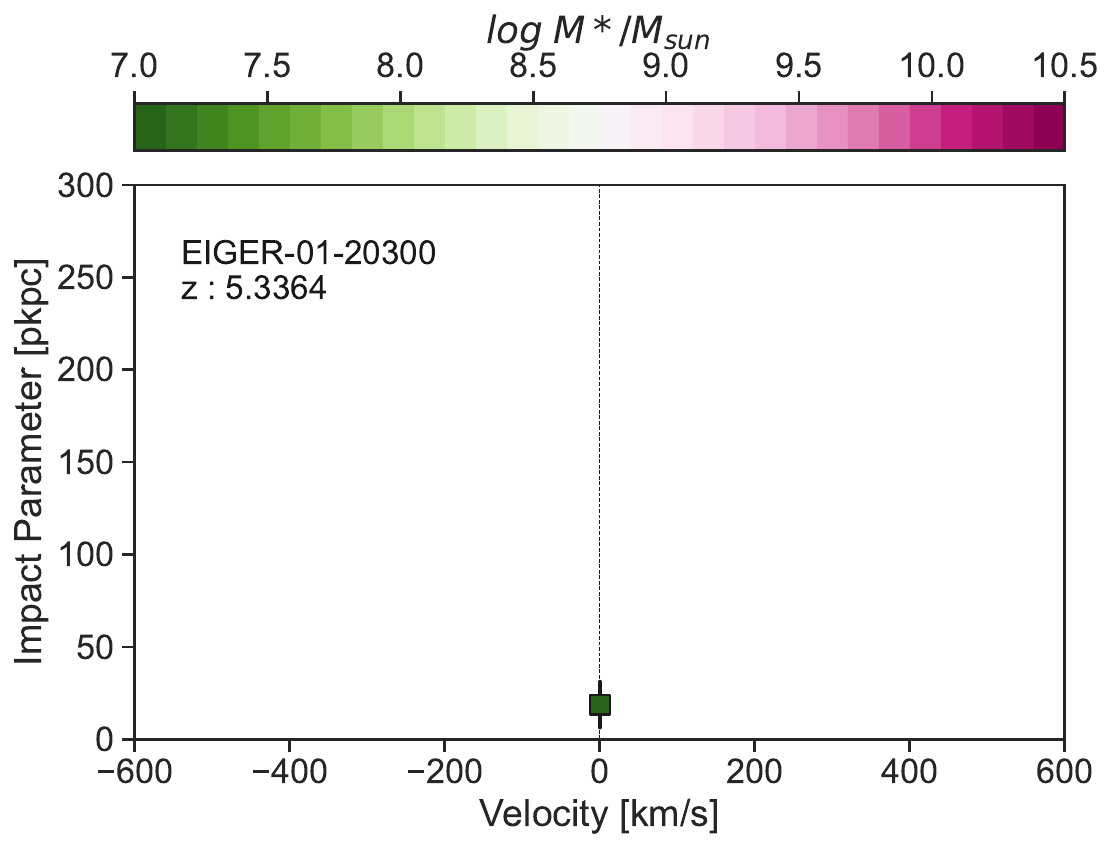}
\caption{Environment around galaxies at closest impact parameter to a {\mgii} absorption system, as a function of impact parameter. The vertical ranges mark the inferred virial radius of individual galaxies. Only two systems are within the virial radius of the associated galaxies. Galaxies are color coded as a function of their stellar masses. The gold stars in the top left panel show the velocity components of three distinct components that are merging to form the full system. Galaxy EIGER-01-10424, although marked as a single system, the imaging reveals two components are merging to form a single system. 
\label{fig:galaxy_environment}}
\end{figure*}

\section{Results}
In this section we present the variation of {\mgii} absorption strength with galaxy properties, their radial profiles, and the absorber kinematics.
\subsection{Distribution of {\mgii} absorption around galaxies}

We first characterize the spatial extent of {\mgii} absorption around the EIGER galaxies. We examine how the {\mgii} rest frame equivalent width ($W_{MgII2796}$) varies as a function of impact parameter ($R$) and the normalized virial radius ($R/R_{200}$) around the EIGER galaxies. Figure \ref{fig:radial_profile} shows the 1-D {\mgii} radial absorption profile around 29 galaxies as function of $R$ (left panel) and $R/R_{200}$ (right panel), respectively. Seven unique {\mgii} absorption system associated with host galaxies are detected and the galaxy at the closest impact parameter to the absorber is assigned as the host galaxy (gray circles). No absorption is detected around 12 galaxies and are marked with downward arrows. These measurements show the 2-$\sigma$ limit on non-detection. Seven  galaxies associated with non-detection are at $R<200$ kpc, suggesting that {\mgii} absorption is patchy at $z>4$. All galaxies are color coded as a function of their redshifts. In particular, for {\mgii} absorption systems at $z \sim$ 4, multiple galaxies are detected within $\pm$400 {\kms} of the absorber redshift. All these associated galaxies are also presented in Figure \ref{fig:radial_profile}. We will discuss the environment of {\mgii} absorbers in the next Section. The error bar on x-axis (right panel) denotes the uncertainty on $R_{200}$ estimates.

Focusing on the closest galaxy (gray circles), two immediate observations stand out in Figure \ref{fig:radial_profile}, left panel. The strongest absorbers are detected at close impact parameters ($R <$ 100 kpc) of their host galaxies. Further, there are several absorbers detected at high impact parameters ($R>$ 150 kpc). These absorbers are at the outer edge of the radial profile for {\mgii} absorption systems observed at $z\approx $2 (gray shaded region, \citealt{Dutta2021}). However, this trend is not taking into consideration that these galaxies all have different masses and virial radii, or the possibility that even fainter galaxies below our detection limit exist at closer distances.

False-color \textit{JWST}/NIRCam  F115W/F200W/F356W broad band images of each galaxy is presented in Figure \ref{fig:galaxy_stamps}. Individual galaxies exhibit diverse morphology. Galaxies at $z<3$, typically show well-formed disks and prominent inner bulge. Most of the $z>3$ galaxies exhibit complex morphology, with several of them exhibiting tidally disturbed features and several individual knots indicating either merger or discrete star-formation events along the galaxy disk. In particular, the stamp of galaxy EIGER-01-10308 stands out (Figure \ref{fig:galaxy_stamps}) as a merging galaxy with tidal streams clearly visible within 2 arcsec of the galaxy. This galaxy is associated with the strongest {\mgii} absorption system reported in this work (Appendix \ref{fig:appendix1}), and we discuss this system in detail in the next section.

As a fraction of their inferred virial radii (Figure \ref{fig:radial_profile}, right panel), most {\mgii} absorption extends out to $\sim$2-3 $R/R_{200}$ of the host galaxies. This spatial extent is much larger than typically observed around galaxies at $z<1$ \citep{Chen2010,Bordoloi2011,Churchill2013,Nielsen2013,Werk2013}. {\mgii} absorption strength also show a steep decline with $R/R_{200}$.  We quantify this radial fall off, by considering only the closest galaxies with detected {\mgii} absorption, and fit a power law: $\log W_{MgII2796}\; = \; (2.403 \pm 0.046) \times (R/R_{200})^{-0.52 \pm 0.034}$. The best fit power law with 68\% confidence interval is presented as the dashed line with gray shaded region in Figure \ref{fig:radial_profile}, right panel). When non-detections are included in the fit, the radial profile becomes slightly steeper but not too different than the previous fit ($\log W_{MgII2796}\; = \; (2.201 \pm 0.046) \times (R/R_{200})^{-0.59 \pm 0.028}$). We note that $71^{+13}_{-19}$\% (5/7) of the {\mgii} absorbers are detected outside the inferred virial radii of the host galaxies.

Figure \ref{fig:CGM_galaxy_property} presents the variation of {\mgii} absorption as a function of stellar mass (top left panel), SFR (top right panel) and specific star-formation rate (sSFR; bottom panel)  respectively. The symbols are color coded to be consistent with Figure \ref{fig:radial_profile}. We note that 86$^{+9}_{-18}$\% (6/7) of the detected {\mgii} absorption systems are associated with higher stellar mass galaxies ($\log M_{*}/M_{\odot} >9$), with strongest absorption systems associated with the highest stellar mass systems.  The two highest redshift {\mgii} absorption systems are associated with lower mass $\log M_{*}/M_{sun} \approx$ 7.1 ($z\sim 5.33$) and $\log M_{*}/M_{sun} \approx$ 9.1 ($z\sim 6.01$) galaxies, respectively. Further, the strongest {\mgii} absorption systems are associated with galaxies with the highest SFR. These results tentatively suggest a correlation between the absorption strength of these strong high-$z$ {\mgii} absorption systems and star-formation activity in their host galaxies, similar to what is observed for strong {\mgii} absorption at $z\sim 1$ galaxies \citep{Bordoloi2011}. However, the variation of  {\mgii} absorption with sSFR of the host galaxies (Figure \ref{fig:CGM_galaxy_property}; bottom panel) doesn't show a strong trend with increasing sSFR. Although the strongest {\mgii} absorption is detected around the most massive and vigorously star-forming galaxies; these are not the galaxies with the highest sSFR in this sample. This suggests that within this highly star-forming galaxy sample; galaxies with higher stellar masses host the strongest {\mgii} absorption. We add the caveat that the sample presented in this work, exclusively consist of high-star forming galaxies ($\log \; sSFR > -9$). Therefore, detailed comparison of galaxy sSFR and CGM absorption would require a sample of galaxies with $log sSFR < -9$. This will be explored in a future work with the once the full EIGER survey is complete (Bordoloi et al. in prep). A unique facet of these high-$z$ {\mgii} absorption systems is that majority of them are beyond the inferred virial radii of their host galaxies. At these distances the {\mgii} absorbing gas may not be bound to the gravitational potential of their host galaxies. We explore this in the next section.

\subsection{CGM kinematics and environments at high-\textit{z}}

In this section we focus on the kinematics and environment of the seven observed {\mgii} absorption profiles and discuss whether the observed absorption profiles are consistent with being bound to the dark matter halo of the host galaxies. We obtain the {\mgii} column densities and the best fit Voigt profiles as described in Section 3.3. Figure \ref{fig:absorption_profile} shows the best fit absorption profiles for both {\mgii} transitions. Different panels show {\mgii} absorption associated with each system. The vertical red ticks mark the position of individual Voigt profile components. We use these components to quantify the velocity distribution of the {\mgii} absorption systems.

Figure \ref{fig:galaxy_kinematics}, left panel shows the velocity centroids of individual Voigt profile fitted {\mgii} absorption components as a function of $R/R_{200}$ from their respective host galaxies. The symbols are color coded to show the {\mgii} column density of each component. The vertical range bars show the velocity range over which the equivalent width of the system is calculated. They are effectively the full width at zero optical depth of each {\mgii} absorption system. Both the thermal and bulk motion associated with the {\mgii} absorption systems are incorporated within the full velocity widths and therefore represents the maximum projected velocity of the absorption systems. The horizontal range bars represent the uncertainty in estimating $R/R_{200}$. Figure \ref{fig:galaxy_kinematics}, right panel shows the distribution of {\mgii} absorption components relative to the host galaxy redshift. The distribution of absorption components show a large velocity spread from -400 {\kms} to 300 {\kms}. The absorption component velocities are offset from the systemic zero velocities with a median absolute velocity of 135 {\kms} and a standard deviation of 85 {\kms}, respectively. This is different than what is observed at the CGM of low-$z$ galaxies, where most of the CGM absorption systems cluster around the systemic zero velocity of their host galaxies and their velocities are almost always consistent with being less than the associated virial velocities \citep{Tumlinson2013,Bordoloi2014a,Huang2016}.

We further investigate if the {\mgii} absorption detected around EIGER galaxies are consistent with being bound to the dark matter halos of the host galaxies in Figure \ref{fig:galaxy_kinematics_escape}. We present the {\mgii} absorption component velocities normalized to the escape velocity associated with the host galaxy at that impact parameter as a function of $R/R_{200}$. The horizontal error bars represent the uncertainty associated with the inferred $R_{200}$. The vertical range bars denote the velocity range over which equivalent width of the system is calculated (normalized to the escape velocity of the system). It is clearly seen that 5 out of 7 {\mgii} absorption systems are detected at $R> 1.3R_{200}$. Two absorption components have velocities higher than the projected escape velocities of these systems.  This suggests that these absorption systems are not consistent with being  bound to the dark matter halos of the host galaxies. Only two absorption systems (galaxies: EIGER-01-10308, EIGER-01-09950) are at $R < R_{200}$, and their component velocities are less than escape velocities associated at these impact parameters. Only these two $z\sim 4$ systems are consistent with being bound to the dark matter halo of their host galaxies. These two systems are also associated with higher galaxy over-densities around them (Figure \ref{fig:galaxy_environment}). Looking at both the absorption component kinematics and the $R/R_{200}$ distribution of these absorption systems, we conclude that the CGM kinematics at high-$z$ is significantly different than what is observed at $z < 1$. At low-$z$ bulk of the CGM absorption systems are consistent with being bound to the dark matter halos of their host galaxies, unlike the CGM of EIGER galaxies. This suggests an evolution in CGM gas kinematics as a galaxy evolves from early Universe to today. In the early Universe, the CGM gas could easily escape from individual galaxy halos and chemically enrich the IGM around galaxies. But as the galaxies became larger at low-$z$, the CGM becomes increasingly bound to the host galaxies. 

We finally explore the environment around the seven EIGER CGM host galaxies and quantify if these are isolated host galaxies or if they have companion galaxies. Figure \ref{fig:galaxy_environment} show the impact parameter to each galaxy at the systemic redshift of the galaxy noted in each panel. We plot all galaxies within 300 kpc from the  J0100+2802 sightline, and within $\pm$600 {\kms} of the host galaxies. Each galaxy is color coded as a function of its stellar mass. The vertical range bars show the associated $R_{200}$ of each galaxy. We describe the environment of each system below.

\textbf{EIGER-01-10308:} This galaxy at $z\sim$ 4.22, is a tidally disturbed merging system. The tidal streams and different components of the merger can be clearly seen in high-resolution \textit{JWST} imaging (Appendix \ref{fig:appendix1}, panel a). The merging system has an integrated stellar mass of $\log M_{*}/M_{sun} =$ 10.17, and a SFR $\sim$ 70 {\msunyr}. We extract individual NIRCam spectra of the two smaller merging components (Appendix \ref{fig:appendix1}, panel b) and compute their individual redshifts.  These two components are at $\sim$ -23 and -80 {\kms} from the main galaxy, respectively. These individual components are marked as gold stars in Figure \ref{fig:galaxy_environment}, top left panel. The galaxy EIGER-01-10308 resides in a local over-density and there are seven additional galaxies within 200 kpc from the quasar sightline. The galaxies are within 220 kpc of each other and have a velocity dispersion of 228 {\kms}. This large over-density of galaxies at close physical and kinematic separation may be part of a proto-group at $\langle z \rangle \approx$ 4.2234. The additional seven galaxies are at higher impact parameter than EIGER-01-10308 and at $R > R_{200}$. The {\mgii} absorption associated with EIGER-01-10308 show complex kinematics with six distinct individual absorption components identified with a velocity spread of $\approx$ 500 {\kms} (Figure \ref{fig:absorption_profile}). Absorption is also detected in {\feii}, {\mgi}, {\aliii} and {\siii} transitions (Appendix \ref{fig:appendix1}, panel c). The galaxy is at low impact parameter with $R/R_{200} <1$ and the velocity of the absorption components are less than the escape velocity associated at this impact parameter. This system is one of the two absorption systems in this work, that has {\mgii} absorption kinematics consistent with being bound to the dark matter halo of the host galaxy.

\textbf{EIGER-01-09950:} At $z\sim$ 4.5, there are three additional galaxies within 300 kpc of this galaxy (Appendix \ref{fig:appendix2}). The galaxy EIGER-01-09950 has a stellar mass of $\log M_{*}/M_{sun} =$ 9.88, and a SFR $\sim$ 29 {\msunyr}, respectively. There are two other galaxies at very close kinematic separations: EIGER-01-09078 at a separation of 16 kpc from EIGER-01-09950 and a velocity offset of 368 {\kms}, and galaxy EIGER-01-08811, at a separation of 22 kpc from EIGER-01-09950 and a velocity offset of 122 {\kms}. Both galaxies EIGER-01-09078 and EIGER-01-08811 are 40 and 50 kpc away from the quasar sightline J0100+2802. A third more massive galaxy (EIGER-01-06193) is detected at 235 kpc away from EIGER-01-09950 and at a velocity separation of 232 {\kms}. Galaxy EIGER-01-06193 is at an impact parameter of 260 kpc from the quasar sightline J0100+2802. These galaxies all lie close to each other (within 232 kpc and 166 {\kms}) and could form a proto-group at $z\sim 4.5192$. The {\mgii} absorption profile is again complex for this system (Figure \ref{fig:absorption_profile}), with five identified distinct absorption components. The absorption spans a velocity range of $\approx$ 350 {\kms}. We also detect absorption in {\feii}, {\mgi}, {\aliii} and {\siii} transitions in this system (Appendix \ref{fig:appendix2}). The strongest {\mgii} absorption component is offset from the systemic redshift galaxy EIGER-01-09950 by $\sim$ 95 {\kms}, and kinematically lines up with the nearby galaxy EIGER-01-09078. But EIGER-01-09078 has a much lower stellar masses (see Table \ref{tab:galaxy_info}), and the impact parameter to it is higher than the inferred virial radius associated with it (Figure \ref{fig:galaxy_environment}). Since the galaxy EIGER-01-09950 is the closest galaxy to the line of sight, and kinematically the projected {\mgii} velocity is lower than the escape velocity of the host galaxy, we conclude that the {\mgii} absorption is consistent with being bound to the dark matter halo of the host galaxy.

\textbf{EIGER-01-10424:} At $z\sim$ 4.64, a faint galaxy is detected at an impact parameter of 72 kpc from the J0100+2802 quasar line of sight. This galaxy has a stellar mass of $\log M_{*}/M_{sun} =$ 9.18, and a SFR $\sim$ 13 {\msunyr}, respectively. The galaxy is next to a bright $z\sim$ 1 foreground galaxy detected in Pa-$\alpha$ emission (see Figure \ref{fig:galaxy_stamps}). Ground based MUSE spectrum of the bright foreground galaxy show [OII] emission doublet, confirming it as a low-$z$ galaxy. There are two faint emission components detected for the target galaxy, suggesting that it is a merging system at $z\sim$ 4.64, however, owing to the position angle of the NIRCam grism spectra, the two components cannot be spatially resolved. We note that since the emission line redshift for this system is estimated from a single H-$\alpha$ emission line, it is possible that this emission is associated with the foreground bright galaxy at $z\sim$ 1. However, ground based seeing does not allow us to check for [OII] emission associated with the fainter components in the MUSE datacube. The presence of metal absorption and the emission lines in the NIRCam grism spectra has led us to conclude that EIGER-01-10424 is the host galaxy for {\mgii} absorption detected at $z\sim$ 4.6.  This system only shows weak {\mgii} absorption doublet, kinematically offset from the host galaxy's systemic redshift by $\sim$ -350 {\kms} (Figure \ref{fig:absorption_profile}), and beyond the inferred virial radius of the host galaxy (Figure \ref{fig:galaxy_environment}). We therefore conclude that the CGM absorption is not consistent with being bound to the host galaxy.

\textbf{Galaxies EIGER-01-06898, EIGER-01-20300:} These galaxies are associated with {\mgii} absorption at $z\sim$ 6.015 and 5.33, respectively. EIGER-01-06898 is at an impact parameter of 201 kpc and has a stellar mass of $\log M_{*}/M_{sun} =$ 9.1, and a SFR $\sim$ 8.4 {\msunyr}, respectively. EIGER-01-20300 is at an impact parameter of 19 kpc and has a stellar mass of $\log M_{*}/M_{sun} =$ 7.1, and a SFR $\sim$ 1.1 {\msunyr}, respectively. Both these galaxies are ``isolated", in the sense that no companion galaxy within $\pm$600 {\kms} and 300 kpc of them is detected. Both galaxies are at an impact parameter beyond the $R_{200}$ radii of their host galaxies (Figure \ref{fig:galaxy_environment}). For EIGER-01-06898, we detect {\mgii} and {\civ} absorption offset from the systemic redshift of the galaxy by $\approx$ -136 {\kms}.  Around EIGER-01-20300, we detect {\mgii}, {\civ}, {\siiv} and {\siii} absorption, kinematically offset from the systemic redshift of the host galaxy by $\approx$ 120 {\kms}. We will report and quantify the high-ionization {\civ} absorption profiles in an upcoming publication (Simcoe et al. in prep). For both these systems, the absorption is kinematically offset from the systemic galaxy redshift and is detected beyond the inferred virial radii of their corresponding host galaxies. We therefore conclude that the {\mgii} absorption detected around these galaxies is not bound to the host galaxy's dark matter halo.

\textbf{Galaxies EIGER-01-06569, EIGER-01-09351:} These two galaxies are associated with {\mgii} absorption at $z\sim$ 2.67 and 2.31, respectively. The galaxy 6569 is at an impact parameter of 270 kpc and has a stellar mass of $\log M_{*}/M_{sun} =$ 10.17, and a SFR $\sim$ 75 {\msunyr}, respectively. EIGER-01-09351 is at an impact parameter of 172 kpc and has a stellar mass of $\log M_{*}/M_{sun} =$ 10.64, and a SFR $\sim$ 75 {\msunyr}, respectively. In both these systems, {\mgii} absorption doublet is detected as any line with bluer rest frame wavelength will be in the Ly-$\alpha$ forest and not observable along this high-$z$ quasar sightline. Both these {\mgii} systems are weak absorption systems (Figure \ref{fig:absorption_profile}) and are beyond the inferred virial radii of their host galaxies. In both cases, the {\mgii} absorption is consistent with not being bound to the dark matter halos of the host galaxies.

In all the detected systems, only two galaxies EIGER-01-10308 and EIGER-01-09950 have associated {\mgii} absorption consistent with being bound to the host galaxy dark matter halos. In both cases, there is a galaxy overdensity suggesting that these galaxies reside in two galaxy proto-groups. In all other cases, where no galaxy overdensity is seen, most of the {\mgii} absorption is consistent with not being bound to the dark matter halo of their host galaxies. This is significantly different than the cool CGM detected at $z<1$, where most of the CGM gas is consistent with being bound to their host galaxies.

\section{Discussion and Conclusions}

The commissioning of \textit{JWST} has opened a new discovery space to study the circumgalactic medium of high-$z$ galaxies. In this work, we present the deep NIR (3.5$\mu$m) WFSS \textit{JWST} NIRCam spectroscopic observations of the $z\sim$6.33 quasar field J0100+2802 from the EIGER survey to characterize the cool CGM (traced by {\mgii} absorption) around 29 $2.3<z<6.3$ galaxies. The \textit{JWST} WFSS spectroscopy is accompanied by deep \textit{JWST}/NIR and HST/Optical broad band imaging and deep ground based high resolution spectroscopy of the quasar. This work builds on the initial EIGER survey papers that characterized the properties of a large sample of [OIII] emitting galaxies at $z=$5.33--6.93 \citep{Kashino2023,Matthee2023}. Our main conclusions are summarized as follows:

\begin{itemize}
    \item Using \textit{JWST}/NIRCam 3.5$\mu$m grism spectroscopy, we discover 29 galaxies within 300 kpc from the quasar sightline J0100+2802 in three redshift windows $2.3 <z<2.7$, $4<z<5.1$, and $5.3<z<6.3$, respectively. Accurate spectroscopic redshifts are measured using strong rest frame optical emission lines (e.g., [SIII], He-I 10830, Pa-$\gamma$, H-$\alpha$, H-$\beta$, [OIII]).

    \item The galaxies span a stellar mass range of $7.1 \leq \log M_{*}/M_{sun} \leq 10.7$, and exhibit strong correlation between star-formation rates and stellar mass of the galaxies. All the galaxies presented in this work are star-forming.

    \item Galaxies identified show a diverse morphology, from tidally disturbed mergers, to well-formed disks. Most of the $z>3$ galaxies show complex morphology of either several clumps or tidally disturbed features.

    \item We identify the CGM host galaxies of seven {\mgii} absorption systems within an impact parameter of 300 kpc. Identifying the closest galaxy to the quasar line of sight as the host, we find that strongest {\mgii} absorption is detected at close impact parameters ($R< 100$ kpc). The {\mgii} absorption strength drops off as a function of galactocentric radii from the host galaxies, characterized by a power law fall off. This radial fall off is slightly shallower than the {\mgii} radial absorption profile observed for $z<1$ galaxies.

    \item There are 12 galaxies within 300 kpc, for which no {\mgii} absorption is detected at a mean detection threshold of 10-20 m{\AA}. This shows that at high-$z$, cool CGM traced by {\mgii} absorption is patchy.

    \item The {\mgii} absorption radial profile normalized to the host galaxy virial radius ($R/R_{200}$), show a steep decline with impact parameter. The radial profile is quantified as a power law fit: $\log W_{MgII2796} = (2.403 \pm 0.046) \times (R/R_{200})^{-0.52 \pm 0.034}$. Most of the CGM host galaxies are detected at $R <2-3 R_{200}$. 

    \item 71$^{+13}_{-19}$\% (5/7) of the {\mgii} absorption systems are detected outside the virial radii of their host galaxies. Two {\mgii} absorption systems are detected within the virial radii of their host galaxies, and both these galaxies reside in local galaxy overdensities. 

    \item 86$^{+9}_{-18}$\% (6/7) of the {\mgii} absorption systems are detected around host galaxies with $\log M_{*}/M_{sun} >$9, with strongest absorption systems associated with the highest stellar mass systems. The two $z>6$ {\mgii} absorption system are associated with lower mass $\log M_{*}/M_{sun} \approx$7 and $\log M_{*}/M_{sun} \approx$9 galaxies, respectively. Similarly, strongest {\mgii} absorption systems are also associated with the most star-forming galaxies.

    \item The {\mgii} absorption kinematics is not symmetrically clustered around the systemic zero velocity of their host galaxies. The absorption components velocities have a large velocity spread (from -400 {\kms} to 300 {\kms}) around the systemic redshift of the host galaxies. The absorption components show a median absolute velocity of 135 {\kms} and a standard deviation of 85 {\kms}.

    \item Five out of the seven {\mgii} absorption systems are associated with host galaxies at $R>1.3R_{200}$. Moreover, two absorption components show projected velocities higher than the escape velocity of the host galaxies. We conclude that five out of seven absorption systems have cool CGM gas, consistent with being unbound to their host dark matter halos. 

    \item We highlight the CGM around two particular {\mgii} absorption systems ($z\sim$ 4.2 and 4.5) because they are associated with host galaxies at $R < R_{200}$, and with absorption gas kinematics consistent with being bound to the dark matter halos of the host galaxies. Both these $z\sim$ 4.22 and $z\sim$ 4.5  {\mgii} absorption systems exhibit complex kinematics spanning $\approx$ 500 {\kms} and 350 {\kms}, respectively. 

    \item The {\mgii} absorption system at $z\sim$ 4.22 is associated with a morphologically disturbed merging galaxy with three distinct merging components within 80 {\kms} of each other. This galaxy is within a local galaxy over-density where seven additional galaxies are observed within 200 kpc of the quasar sightline and within $\pm$500 {\kms} of the host galaxy. These galaxies might be part of a galaxy proto-group at $z\sim$ 4.22.

    \item The {\mgii} absorption system at $z\sim$ 4.5 is associated with a galaxy with two close kinematic companions within 16-22 kpc of the host galaxy. Both these companion galaxies are within velocity separation of $<$ 370 {\kms} from the host galaxy. A third massive companion galaxy is detected 235 kpc from the CGM host galaxy. These galaxies might be part of a galaxy proto-group at $z\sim$ 4.5.

    \item The two strongest and kinematically most complex {\mgii} absorption systems (at  $z\sim$ 4.22 and $z\sim$ 4.5) are both part of two local galaxy over-densities. The {\mgii} absorption detected in these systems may be part of the intra-group gas associated with these two ``proto-group" galaxies at high-$z$.
        
\end{itemize}

In summary, we present the first results characterizing the cool CGM around $2.3<z<6.3$ galaxies in the first field of the EIGER survey. We examine CGM hosts of seven {\mgii} absorption systems and find that most of the high-$z$ {\mgii} absorption is not consistent with being bound to the dark matter halos of the host galaxies. This is in contrast to what is seen for CGM of $z<1$ galaxies \citep{Tumlinson2017}. In particular, extensive HST/COS CGM surveys of $z<0.2$, L* and sub-L* galaxies show that at low-$z$ most of the CGM is kinematically consistent with being bound to their host galaxies. These differences arise owing to a combination of lower gravitational potential of high-$z$ galaxies and a much higher Hubble parameter in the earlier Universe. These findings indicate that the galaxies in the early Universe were much more efficient in distributing metals produced in stars out of galaxies and chemically enrich their IGM. Such chemically enriched gas could be deposited to nearby galaxies at a later time, providing fuel for next generation of stars in them.

These observations will enable direct comparison of CGM-galaxy correlation in next generation simulations. Several simulations have looked into statistical properties of metal absorption line systems at high-$z$. These works reproduce the absorber statistics of strong absorption systems but significantly overproduce weak metal absorption systems \citep{Hasan2020,Finlator2020,Rahmati2016}, perhaps suggesting that the feedback prescription used in these simulations were too strong at high-$z$. There are tensions between the observed metal absorption line statistics of cool {\mgii} gas and those produced in simulations at high-$z$ \citep{Keating2016}.

In the \texttt{TECHNICOLOR DAWN} simulation, a correlation between high-$z$ absorbers and galaxies within a few hundred physical kpc was found \citep{Finlator2020,Doughty2023}. A two-point correlation function analysis between metal absorption systems and galaxies show a strong correlation at small impact parameters ($R < $ 100 pkpc), that qualitatively trace the {\mgii} absorption strength- $R$ anti-correlation presented in this work. Interestingly, \citealt{Doughty2023} find that there is a strong correlation between metal absorber type/strengths with the local galaxy over density than the stellar mass of the host galaxies. Several strong systems in this work are also observed in local overdensities and better statistics is needed to explore these correlations. With new \textit{JWST} observations providing direct observational constraints on CGM-galaxy correlation, quantitative comparison in future would enable direct constraints on the feedback prescriptions being implemented in these simulations. 

These results reinforce the power of \textit{JWST}/NIRCam grism observations to efficiently conduct high-$z$ galaxy spectroscopy campaigns. By combining the high fidelity \textit{JWST} spectroscopic campaign with deep group based NIR spectroscopy of $z>6$ quasars, we demonstrate an efficient program design to census the cool CGM around high-$z$ galaxies in the EIGER survey. In an upcoming paper, we will focus on detailed properties of the CGM of high-$z$ {\oi} and {\civ} absorption systems (Simcoe et al in prep.). We will further extend this work to the full six quasar fields of the EIGER survey (Bordoloi et al in prep) to better quantify the CGM-galaxy correlation in the EIGER survey.

\begin{acknowledgments}
This work is based on observations made with the
NASA/ESA/CSA James Webb Space Telescope. The
data were obtained from the Mikulski Archive for Space
Telescopes at the Space Telescope Science Institute,
which is operated by the Association of Universities for
Research in Astronomy, Inc., under NASA contract NAS
5-03127 for \textit{JWST}. These observations are associated
with program \# 1243.  This paper includes data gathered with the 6.5 meter Magellan Telescopes located at Las Campanas Observatory, Chile. This work has been supported by JSPS KAKENHI
Grant Number JP21K13956 (DK). RS acknowledges
support from NASA award number HST-GO-15085.001. Some/all of the data presented in this paper were obtained from the Mikulski Archive for Space Telescopes (MAST) at the Space Telescope Science Institute. The specific observations analyzed can be accessed via \dataset[http://dx.doi.org/10.17909/qhnr-e585]{http://dx.doi.org/10.17909/qhnr-e585}.

\end{acknowledgments}

\begin{deluxetable*}{ccccccccc}
\tablecolumns{9}
\tablewidth{0pt}
\tablecaption{EIGER galaxy properties along with {\mgii} absorption measurements\label{tab:galaxy_info}}
\tablehead{
\colhead{ID\tablenotemark{a}}&
\colhead{galaxy}&
\colhead{galaxy}&
\colhead{$z_{\rm{sys}}$\tablenotemark{b}}&
\colhead{$R$\tablenotemark{c}}&
\colhead{$\log M_{*}$}&
\colhead{$\rm{R_{vir}}$\tablenotemark{d}}&
\colhead{$\log$ SFR}&
\colhead{$\rm{W_r}$\tablenotemark{e}}\\
\colhead{}&
\colhead{$\alpha$  [J2000]}&
\colhead{$\delta$  [J2000]}&
\colhead{}&
\colhead{[pkpc]}&
\colhead{$\rm{M_{sun}}$}&
\colhead{[pkpc]}&
\colhead{$\rm{M_{sun}yr^{-1}}$}&
\colhead{[m\AA]}
}
\startdata
EIGER-01-09351 & 01:00:14.38 & 28:02:16.42 & 2.3087$\pm$0.0001 & 171.9 & 10.64$^{+0.08}_{-0.14}$ & 133$^{+18}_{-15}$ & 1.88$^{+0.11}_{-0.09}$ & 34$\pm$3 \\
EIGER-01-09138 & 01:00:14.28 & 28:02:14.86 & 2.3537$\pm$0.0001 & 168.2 & 9.87$^{+0.25}_{-0.16}$ & 87$^{+11}_{-9}$ & 1.76$^{+0.04}_{-0.06}$ & $>$27 \\
EIGER-01-08219 & 01:00:14.72 & 28:02:18.91 & 2.3571$\pm$0.0000 & 198.0 & 9.64$^{+0.22}_{-0.20}$ & 79$^{+9}_{-9}$ & 1.96$^{+0.05}_{-0.06}$ & $>$8 \\
EIGER-01-06569 & 01:00:14.31 & 28:02:54.32 & 2.6783$\pm$0.0001 & 270.4 & 10.17$^{+0.21}_{-0.36}$ & 92$^{+17}_{-12}$ & 1.87$^{+0.20}_{-0.14}$ & 29$\pm$6 \\
EIGER-01-06330 & 01:00:14.76 & 28:02:45.49 & 4.2156$\pm$0.0003 & 210.8 & 8.56$^{+0.20}_{-0.22}$ & 31$^{+4}_{-3}$ & 0.18$^{+0.12}_{-0.09}$ & $>$4 \\
EIGER-01-07514 & 01:00:13.80 & 28:02:44.01 & 4.2208$\pm$0.0002 & 145.7 & 9.00$^{+0.23}_{-0.22}$ & 38$^{+5}_{-4}$ & 0.85$^{+0.06}_{-0.05}$ & 2118$\pm$52 \\
EIGER-01-07870 & 01:00:14.00 & 28:02:34.76 & 4.2220$\pm$0.0003 & 109.8 & 8.75$^{+0.15}_{-0.18}$ & 34$^{+4}_{-4}$ & 0.36$^{+0.12}_{-0.11}$ & 2119$\pm$52 \\
EIGER-01-07143 & 01:00:14.47 & 28:02:37.92 & 4.2234$\pm$0.0001 & 158.1 & 9.16$^{+0.14}_{-0.20}$ & 40$^{+5}_{-4}$ & 1.03$^{+0.05}_{-0.04}$ & 2127$\pm$52 \\
EIGER-01-08125 & 01:00:13.89 & 28:02:33.98 & 4.2236$\pm$0.0002 & 98.6 & 8.88$^{+0.17}_{-0.39}$ & 36$^{+5}_{-4}$ & 0.53$^{+0.10}_{-0.09}$ & 2116$\pm$52 \\
EIGER-01-07015 & 01:00:14.47 & 28:02:38.32 & 4.2237$\pm$0.0004 & 159.4 & 8.52$^{+0.19}_{-0.20}$ & 30$^{+4}_{-3}$ & 0.72$^{+0.05}_{-0.06}$ & 2129$\pm$53 \\
EIGER-01-18550 & 01:00:11.50 & 28:02:26.27 & 4.2256$\pm$0.0000 & 139.0 & 9.09$^{+0.19}_{-0.22}$ & 39$^{+5}_{-5}$ & 0.83$^{+0.08}_{-0.08}$ & 2118$\pm$53 \\
EIGER-01-10308 & 01:00:12.71 & 28:02:29.24 & 4.2263$\pm$0.0003 & 36.3 & 10.11$^{+0.16}_{-0.17}$ & 62$^{+7}_{-6}$ & 1.84$^{+0.08}_{-0.08}$ & 2122$\pm$52 \\
EIGER-01-07059 & 01:00:14.30 & 28:02:42.43 & 4.2272$\pm$0.0006 & 165.0 & 8.81$^{+0.14}_{-0.26}$ & 35$^{+5}_{-4}$ & 0.41$^{+0.12}_{-0.10}$ & 2121$\pm$52 \\
EIGER-01-19137 & 01:00:11.88 & 28:02:27.81 & 4.4245$\pm$0.0001 & 103.3 & 9.17$^{+0.19}_{-0.14}$ & 39$^{+4}_{-4}$ & 1.83$^{+0.09}_{-0.07}$ & $>$5 \\
EIGER-01-09950 & 01:00:13.28 & 28:02:28.54 & 4.5159$\pm$0.0001 & 30.4 & 9.88$^{+0.15}_{-0.20}$ & 53$^{+6}_{-5}$ & 1.46$^{+0.11}_{-0.10}$ & 891$\pm$30 \\
EIGER-01-08811 & 01:00:13.53 & 28:02:28.89 & 4.5182$\pm$0.0007 & 50.9 & 8.30$^{+0.17}_{-0.27}$ & 26$^{+4}_{-3}$ & -0.05$^{+0.12}_{-0.08}$ & 902$\pm$30 \\
EIGER-01-06193 & 01:00:15.92 & 28:02:28.58 & 4.5202$\pm$0.0003 & 260.3 & 10.70$^{+0.03}_{-0.03}$ & 120$^{+79}_{-26}$ & 2.15$^{+0.06}_{-0.05}$ & 890$\pm$30 \\
EIGER-01-09078 & 01:00:13.45 & 28:02:27.48 & 4.5227$\pm$0.0001 & 40.3 & 8.79$^{+0.25}_{-0.35}$ & 32$^{+5}_{-5}$ & 0.61$^{+0.07}_{-0.07}$ & 878$\pm$30 \\
EIGER-01-08183 & 01:00:12.91 & 28:02:50.56 & 4.5466$\pm$0.0002 & 166.4 & 8.24$^{+0.26}_{-0.35}$ & 25$^{+4}_{-4}$ & 0.64$^{+0.08}_{-0.06}$ & $>$20 \\
EIGER-01-10424 & 01:00:13.07 & 28:02:36.60 & 4.6491$\pm$0.0003 & 71.7 & 9.18$^{+0.18}_{-0.18}$ & 37$^{+4}_{-4}$ & 1.11$^{+0.07}_{-0.07}$ & 109$\pm$4 \\
EIGER-01-16490 & 01:00:09.55 & 28:02:26.67 & 4.9422$\pm$0.0005 & 295.7 & 9.00$^{+0.17}_{-0.17}$ & 32$^{+4}_{-3}$ & 0.90$^{+0.07}_{-0.06}$ & $>$9 \\
EIGER-01-20300 & 01:00:12.88 & 28:02:23.44 & 5.3364$\pm$0.0021 & 18.8 & 7.10$^{+0.17}_{-0.10}$ & 12$^{+1}_{-1}$ & 0.05$^{+0.42}_{-0.18}$ & 167$\pm$3 \\
EIGER-01-19002 & 01:00:10.51 & 28:02:51.33 & 5.9079$\pm$0.0023 & 246.0 & 8.30$^{+0.20}_{-0.28}$ & 19$^{+3}_{-2}$ & 0.21$^{+0.17}_{-0.12}$ & $>$76 \\
EIGER-01-06027 & 01:00:15.74 & 28:02:30.31 & 5.9400$\pm$0.0023 & 213.7 & 7.83$^{+0.43}_{-0.37}$ & 15$^{+3}_{-3}$ & 0.58$^{+0.37}_{-0.19}$ & $>$9 \\
EIGER-01-08200 & 01:00:14.47 & 28:02:19.32 & 5.9417$\pm$0.0004 & 119.6 & 8.55$^{+0.26}_{-0.28}$ & 21$^{+3}_{-3}$ & 0.98$^{+0.14}_{-0.13}$ & $>$9 \\
EIGER-01-06898 & 01:00:15.58 & 28:02:19.91 & 6.0154$\pm$0.0001 & 200.7 & 9.10$^{+0.19}_{-0.24}$ & 27$^{+4}_{-3}$ & 0.93$^{+0.23}_{-0.19}$ & 46$\pm$6 \\
EIGER-01-07979 & 01:00:13.96 & 28:02:32.78 & 6.1883$\pm$0.0024 & 82.4 & 7.70$^{+0.24}_{-0.18}$ & 14$^{+2}_{-2}$ & 0.36$^{+0.34}_{-0.09}$ & $>$14 \\
EIGER-01-16842 & 01:00:09.74 & 28:02:29.50 & 6.2051$\pm$0.0024 & 249.3 & 8.13$^{+0.34}_{-0.29}$ & 17$^{+3}_{-2}$ & 0.69$^{+0.20}_{-0.14}$ & $>$19 \\
EIGER-01-10430 & 01:00:12.79 & 28:02:25.55 & 6.3288$\pm$0.0002 & 16.9 & 7.84$^{+0.11}_{-0.09}$ & 14$^{+1}_{-1}$ & 1.36$^{+0.29}_{-0.19}$ & $>$6 \\
  \enddata
  \vspace{-0.2cm}
\tablenotetext{a}{Galaxies within 300 kpc of the quasar J0100+2802.}
\tablenotetext{b}{Uncertainties listed are line centroiding errors.}
\tablenotetext{c}{Impact parameter in physical kpc.}
\tablenotetext{d}{Virial radius in kpc.}
\tablenotetext{e}{{\mgii} rest frame equivalent widths. Limits on $W_r$ are 2$\sigma$.}
\end{deluxetable*}

\begin{deluxetable*}{ccccc}
\tablecolumns{5}
\tablewidth{0pt}
\tablecaption{Voigt profile fit parameters for {\mgii} absorption\label{tab:vfit_info}}
\tablehead{
\colhead{ID}&
\colhead{$z_{\rm{sys}}$}&
\colhead{$\rm{\log N/cm^{-2}}$}&
\colhead{$\rm{b \;[kms^{-1}]}$}&
\colhead{$\rm{v \;[kms^{-1}]}$}
}
\startdata
EIGER-01-09351 & 2.3087 & 11.84$^{+0.04}_{-0.05}$ & 40$^{+4.7}_{-4.3}$ & -48$^{+4.2}_{-3.9}$ \\
-- & -- & 11.81$^{+0.04}_{-0.04}$ & 11$^{+1.7}_{-1.7}$ & 17$^{+0.9}_{-0.9}$ \\
\hline
EIGER-01-06569 & 2.6783 & 11.71$^{+0.17}_{-0.21}$ & 31$^{+19.9}_{-27.5}$ & -93$^{+10.1}_{-12.5}$ \\
\hline
EIGER-01-10308 & 4.2263 & 12.70$^{+0.04}_{-0.05}$ & 20$^{+2.0}_{-2.0}$ & -278$^{+2.1}_{-2.2}$ \\
-- & -- & 13.28$^{+0.03}_{-0.02}$ & 16$^{+2.8}_{-1.9}$ & -233$^{+1.4}_{-1.0}$ \\
-- & -- & 13.48$^{+0.41}_{-0.30}$ & 25$^{+12.1}_{-12.2}$ & -172$^{+20.4}_{-10.6}$ \\
-- & -- & 14.05$^{+0.10}_{-0.19}$ & 22$^{+1.7}_{-6.5}$ & -134$^{+4.4}_{-3.6}$ \\
-- & -- & 12.78$^{+0.01}_{-0.02}$ & 36$^{+1.7}_{-2.1}$ & -29$^{+1.5}_{-1.1}$ \\
-- & -- & 13.50$^{+0.01}_{-0.01}$ & 16$^{+0.2}_{-0.2}$ & 127$^{+0.2}_{-0.2}$ \\
\hline
EIGER-01-09950 & 4.5159 & 12.52$^{+0.01}_{-0.01}$ & 14$^{+1.0}_{-1.0}$ & 44$^{+0.5}_{-0.5}$ \\
-- & -- & 13.29$^{+0.02}_{-0.02}$ & 9$^{+0.3}_{-0.3}$ & 95$^{+0.3}_{-0.3}$ \\
-- & -- & 12.25$^{+0.15}_{-0.13}$ & 27$^{+12.9}_{-12.3}$ & 134$^{+5.3}_{-9.2}$ \\
-- & -- & 12.90$^{+0.02}_{-0.02}$ & 15$^{+1.7}_{-1.7}$ & 211$^{+0.8}_{-0.8}$ \\
-- & -- & 13.04$^{+0.01}_{-0.01}$ & 17$^{+0.8}_{-0.8}$ & 258$^{+0.6}_{-0.6}$ \\
\hline
EIGER-01-10424 & 4.6491 & 12.55$^{+0.01}_{-0.01}$ & 12$^{+0.8}_{-0.8}$ & -310$^{+0.4}_{-0.4}$ \\
\hline
EIGER-01-20300 & 5.3364 & 12.11$^{+0.10}_{-0.10}$ & 25$^{+5.4}_{-4.6}$ & 95$^{+6.0}_{-4.8}$ \\
-- & -- & 12.41$^{+0.05}_{-0.11}$ & 9$^{+1.8}_{-1.6}$ & 126$^{+1.2}_{-1.8}$ \\
-- & -- & 11.87$^{+0.28}_{-0.07}$ & 13$^{+12.7}_{-6.9}$ & 158$^{+2.2}_{-12.6}$ \\
\hline
EIGER-01-06898 & 6.0154 & 12.43$^{+0.12}_{-0.07}$ & 4$^{+0.7}_{-1.3}$ & -136$^{+0.8}_{-1.2}$ \\
\enddata
\vspace{-0.5cm}
\end{deluxetable*}

%

\facilities{JWST (NIRCam), Magellan (FIRE), VLT (X-shooter), Keck (HIRES)}


\software{astropy \citep{2013A&A...558A..33A,2018AJ....156..123A},  
          Cloudy \citep{2013RMxAA..49..137F}, 
          Source Extractor \citep{1996A&AS..117..393B}
          rbcodes \citep{rbcodes}
          rbvfit \citep{rongmon_bordoloi_2023_10403232}.
          }




\bibliography{EIGER_CGM}{}

\begin{thebibliography}{}
\expandafter\ifx\csname natexlab\endcsname\relax\def\natexlab#1{#1}\fi
\providecommand{\url}[1]{\href{#1}{#1}}
\providecommand{\dodoi}[1]{doi:~\href{http://doi.org/#1}{\nolinkurl{#1}}}
\providecommand{\doeprint}[1]{\href{http://ascl.net/#1}{\nolinkurl{http://ascl.net/#1}}}
\providecommand{\doarXiv}[1]{\href{https://arxiv.org/abs/#1}{\nolinkurl{https://arxiv.org/abs/#1}}}

\bibitem[{{Astropy Collaboration} {et~al.}(2013){Astropy Collaboration},
  {Robitaille}, {Tollerud}, {Greenfield}, {Droettboom}, {Bray}, {Aldcroft},
  {Davis}, {Ginsburg}, {Price-Whelan}, {Kerzendorf}, {Conley}, {Crighton},
  {Barbary}, {Muna}, {Ferguson}, {Grollier}, {Parikh}, {Nair}, {Unther},
  {Deil}, {Woillez}, {Conseil}, {Kramer}, {Turner}, {Singer}, {Fox}, {Weaver},
  {Zabalza}, {Edwards}, {Azalee Bostroem}, {Burke}, {Casey}, {Crawford},
  {Dencheva}, {Ely}, {Jenness}, {Labrie}, {Lim}, {Pierfederici}, {Pontzen},
  {Ptak}, {Refsdal}, {Servillat}, \& {Streicher}}]{2013A&A...558A..33A}
{Astropy Collaboration}, {Robitaille}, T.~P., {Tollerud}, E.~J., {et~al.} 2013,
  \aap, 558, A33, \dodoi{10.1051/0004-6361/201322068}

\bibitem[{{Astropy Collaboration} {et~al.}(2018){Astropy Collaboration},
  {Price-Whelan}, {Sip{\H{o}}cz}, {G{\"u}nther}, {Lim}, {Crawford}, {Conseil},
  {Shupe}, {Craig}, {Dencheva}, {Ginsburg}, {VanderPlas}, {Bradley},
  {P{\'e}rez-Su{\'a}rez}, {de Val-Borro}, {Aldcroft}, {Cruz}, {Robitaille},
  {Tollerud}, {Ardelean}, {Babej}, {Bach}, {Bachetti}, {Bakanov}, {Bamford},
  {Barentsen}, {Barmby}, {Baumbach}, {Berry}, {Biscani}, {Boquien}, {Bostroem},
  {Bouma}, {Brammer}, {Bray}, {Breytenbach}, {Buddelmeijer}, {Burke},
  {Calderone}, {Cano Rodr{\'\i}guez}, {Cara}, {Cardoso}, {Cheedella}, {Copin},
  {Corrales}, {Crichton}, {D'Avella}, {Deil}, {Depagne}, {Dietrich}, {Donath},
  {Droettboom}, {Earl}, {Erben}, {Fabbro}, {Ferreira}, {Finethy}, {Fox},
  {Garrison}, {Gibbons}, {Goldstein}, {Gommers}, {Greco}, {Greenfield},
  {Groener}, {Grollier}, {Hagen}, {Hirst}, {Homeier}, {Horton}, {Hosseinzadeh},
  {Hu}, {Hunkeler}, {Ivezi{\'c}}, {Jain}, {Jenness}, {Kanarek}, {Kendrew},
  {Kern}, {Kerzendorf}, {Khvalko}, {King}, {Kirkby}, {Kulkarni}, {Kumar},
  {Lee}, {Lenz}, {Littlefair}, {Ma}, {Macleod}, {Mastropietro}, {McCully},
  {Montagnac}, {Morris}, {Mueller}, {Mumford}, {Muna}, {Murphy}, {Nelson},
  {Nguyen}, {Ninan}, {N{\"o}the}, {Ogaz}, {Oh}, {Parejko}, {Parley}, {Pascual},
  {Patil}, {Patil}, {Plunkett}, {Prochaska}, {Rastogi}, {Reddy Janga},
  {Sabater}, {Sakurikar}, {Seifert}, {Sherbert}, {Sherwood-Taylor}, {Shih},
  {Sick}, {Silbiger}, {Singanamalla}, {Singer}, {Sladen}, {Sooley},
  {Sornarajah}, {Streicher}, {Teuben}, {Thomas}, {Tremblay}, {Turner},
  {Terr{\'o}n}, {van Kerkwijk}, {de la Vega}, {Watkins}, {Weaver}, {Whitmore},
  {Woillez}, {Zabalza}, \& {Astropy Contributors}}]{2018AJ....156..123A}
{Astropy Collaboration}, {Price-Whelan}, A.~M., {Sip{\H{o}}cz}, B.~M., {et~al.}
  2018, \aj, 156, 123, \dodoi{10.3847/1538-3881/aabc4f}

\bibitem[{{Ba{\~n}ados} {et~al.}(2018){Ba{\~n}ados}, {Venemans},
  {Mazzucchelli}, {Farina}, {Walter}, {Wang}, {Decarli}, {Stern}, {Fan},
  {Davies}, {Hennawi}, {Simcoe}, {Turner}, {Rix}, {Yang}, {Kelson}, {Rudie}, \&
  {Winters}}]{Banados2018}
{Ba{\~n}ados}, E., {Venemans}, B.~P., {Mazzucchelli}, C., {et~al.} 2018, \nat,
  553, 473, \dodoi{10.1038/nature25180}

\bibitem[{{Becker} {et~al.}(2019){Becker}, {Pettini}, {Rafelski}, {D'Odorico},
  {Boera}, {Christensen}, {Cupani}, {Ellison}, {Farina}, {Fumagalli},
  {L{\'o}pez}, {Neeleman}, {Ryan-Weber}, \& {Worseck}}]{Becker2019}
{Becker}, G.~D., {Pettini}, M., {Rafelski}, M., {et~al.} 2019, \apj, 883, 163,
  \dodoi{10.3847/1538-4357/ab3eb5}

\bibitem[{{Becker} {et~al.}(2001){Becker}, {Fan}, {White}, {Strauss},
  {Narayanan}, {Lupton}, {Gunn}, {Annis}, {Bahcall}, {Brinkmann}, {Connolly},
  {Csabai}, {Czarapata}, {Doi}, {Heckman}, {Hennessy}, {Ivezi{\'c}}, {Knapp},
  {Lamb}, {McKay}, {Munn}, {Nash}, {Nichol}, {Pier}, {Richards}, {Schneider},
  {Stoughton}, {Szalay}, {Thakar}, \& {York}}]{Becker2001}
{Becker}, R.~H., {Fan}, X., {White}, R.~L., {et~al.} 2001, \aj, 122, 2850,
  \dodoi{10.1086/324231}

\bibitem[{{Behroozi} {et~al.}(2019){Behroozi}, {Wechsler}, {Hearin}, \&
  {Conroy}}]{Behroozi2019}
{Behroozi}, P., {Wechsler}, R.~H., {Hearin}, A.~P., \& {Conroy}, C. 2019,
  \mnras, 488, 3143, \dodoi{10.1093/mnras/stz1182}

\bibitem[{{Bertin} \& {Arnouts}(1996)}]{1996A&AS..117..393B}
{Bertin}, E., \& {Arnouts}, S. 1996, \aaps, 117, 393,
  \dodoi{10.1051/aas:1996164}

\bibitem[{Bordoloi {et~al.}(2023)Bordoloi, jbhiggi, Manoranjan, \&
  Ribaudo}]{rongmon_bordoloi_2023_10403232}
Bordoloi, R., jbhiggi, Manoranjan, S., \& Ribaudo, J. 2023, rongmon/rbvfit:
  rbvfit v0.01, alpha1,  Zenodo, \dodoi{10.5281/zenodo.10403232}

\bibitem[{{Bordoloi} {et~al.}(2014{\natexlab{a}}){Bordoloi}, {Lilly},
  {Kacprzak}, \& {Churchill}}]{Bordoloi2014c}
{Bordoloi}, R., {Lilly}, S.~J., {Kacprzak}, G.~G., \& {Churchill}, C.~W.
  2014{\natexlab{a}}, \apj, 784, 108, \dodoi{10.1088/0004-637X/784/2/108}

\bibitem[{{Bordoloi} {et~al.}(2022{\natexlab{a}}){Bordoloi}, {Liu}, \&
  {Clark}}]{rbcodes}
{Bordoloi}, R., {Liu}, B., \& {Clark}, S. 2022{\natexlab{a}}, rongmon/rbcodes:
  rbcodes v0.2, v0.2,  Zenodo, \dodoi{10.5281/zenodo.7226204}

\bibitem[{{Bordoloi} {et~al.}(2018){Bordoloi}, {Prochaska}, {Tumlinson},
  {Werk}, {Tripp}, \& {Burchett}}]{Bordoloi2018}
{Bordoloi}, R., {Prochaska}, J.~X., {Tumlinson}, J., {et~al.} 2018, \apj, 864,
  132, \dodoi{10.3847/1538-4357/aad8ac}

\bibitem[{{Bordoloi} {et~al.}(2011){Bordoloi}, {Lilly}, {Knobel}, {Bolzonella},
  {Kampczyk}, {Carollo}, {Iovino}, {Zucca}, {Contini}, {Kneib}, {Le Fevre},
  {Mainieri}, {Renzini}, {Scodeggio}, {Zamorani}, {Balestra}, {Bardelli},
  {Bongiorno}, {Caputi}, {Cucciati}, {de la Torre}, {de Ravel}, {Garilli},
  {Kova{\v{c}}}, {Lamareille}, {Le Borgne}, {Le Brun}, {Maier}, {Mignoli},
  {Pello}, {Peng}, {Perez Montero}, {Presotto}, {Scarlata}, {Silverman},
  {Tanaka}, {Tasca}, {Tresse}, {Vergani}, {Barnes}, {Cappi}, {Cimatti},
  {Coppa}, {Diener}, {Franzetti}, {Koekemoer}, {L{\'o}pez-Sanjuan},
  {McCracken}, {Moresco}, {Nair}, {Oesch}, {Pozzetti}, \&
  {Welikala}}]{Bordoloi2011}
{Bordoloi}, R., {Lilly}, S.~J., {Knobel}, C., {et~al.} 2011, \apj, 743, 10,
  \dodoi{10.1088/0004-637X/743/1/10}

\bibitem[{{Bordoloi} {et~al.}(2014{\natexlab{b}}){Bordoloi}, {Lilly},
  {Hardmeier}, {Contini}, {Kneib}, {Le Fevre}, {Mainieri}, {Renzini},
  {Scodeggio}, {Zamorani}, {Bardelli}, {Bolzonella}, {Bongiorno}, {Caputi},
  {Carollo}, {Cucciati}, {de la Torre}, {de Ravel}, {Garilli}, {Iovino},
  {Kampczyk}, {Kova{\v{c}}}, {Knobel}, {Lamareille}, {Le Borgne}, {Le Brun},
  {Maier}, {Mignoli}, {Oesch}, {Pello}, {Peng}, {Perez Montero}, {Presotto},
  {Silverman}, {Tanaka}, {Tasca}, {Tresse}, {Vergani}, {Zucca}, {Cappi},
  {Cimatti}, {Coppa}, {Franzetti}, {Koekemoer}, {Moresco}, {Nair}, \&
  {Pozzetti}}]{Bordoloi2014b}
{Bordoloi}, R., {Lilly}, S.~J., {Hardmeier}, E., {et~al.} 2014{\natexlab{b}},
  \apj, 794, 130, \dodoi{10.1088/0004-637X/794/2/130}

\bibitem[{{Bordoloi} {et~al.}(2014{\natexlab{c}}){Bordoloi}, {Tumlinson},
  {Werk}, {Oppenheimer}, {Peeples}, {Prochaska}, {Tripp}, {Katz}, {Dav{\'e}},
  {Fox}, {Thom}, {Ford}, {Weinberg}, {Burchett}, \&
  {Kollmeier}}]{Bordoloi2014a}
{Bordoloi}, R., {Tumlinson}, J., {Werk}, J.~K., {et~al.} 2014{\natexlab{c}},
  \apj, 796, 136, \dodoi{10.1088/0004-637X/796/2/136}

\bibitem[{{Bordoloi} {et~al.}(2022{\natexlab{b}}){Bordoloi}, {O'Meara},
  {Sharon}, {Rigby}, {Cooke}, {Shaban}, {Matuszewski}, {Rizzi}, {Doppmann},
  {Martin}, {Moore}, {Morrissey}, \& {Neill}}]{Bordoloi2022}
{Bordoloi}, R., {O'Meara}, J.~M., {Sharon}, K., {et~al.} 2022{\natexlab{b}},
  \nat, 606, 59, \dodoi{10.1038/s41586-022-04616-1}

\bibitem[{{Bouch{\'e}} {et~al.}(2012){Bouch{\'e}}, {Hohensee}, {Vargas},
  {Kacprzak}, {Martin}, {Cooke}, \& {Churchill}}]{Bouche2012}
{Bouch{\'e}}, N., {Hohensee}, W., {Vargas}, R., {et~al.} 2012, \mnras, 426,
  801, \dodoi{10.1111/j.1365-2966.2012.21114.x}

\bibitem[{{Bryan} \& {Norman}(1998)}]{Bryan1998}
{Bryan}, G.~L., \& {Norman}, M.~L. 1998, \apj, 495, 80, \dodoi{10.1086/305262}

\bibitem[{{Burchett} {et~al.}(2016){Burchett}, {Tripp}, {Bordoloi}, {Werk},
  {Prochaska}, {Tumlinson}, {Willmer}, {O'Meara}, \& {Katz}}]{Burchett2016}
{Burchett}, J.~N., {Tripp}, T.~M., {Bordoloi}, R., {et~al.} 2016, \apj, 832,
  124, \dodoi{10.3847/0004-637X/832/2/124}

\bibitem[{Bushouse {et~al.}(2022{\natexlab{a}})Bushouse, Eisenhamer, Dencheva,
  Davies, Greenfield, Morrison, Hodge, Simon, Grumm, Droettboom, Slavich,
  Sosey, Pauly, Miller, Jedrzejewski, Hack, Davis, Crawford, Law, Gordon,
  Regan, Cara, MacDonald, Bradley, Shanahan, Jamieson, Teodoro, \&
  Williams}]{bushouse_howard_2022_7325378}
Bushouse, H., Eisenhamer, J., Dencheva, N., {et~al.} 2022{\natexlab{a}}, JWST
  Calibration Pipeline, 1.8.2,  Zenodo, \dodoi{10.5281/zenodo.7325378}

\bibitem[{Bushouse {et~al.}(2022{\natexlab{b}})Bushouse, Eisenhamer, Dencheva,
  Davies, Greenfield, Morrison, Hodge, Simon, Grumm, Droettboom, Slavich,
  Sosey, Pauly, Miller, Jedrzejewski, Hack, Davis, Crawford, Law, Gordon,
  Regan, Cara, MacDonald, Bradley, Shanahan, Jamieson, Teodoro, \&
  Williams}]{bushouse_howard_2022_7071140}
---. 2022{\natexlab{b}}, JWST Calibration Pipeline, 1.7.0,  Zenodo,
  \dodoi{10.5281/zenodo.7071140}

\bibitem[{{Calzetti} {et~al.}(2000){Calzetti}, {Armus}, {Bohlin}, {Kinney},
  {Koornneef}, \& {Storchi-Bergmann}}]{Calzetti2000}
{Calzetti}, D., {Armus}, L., {Bohlin}, R.~C., {et~al.} 2000, \apj, 533, 682,
  \dodoi{10.1086/308692}

\bibitem[{{Chabrier}(2003)}]{Chabrier2003}
{Chabrier}, G. 2003, \pasp, 115, 763, \dodoi{10.1086/376392}

\bibitem[{{Chen} {et~al.}(2010){Chen}, {Helsby}, {Gauthier}, {Shectman},
  {Thompson}, \& {Tinker}}]{Chen2010}
{Chen}, H.-W., {Helsby}, J.~E., {Gauthier}, J.-R., {et~al.} 2010, \apj, 714,
  1521, \dodoi{10.1088/0004-637X/714/2/1521}

\bibitem[{{Chen} {et~al.}(2001){Chen}, {Lanzetta}, \& {Webb}}]{Chen2001}
{Chen}, H.-W., {Lanzetta}, K.~M., \& {Webb}, J.~K. 2001, \apj, 556, 158,
  \dodoi{10.1086/321537}

\bibitem[{{Chen} {et~al.}(2020){Chen}, {Zahedy}, {Boettcher}, {Cooper},
  {Johnson}, {Rudie}, {Chen}, {Walth}, {Cantalupo}, {Cooksey},
  {Faucher-Gigu{\`e}re}, {Greene}, {Lopez}, {Mulchaey}, {Penton}, {Petitjean},
  {Putman}, {Rafelski}, {Rauch}, {Schaye}, {Simcoe}, \& {Weiner}}]{Chen2020}
{Chen}, H.-W., {Zahedy}, F.~S., {Boettcher}, E., {et~al.} 2020, \mnras, 497,
  498, \dodoi{10.1093/mnras/staa1773}

\bibitem[{{Choi} {et~al.}(2016){Choi}, {Dotter}, {Conroy}, {Cantiello},
  {Paxton}, \& {Johnson}}]{Choi2016}
{Choi}, J., {Dotter}, A., {Conroy}, C., {et~al.} 2016, \apj, 823, 102,
  \dodoi{10.3847/0004-637X/823/2/102}

\bibitem[{{Churchill} {et~al.}(2013){Churchill}, {Nielsen}, {Kacprzak}, \&
  {Trujillo-Gomez}}]{Churchill2013}
{Churchill}, C.~W., {Nielsen}, N.~M., {Kacprzak}, G.~G., \& {Trujillo-Gomez},
  S. 2013, \apjl, 763, L42, \dodoi{10.1088/2041-8205/763/2/L42}

\bibitem[{{Cooper} {et~al.}(2019){Cooper}, {Simcoe}, {Cooksey}, {Bordoloi},
  {Miller}, {Furesz}, {Turner}, \& {Ba{\~n}ados}}]{Cooper2019}
{Cooper}, T.~J., {Simcoe}, R.~A., {Cooksey}, K.~L., {et~al.} 2019, \apj, 882,
  77, \dodoi{10.3847/1538-4357/ab3402}

\bibitem[{{Davies} {et~al.}(2023){Davies}, {Ryan-Weber}, {D'Odorico}, {Bosman},
  {Meyer}, {Becker}, {Cupani}, {Bischetti}, {Sebastian}, {Eilers}, {Farina},
  {Wang}, {Yang}, \& {Zhu}}]{Davies2023}
{Davies}, R.~L., {Ryan-Weber}, E., {D'Odorico}, V., {et~al.} 2023, \mnras, 521,
  289, \dodoi{10.1093/mnras/stac3662}

\bibitem[{{D{\'\i}az} {et~al.}(2021){D{\'\i}az}, {Ryan-Weber}, {Karman},
  {Caputi}, {Salvadori}, {Crighton}, {Ouchi}, \& {Vanzella}}]{Diaz2021}
{D{\'\i}az}, C.~G., {Ryan-Weber}, E.~V., {Karman}, W., {et~al.} 2021, \mnras,
  502, 2645, \dodoi{10.1093/mnras/staa3129}

\bibitem[{{Dotter}(2016)}]{Dotter2016}
{Dotter}, A. 2016, \apjs, 222, 8, \dodoi{10.3847/0067-0049/222/1/8}

\bibitem[{{Doughty} \& {Finlator}(2023)}]{Doughty2023}
{Doughty}, C.~C., \& {Finlator}, K.~M. 2023, \mnras, 518, 4159,
  \dodoi{10.1093/mnras/stac3342}

\bibitem[{{Dutta} {et~al.}(2021){Dutta}, {Fumagalli}, {Fossati}, {Bielby},
  {Stott}, {Lofthouse}, {Cantalupo}, {Cullen}, {Crain}, {Tripp}, {Prochaska},
  {Arrigoni Battaia}, {Burchett}, {Fynbo}, {Murphy}, {Schaye}, {Tejos}, \&
  {Theuns}}]{Dutta2021}
{Dutta}, R., {Fumagalli}, M., {Fossati}, M., {et~al.} 2021, \mnras, 508, 4573,
  \dodoi{10.1093/mnras/stab2752}

\bibitem[{{Eilers} {et~al.}(2023){Eilers}, {Simcoe}, {Yue}, {Mackenzie},
  {Matthee}, {{\v{D}}urov{\v{c}}{\'\i}kov{\'a}}, {Kashino}, {Bordoloi}, \&
  {Lilly}}]{Eilers2023}
{Eilers}, A.-C., {Simcoe}, R.~A., {Yue}, M., {et~al.} 2023, \apj, 950, 68,
  \dodoi{10.3847/1538-4357/acd776}

\bibitem[{{Ferland} {et~al.}(2013){Ferland}, {Porter}, {van Hoof}, {Williams},
  {Abel}, {Lykins}, {Shaw}, {Henney}, \& {Stancil}}]{2013RMxAA..49..137F}
{Ferland}, G.~J., {Porter}, R.~L., {van Hoof}, P.~A.~M., {et~al.} 2013, \rmxaa,
  49, 137.
\newblock \doarXiv{1302.4485}

\bibitem[{{Finlator} {et~al.}(2020){Finlator}, {Doughty}, {Cai}, \&
  {D{\'\i}az}}]{Finlator2020}
{Finlator}, K., {Doughty}, C., {Cai}, Z., \& {D{\'\i}az}, G. 2020, \mnras, 493,
  3223, \dodoi{10.1093/mnras/staa377}

\bibitem[{{Ford} {et~al.}(2014){Ford}, {Dav{\'e}}, {Oppenheimer}, {Katz},
  {Kollmeier}, {Thompson}, \& {Weinberg}}]{Ford2014}
{Ford}, A.~B., {Dav{\'e}}, R., {Oppenheimer}, B.~D., {et~al.} 2014, \mnras,
  444, 1260, \dodoi{10.1093/mnras/stu1418}

\bibitem[{{Gaia Collaboration} {et~al.}(2018){Gaia Collaboration}, {Brown},
  {Vallenari}, {Prusti}, {de Bruijne}, {Babusiaux}, {Bailer-Jones}, {Biermann},
  {Evans}, {Eyer}, {Jansen}, {Jordi}, {Klioner}, {Lammers}, {Lindegren},
  {Luri}, {Mignard}, {Panem}, {Pourbaix}, {Randich}, {Sartoretti}, {Siddiqui},
  {Soubiran}, {van Leeuwen}, {Walton}, {Arenou}, {Bastian}, {Cropper},
  {Drimmel}, {Katz}, {Lattanzi}, {Bakker}, {Cacciari}, {Casta{\~n}eda},
  {Chaoul}, {Cheek}, {De Angeli}, {Fabricius}, {Guerra}, {Holl}, {Masana},
  {Messineo}, {Mowlavi}, {Nienartowicz}, {Panuzzo}, {Portell}, {Riello},
  {Seabroke}, {Tanga}, {Th{\'e}venin}, {Gracia-Abril}, {Comoretto},
  {Garcia-Reinaldos}, {Teyssier}, {Altmann}, {Andrae}, {Audard},
  {Bellas-Velidis}, {Benson}, {Berthier}, {Blomme}, {Burgess}, {Busso},
  {Carry}, {Cellino}, {Clementini}, {Clotet}, {Creevey}, {Davidson}, {De
  Ridder}, {Delchambre}, {Dell'Oro}, {Ducourant},
  {Fern{\'a}ndez-Hern{\'a}ndez}, {Fouesneau}, {Fr{\'e}mat}, {Galluccio},
  {Garc{\'\i}a-Torres}, {Gonz{\'a}lez-N{\'u}{\~n}ez}, {Gonz{\'a}lez-Vidal},
  {Gosset}, {Guy}, {Halbwachs}, {Hambly}, {Harrison}, {Hern{\'a}ndez},
  {Hestroffer}, {Hodgkin}, {Hutton}, {Jasniewicz}, {Jean-Antoine-Piccolo},
  {Jordan}, {Korn}, {Krone-Martins}, {Lanzafame}, {Lebzelter}, {L{\"o}ffler},
  {Manteiga}, {Marrese}, {Mart{\'\i}n-Fleitas}, {Moitinho}, {Mora}, {Muinonen},
  {Osinde}, {Pancino}, {Pauwels}, {Petit}, {Recio-Blanco}, {Richards},
  {Rimoldini}, {Robin}, {Sarro}, {Siopis}, {Smith}, {Sozzetti}, {S{\"u}veges},
  {Torra}, {van Reeven}, {Abbas}, {Abreu Aramburu}, {Accart}, {Aerts},
  {Altavilla}, {{\'A}lvarez}, {Alvarez}, {Alves}, {Anderson}, {Andrei},
  {Anglada Varela}, {Antiche}, {Antoja}, {Arcay}, {Astraatmadja}, {Bach},
  {Baker}, {Balaguer-N{\'u}{\~n}ez}, {Balm}, {Barache}, {Barata}, {Barbato},
  {Barblan}, {Barklem}, {Barrado}, {Barros}, {Barstow}, {Bartholom{\'e}
  Mu{\~n}oz}, {Bassilana}, {Becciani}, {Bellazzini}, {Berihuete}, {Bertone},
  {Bianchi}, {Bienaym{\'e}}, {Blanco-Cuaresma}, {Boch}, {Boeche}, {Bombrun},
  {Borrachero}, {Bossini}, {Bouquillon}, {Bourda}, {Bragaglia}, {Bramante},
  {Breddels}, {Bressan}, {Brouillet}, {Br{\"u}semeister}, {Brugaletta},
  {Bucciarelli}, {Burlacu}, {Busonero}, {Butkevich}, {Buzzi}, {Caffau},
  {Cancelliere}, {Cannizzaro}, {Cantat-Gaudin}, {Carballo}, {Carlucci},
  {Carrasco}, {Casamiquela}, {Castellani}, {Castro-Ginard}, {Charlot},
  {Chemin}, {Chiavassa}, {Cocozza}, {Costigan}, {Cowell}, {Crifo}, {Crosta},
  {Crowley}, {Cuypers}, {Dafonte}, {Damerdji}, {Dapergolas}, {David}, {David},
  {de Laverny}, {De Luise}, {De March}, {de Martino}, {de Souza}, {de Torres},
  {Debosscher}, {del Pozo}, {Delbo}, {Delgado}, {Delgado}, {Di Matteo},
  {Diakite}, {Diener}, {Distefano}, {Dolding}, {Drazinos}, {Dur{\'a}n},
  {Edvardsson}, {Enke}, {Eriksson}, {Esquej}, {Eynard Bontemps}, {Fabre},
  {Fabrizio}, {Faigler}, {Falc{\~a}o}, {Farr{\`a}s Casas}, {Federici},
  {Fedorets}, {Fernique}, {Figueras}, {Filippi}, {Findeisen}, {Fonti},
  {Fraile}, {Fraser}, {Fr{\'e}zouls}, {Gai}, {Galleti}, {Garabato},
  {Garc{\'\i}a-Sedano}, {Garofalo}, {Garralda}, {Gavel}, {Gavras}, {Gerssen},
  {Geyer}, {Giacobbe}, {Gilmore}, {Girona}, {Giuffrida}, {Glass}, {Gomes},
  {Granvik}, {Gueguen}, {Guerrier}, {Guiraud}, {Guti{\'e}rrez-S{\'a}nchez},
  {Haigron}, {Hatzidimitriou}, {Hauser}, {Haywood}, {Heiter}, {Helmi}, {Heu},
  {Hilger}, {Hobbs}, {Hofmann}, {Holland}, {Huckle}, {Hypki}, {Icardi},
  {Jan{\ss}en}, {Jevardat de Fombelle}, {Jonker}, {Juh{\'a}sz}, {Julbe},
  {Karampelas}, {Kewley}, {Klar}, {Kochoska}, {Kohley}, {Kolenberg},
  {Kontizas}, {Kontizas}, {Koposov}, {Kordopatis}, {Kostrzewa-Rutkowska},
  {Koubsky}, {Lambert}, {Lanza}, {Lasne}, {Lavigne}, {Le Fustec}, {Le
  Poncin-Lafitte}, {Lebreton}, {Leccia}, {Leclerc}, {Lecoeur-Taibi},
  {Lenhardt}, {Leroux}, {Liao}, {Licata}, {Lindstr{\o}m}, {Lister}, {Livanou},
  {Lobel}, {L{\'o}pez}, {Managau}, {Mann}, {Mantelet}, {Marchal}, {Marchant},
  {Marconi}, {Marinoni}, {Marschalk{\'o}}, {Marshall}, {Martino}, {Marton},
  {Mary}, {Massari}, {Matijevi{\v{c}}}, {Mazeh}, {McMillan}, {Messina},
  {Michalik}, {Millar}, {Molina}, {Molinaro}, {Moln{\'a}r}, {Montegriffo},
  {Mor}, {Morbidelli}, {Morel}, {Morris}, {Mulone}, {Muraveva}, {Musella},
  {Nelemans}, {Nicastro}, {Noval}, {O'Mullane}, {Ord{\'e}novic},
  {Ord{\'o}{\~n}ez-Blanco}, {Osborne}, {Pagani}, {Pagano}, {Pailler},
  {Palacin}, {Palaversa}, {Panahi}, {Pawlak}, {Piersimoni}, {Pineau}, {Plachy},
  {Plum}, {Poggio}, {Poujoulet}, {Pr{\v{s}}a}, {Pulone}, {Racero}, {Ragaini},
  {Rambaux}, {Ramos-Lerate}, {Regibo}, {Reyl{\'e}}, {Riclet}, {Ripepi}, {Riva},
  {Rivard}, {Rixon}, {Roegiers}, {Roelens}, {Romero-G{\'o}mez}, {Rowell},
  {Royer}, {Ruiz-Dern}, {Sadowski}, {Sagrist{\`a} Sell{\'e}s}, {Sahlmann},
  {Salgado}, {Salguero}, {Sanna}, {Santana-Ros}, {Sarasso}, {Savietto},
  {Schultheis}, {Sciacca}, {Segol}, {Segovia}, {S{\'e}gransan}, {Shih},
  {Siltala}, {Silva}, {Smart}, {Smith}, {Solano}, {Solitro}, {Sordo}, {Soria
  Nieto}, {Souchay}, {Spagna}, {Spoto}, {Stampa}, {Steele},
  {Steidelm{\"u}ller}, {Stephenson}, {Stoev}, {Suess}, {Surdej}, {Szabados},
  {Szegedi-Elek}, {Tapiador}, {Taris}, {Tauran}, {Taylor}, {Teixeira},
  {Terrett}, {Teyssandier}, {Thuillot}, {Titarenko}, {Torra Clotet}, {Turon},
  {Ulla}, {Utrilla}, {Uzzi}, {Vaillant}, {Valentini}, {Valette}, {van Elteren},
  {Van Hemelryck}, {van Leeuwen}, {Vaschetto}, {Vecchiato}, {Veljanoski},
  {Viala}, {Vicente}, {Vogt}, {von Essen}, {Voss}, {Votruba}, {Voutsinas},
  {Walmsley}, {Weiler}, {Wertz}, {Wevers}, {Wyrzykowski}, {Yoldas},
  {{\v{Z}}erjal}, {Ziaeepour}, {Zorec}, {Zschocke}, {Zucker}, {Zurbach}, \&
  {Zwitter}}]{Gaia2018}
{Gaia Collaboration}, {Brown}, A.~G.~A., {Vallenari}, A., {et~al.} 2018, \aap,
  616, A1, \dodoi{10.1051/0004-6361/201833051}

\bibitem[{{Galbiati} {et~al.}(2023){Galbiati}, {Fumagalli}, {Fossati},
  {Lofthouse}, {Dutta}, {Prochaska}, {Murphy}, \& {Cantalupo}}]{Galbiati2023}
{Galbiati}, M., {Fumagalli}, M., {Fossati}, M., {et~al.} 2023, arXiv e-prints,
  arXiv:2302.00021, \dodoi{10.48550/arXiv.2302.00021}

\bibitem[{{Hasan} {et~al.}(2020){Hasan}, {Churchill}, {Stemock}, {Mathes},
  {Nielsen}, {Finlator}, {Doughty}, {Croom}, {Kacprzak}, \&
  {Murphy}}]{Hasan2020}
{Hasan}, F., {Churchill}, C.~W., {Stemock}, B., {et~al.} 2020, \apj, 904, 44,
  \dodoi{10.3847/1538-4357/abbe0b}

\bibitem[{{Huang} {et~al.}(2016){Huang}, {Chen}, {Johnson}, \&
  {Weiner}}]{Huang2016}
{Huang}, Y.-H., {Chen}, H.-W., {Johnson}, S.~D., \& {Weiner}, B.~J. 2016,
  \mnras, 455, 1713, \dodoi{10.1093/mnras/stv2327}

\bibitem[{{Johnson} {et~al.}(2021){Johnson}, {Leja}, {Conroy}, \&
  {Speagle}}]{Johnson2021}
{Johnson}, B.~D., {Leja}, J., {Conroy}, C., \& {Speagle}, J.~S. 2021, \apjs,
  254, 22, \dodoi{10.3847/1538-4365/abef67}

\bibitem[{{Johnson} {et~al.}(2015){Johnson}, {Chen}, \&
  {Mulchaey}}]{Johnson2015}
{Johnson}, S.~D., {Chen}, H.-W., \& {Mulchaey}, J.~S. 2015, \mnras, 449, 3263,
  \dodoi{10.1093/mnras/stv553}

\bibitem[{{Johnson} {et~al.}(2017){Johnson}, {Chen}, {Mulchaey}, {Schaye}, \&
  {Straka}}]{Johnson2017}
{Johnson}, S.~D., {Chen}, H.-W., {Mulchaey}, J.~S., {Schaye}, J., \& {Straka},
  L.~A. 2017, \apjl, 850, L10, \dodoi{10.3847/2041-8213/aa9370}

\bibitem[{{Kacprzak} {et~al.}(2012){Kacprzak}, {Churchill}, \&
  {Nielsen}}]{Kacprzak2012}
{Kacprzak}, G.~G., {Churchill}, C.~W., \& {Nielsen}, N.~M. 2012, \apjl, 760,
  L7, \dodoi{10.1088/2041-8205/760/1/L7}

\bibitem[{{Kashino} {et~al.}(2023{\natexlab{a}}){Kashino}, {Lilly}, {Matthee},
  {Eilers}, {Mackenzie}, {Bordoloi}, \& {Simcoe}}]{Kashino2023}
{Kashino}, D., {Lilly}, S.~J., {Matthee}, J., {et~al.} 2023{\natexlab{a}},
  \apj, 950, 66, \dodoi{10.3847/1538-4357/acc588}

\bibitem[{{Kashino} {et~al.}(2023{\natexlab{b}}){Kashino}, {Lilly}, {Simcoe},
  {Bordoloi}, {Mackenzie}, {Matthee}, \& {Eilers}}]{Kashino2023b}
{Kashino}, D., {Lilly}, S.~J., {Simcoe}, R.~A., {et~al.} 2023{\natexlab{b}},
  \nat, 617, 261, \dodoi{10.1038/s41586-023-05901-3}

\bibitem[{{Keating} {et~al.}(2016){Keating}, {Puchwein}, {Haehnelt}, {Bird}, \&
  {Bolton}}]{Keating2016}
{Keating}, L.~C., {Puchwein}, E., {Haehnelt}, M.~G., {Bird}, S., \& {Bolton},
  J.~S. 2016, \mnras, 461, 606, \dodoi{10.1093/mnras/stw1306}

\bibitem[{{Krist} {et~al.}(2011){Krist}, {Hook}, \& {Stoehr}}]{Krist2011}
{Krist}, J.~E., {Hook}, R.~N., \& {Stoehr}, F. 2011, in Society of
  Photo-Optical Instrumentation Engineers (SPIE) Conference Series, Vol. 8127,
  Optical Modeling and Performance Predictions V, ed. M.~A. {Kahan}, 81270J,
  \dodoi{10.1117/12.892762}

\bibitem[{{Lan} \& {Mo}(2018)}]{Lan2018}
{Lan}, T.-W., \& {Mo}, H. 2018, \apj, 866, 36, \dodoi{10.3847/1538-4357/aadc08}

\bibitem[{{Lehner} {et~al.}(2013){Lehner}, {Howk}, {Tripp}, {Tumlinson},
  {Prochaska}, {O'Meara}, {Thom}, {Werk}, {Fox}, \& {Ribaudo}}]{Lehner2013}
{Lehner}, N., {Howk}, J.~C., {Tripp}, T.~M., {et~al.} 2013, \apj, 770, 138,
  \dodoi{10.1088/0004-637X/770/2/138}

\bibitem[{{Lehner} {et~al.}(2022){Lehner}, {Kopenhafer}, {O'Meara}, {Howk},
  {Fumagalli}, {Prochaska}, {Acharyya}, {O'Shea}, {Peeples}, {Tumlinson}, \&
  {Hummels}}]{Lehner2022}
{Lehner}, N., {Kopenhafer}, C., {O'Meara}, J.~M., {et~al.} 2022, \apj, 936,
  156, \dodoi{10.3847/1538-4357/ac7400}

\bibitem[{{Liang} \& {Chen}(2014)}]{Liang2014}
{Liang}, C.~J., \& {Chen}, H.-W. 2014, \mnras, 445, 2061,
  \dodoi{10.1093/mnras/stu1901}

\bibitem[{{Lopez} {et~al.}(2018){Lopez}, {Tejos}, {Ledoux}, {Barrientos},
  {Sharon}, {Rigby}, {Gladders}, {Bayliss}, \& {Pessa}}]{Lopez2018}
{Lopez}, S., {Tejos}, N., {Ledoux}, C., {et~al.} 2018, \nat, 554, 493,
  \dodoi{10.1038/nature25436}

\bibitem[{{Lundgren} {et~al.}(2021){Lundgren}, {Creech}, {Brammer}, {Kirse},
  {Peek}, {Wake}, {York}, {Chisholm}, {Erb}, {Kulkarni}, {Straka}, {Tremonti},
  \& {van Dokkum}}]{Lundgren2021}
{Lundgren}, B.~F., {Creech}, S., {Brammer}, G., {et~al.} 2021, \apj, 913, 50,
  \dodoi{10.3847/1538-4357/abef6a}

\bibitem[{{Mack} {et~al.}(2022){Mack}, {Hack}, {Burger}, {White}, {Bajaj},
  {Avila}, {Anand}, \& {de la Pena}}]{Mack2022}
{Mack}, J., {Hack}, W., {Burger}, M., {et~al.} 2022, {Improved Absolute
  Astrometry for ACS and WFC3 Data Products}, Instrument Science Report WFC3
  2022-06

\bibitem[{{Martin} {et~al.}(2019){Martin}, {Ho}, {Kacprzak}, \&
  {Churchill}}]{Martin2019}
{Martin}, C.~L., {Ho}, S.~H., {Kacprzak}, G.~G., \& {Churchill}, C.~W. 2019,
  \apj, 878, 84, \dodoi{10.3847/1538-4357/ab18ac}

\bibitem[{{Matejek} \& {Simcoe}(2012)}]{Matejek2012}
{Matejek}, M.~S., \& {Simcoe}, R.~A. 2012, \apj, 761, 112,
  \dodoi{10.1088/0004-637X/761/2/112}

\bibitem[{{Matthee} {et~al.}(2023{\natexlab{a}}){Matthee}, {Mackenzie},
  {Simcoe}, {Kashino}, {Lilly}, {Bordoloi}, \& {Eilers}}]{Matthee2023}
{Matthee}, J., {Mackenzie}, R., {Simcoe}, R.~A., {et~al.} 2023{\natexlab{a}},
  \apj, 950, 67, \dodoi{10.3847/1538-4357/acc846}

\bibitem[{{Matthee} {et~al.}(2023{\natexlab{b}}){Matthee}, {Naidu}, {Brammer},
  {Chisholm}, {Eilers}, {Goulding}, {Greene}, {Kashino}, {Labbe}, {Lilly},
  {Mackenzie}, {Oesch}, {Weibel}, {Wuyts}, {Xiao}, {Bordoloi}, {Bouwens}, {van
  Dokkum}, {Illingworth}, {Kramarenko}, {Maseda}, {Mason}, {Meyer}, {Nelson},
  {Reddy}, {Shivaei}, {Simcoe}, \& {Yue}}]{Matthee2023b}
{Matthee}, J., {Naidu}, R.~P., {Brammer}, G., {et~al.} 2023{\natexlab{b}},
  arXiv e-prints, arXiv:2306.05448, \dodoi{10.48550/arXiv.2306.05448}

\bibitem[{{McCabe} {et~al.}(2021){McCabe}, {Borthakur}, {Heckman}, {Tumlinson},
  {Bordoloi}, \& {Dave}}]{McCabe2021}
{McCabe}, T., {Borthakur}, S., {Heckman}, T., {et~al.} 2021, \apj, 923, 189,
  \dodoi{10.3847/1538-4357/ac283c}

\bibitem[{{Merlin} {et~al.}(2022){Merlin}, {Bonchi}, {Paris}, {Belfiori},
  {Fontana}, {Castellano}, {Nonino}, {Polenta}, {Santini}, {Yang},
  {Glazebrook}, {Treu}, {Roberts-Borsani}, {Trenti}, {Birrer}, {Brammer},
  {Grillo}, {Calabr{\`o}}, {Marchesini}, {Mason}, {Mercurio}, {Morishita},
  {Strait}, {Boyett}, {Leethochawalit}, {Nanayakkara}, {Vulcani}, {Bradac}, \&
  {Wang}}]{Merlin2022}
{Merlin}, E., {Bonchi}, A., {Paris}, D., {et~al.} 2022, \apjl, 938, L14,
  \dodoi{10.3847/2041-8213/ac8f93}

\bibitem[{{Nielsen} {et~al.}(2013){Nielsen}, {Churchill}, \&
  {Kacprzak}}]{Nielsen2013}
{Nielsen}, N.~M., {Churchill}, C.~W., \& {Kacprzak}, G.~G. 2013, \apj, 776,
  115, \dodoi{10.1088/0004-637X/776/2/115}

\bibitem[{{Oesch} {et~al.}(2023){Oesch}, {Brammer}, {Naidu}, {Bouwens},
  {Chisholm}, {Illingworth}, {Matthee}, {Nelson}, {Qin}, {Reddy}, {Shapley},
  {Shivaei}, {van Dokkum}, {Weibel}, {Whitaker}, {Wuyts}, {Covelo-Paz},
  {Endsley}, {Fudamoto}, {Giovinazzo}, {Herard-Demanche}, {Kerutt},
  {Kramarenko}, {Labbe}, {Leonova}, {Lin}, {Magee}, {Marchesini}, {Maseda},
  {Mason}, {Matharu}, {Meyer}, {Neufeld}, {Prieto Lyon}, {Schaerer}, {Sharma},
  {Shuntov}, {Smit}, {Stefanon}, {Wyithe}, \& {Xiao}}]{Oesch2023}
{Oesch}, P.~A., {Brammer}, G., {Naidu}, R.~P., {et~al.} 2023, arXiv e-prints,
  arXiv:2304.02026, \dodoi{10.48550/arXiv.2304.02026}

\bibitem[{{Peeples} {et~al.}(2014){Peeples}, {Werk}, {Tumlinson},
  {Oppenheimer}, {Prochaska}, {Katz}, \& {Weinberg}}]{Peeples2014}
{Peeples}, M.~S., {Werk}, J.~K., {Tumlinson}, J., {et~al.} 2014, \apj, 786, 54,
  \dodoi{10.1088/0004-637X/786/1/54}

\bibitem[{{P{\'e}roux} \& {Howk}(2020)}]{Peroux2020}
{P{\'e}roux}, C., \& {Howk}, J.~C. 2020, \araa, 58, 363,
  \dodoi{10.1146/annurev-astro-021820-120014}

\bibitem[{{Planck Collaboration} {et~al.}(2020){Planck Collaboration},
  {Aghanim}, {Akrami}, {Ashdown}, {Aumont}, {Baccigalupi}, {Ballardini},
  {Banday}, {Barreiro}, {Bartolo}, {Basak}, {Battye}, {Benabed}, {Bernard},
  {Bersanelli}, {Bielewicz}, {Bock}, {Bond}, {Borrill}, {Bouchet}, {Boulanger},
  {Bucher}, {Burigana}, {Butler}, {Calabrese}, {Cardoso}, {Carron},
  {Challinor}, {Chiang}, {Chluba}, {Colombo}, {Combet}, {Contreras}, {Crill},
  {Cuttaia}, {de Bernardis}, {de Zotti}, {Delabrouille}, {Delouis}, {Di
  Valentino}, {Diego}, {Dor{\'e}}, {Douspis}, {Ducout}, {Dupac}, {Dusini},
  {Efstathiou}, {Elsner}, {En{\ss}lin}, {Eriksen}, {Fantaye}, {Farhang},
  {Fergusson}, {Fernandez-Cobos}, {Finelli}, {Forastieri}, {Frailis},
  {Fraisse}, {Franceschi}, {Frolov}, {Galeotta}, {Galli}, {Ganga},
  {G{\'e}nova-Santos}, {Gerbino}, {Ghosh}, {Gonz{\'a}lez-Nuevo}, {G{\'o}rski},
  {Gratton}, {Gruppuso}, {Gudmundsson}, {Hamann}, {Handley}, {Hansen},
  {Herranz}, {Hildebrandt}, {Hivon}, {Huang}, {Jaffe}, {Jones}, {Karakci},
  {Keih{\"a}nen}, {Keskitalo}, {Kiiveri}, {Kim}, {Kisner}, {Knox},
  {Krachmalnicoff}, {Kunz}, {Kurki-Suonio}, {Lagache}, {Lamarre}, {Lasenby},
  {Lattanzi}, {Lawrence}, {Le Jeune}, {Lemos}, {Lesgourgues}, {Levrier},
  {Lewis}, {Liguori}, {Lilje}, {Lilley}, {Lindholm}, {L{\'o}pez-Caniego},
  {Lubin}, {Ma}, {Mac{\'\i}as-P{\'e}rez}, {Maggio}, {Maino}, {Mandolesi},
  {Mangilli}, {Marcos-Caballero}, {Maris}, {Martin}, {Martinelli},
  {Mart{\'\i}nez-Gonz{\'a}lez}, {Matarrese}, {Mauri}, {McEwen}, {Meinhold},
  {Melchiorri}, {Mennella}, {Migliaccio}, {Millea}, {Mitra},
  {Miville-Desch{\^e}nes}, {Molinari}, {Montier}, {Morgante}, {Moss}, {Natoli},
  {N{\o}rgaard-Nielsen}, {Pagano}, {Paoletti}, {Partridge}, {Patanchon},
  {Peiris}, {Perrotta}, {Pettorino}, {Piacentini}, {Polastri}, {Polenta},
  {Puget}, {Rachen}, {Reinecke}, {Remazeilles}, {Renzi}, {Rocha}, {Rosset},
  {Roudier}, {Rubi{\~n}o-Mart{\'\i}n}, {Ruiz-Granados}, {Salvati}, {Sandri},
  {Savelainen}, {Scott}, {Shellard}, {Sirignano}, {Sirri}, {Spencer},
  {Sunyaev}, {Suur-Uski}, {Tauber}, {Tavagnacco}, {Tenti}, {Toffolatti},
  {Tomasi}, {Trombetti}, {Valenziano}, {Valiviita}, {Van Tent}, {Vibert},
  {Vielva}, {Villa}, {Vittorio}, {Wandelt}, {Wehus}, {White}, {White},
  {Zacchei}, \& {Zonca}}]{Planck2020}
{Planck Collaboration}, {Aghanim}, N., {Akrami}, Y., {et~al.} 2020, \aap, 641,
  A6, \dodoi{10.1051/0004-6361/201833910}

\bibitem[{{Prochaska} {et~al.}(2020){Prochaska}, {Hennawi}, {Westfall},
  {Cooke}, {Wang}, {Hsyu}, {Davies}, {Farina}, \& {Pelliccia}}]{PypeIt2020}
{Prochaska}, J., {Hennawi}, J., {Westfall}, K., {et~al.} 2020, The Journal of
  Open Source Software, 5, 2308, \dodoi{10.21105/joss.02308}

\bibitem[{{Prochaska} {et~al.}(2011){Prochaska}, {Weiner}, {Chen}, {Mulchaey},
  \& {Cooksey}}]{Prochaska2011}
{Prochaska}, J.~X., {Weiner}, B., {Chen}, H.~W., {Mulchaey}, J., \& {Cooksey},
  K. 2011, \apj, 740, 91, \dodoi{10.1088/0004-637X/740/2/91}

\bibitem[{{Prochaska} {et~al.}(2017){Prochaska}, {Werk}, {Worseck}, {Tripp},
  {Tumlinson}, {Burchett}, {Fox}, {Fumagalli}, {Lehner}, {Peeples}, \&
  {Tejos}}]{Prochaska2017}
{Prochaska}, J.~X., {Werk}, J.~K., {Worseck}, G., {et~al.} 2017, \apj, 837,
  169, \dodoi{10.3847/1538-4357/aa6007}

\bibitem[{{Rahmati} {et~al.}(2016){Rahmati}, {Schaye}, {Crain}, {Oppenheimer},
  {Schaller}, \& {Theuns}}]{Rahmati2016}
{Rahmati}, A., {Schaye}, J., {Crain}, R.~A., {et~al.} 2016, \mnras, 459, 310,
  \dodoi{10.1093/mnras/stw453}

\bibitem[{{Rauch} {et~al.}(1999){Rauch}, {Sargent}, \& {Barlow}}]{Rauch1999}
{Rauch}, M., {Sargent}, W. L.~W., \& {Barlow}, T.~A. 1999, \apj, 515, 500,
  \dodoi{10.1086/307060}

\bibitem[{{Rigby} {et~al.}(2022){Rigby}, {Perrin}, {McElwain}, {Kimble},
  {Friedman}, {Lallo}, {Doyon}, {Feinberg}, {Ferruit}, {Glasse}, {Rieke},
  {Rieke}, {Wright}, {Willott}, {Colon}, {Milam}, {Neff}, {Stark}, {Valenti},
  {Abell}, {Abney}, {Abul-Huda}, {Acton}, {Adams}, {Adler}, {Aguilar}, {Ahmed},
  {Albert}, {Alberts}, {Aldridge}, {Allen}, {Altenburg}, {Alvarez Marquez},
  {Alves de Oliveira}, {Andersen}, {Anderson}, {Anderson}, {Argyriou},
  {Armstrong}, {Arribas}, {Artigau}, {Arvai}, {Atkinson}, {Bacon}, {Bair},
  {Banks}, {Barrientes}, {Barringer}, {Bartosik}, {Bast}, {Baudoz}, {Beatty},
  {Bechtold}, {Beck}, {Bergeron}, {Bergkoetter}, {Bhatawdekar}, {Birkmann},
  {Blazek}, {Blome}, {Boccaletti}, {Boeker}, {Boia}, {Bonaventura}, {Bond},
  {Bosley}, {Boucarut}, {Bourque}, {Bouwman}, {Bower}, {Bowers}, {Boyer},
  {Bradley}, {Brady}, {Braun}, {Breda}, {Bresnahan}, {Bright}, {Britt},
  {Bromenschenkel}, {Brooks}, {Brooks}, {Brown}, {Brown}, {Brown}, {Bunker},
  {Burger}, {Bushouse}, {Cale}, {Cameron}, {Cameron}, {Canipe}, {Caplinger},
  {Caputo}, {Cara}, {Carey}, {Carniani}, {Carrasquilla}, {Carruthers}, {Case},
  {Catherine}, {Chance}, {Chapman}, {Charlot}, {Charlow}, {Chayer}, {Chen},
  {Cherinka}, {Chichester}, {Chilton}, {Chonis}, {Clampin}, {Clark}, {Clark},
  {Coe}, {Coleman}, {Comber}, {Comeau}, {Connolly}, {Cooper}, {Cooper},
  {Coppock}, {Correnti}, {Cossou}, {Coulais}, {Coyle}, {Cracraft}, {Curti},
  {Cuturic}, {Davis}, {Davis}, {Dean}, {DeLisa}, {deMeester}, {Dencheva},
  {Dencheva}, {DePasquale}, {Deschenes}, {Hunor Detre}, {Diaz}, {Dicken},
  {DiFelice}, {Dillman}, {Dixon}, {Doggett}, {Donaldson}, {Douglas}, {DuPrie},
  {Dupuis}, {Durning}, {Easmin}, {Eck}, {Edeani}, {Egami}, {Ehrenwinkler},
  {Eisenhamer}, {Eisenhower}, {Elie}, {Elliott}, {Elliott}, {Ellis},
  {Engesser}, {Espinoza}, {Etienne}, {Etxaluze}, {Falini}, {Feeney}, {Ferry},
  {Filippazzo}, {Fincham}, {Fix}, {Flagey}, {Florian}, {Flynn}, {Fontanella},
  {Ford}, {Forshay}, {Fox}, {Franz}, {Fu}, {Fullerton}, {Galkin}, {Galyer},
  {Garcia Marin}, {Gardner}, {Gardner}, {Garland}, {Garrett}, {Gasman},
  {Gaspar}, {Gaudreau}, {Gauthier}, {Geers}, {Geithner}, {Gennaro}, {Giardino},
  {Girard}, {Giuliano}, {Glassmire}, {Glauser}, {Glazer}, {Godfrey},
  {Golimowski}, {Gollnitz}, {Gong}, {Gonzaga}, {Gordon}, {Gordon},
  {Goudfrooij}, {Greene}, {Greenhouse}, {Grimaldi}, {Groebner}, {Grundy},
  {Guillard}, {Gutman}, {Ha}, {Haderlein}, {Hagedorn}, {Hainline}, {Haley},
  {Hami}, {Hamilton}, {Hammel}, {Hansen}, {Harkins}, {Harr}, {Hart}, {Hart},
  {Hartig}, {Hashimoto}, {Haskins}, {Hathaway}, {Havey}, {Hayden}, {Hecht},
  {Heller-Boyer}, {Henriques}, {Henry}, {Hermann}, {Hernandez}, {Hesman},
  {Hicks}, {Hilbert}, {Hines}, {Hoffman}, {Holfeltz}, {Holler}, {Hoppa},
  {Hott}, {Howard}, {Howard}, {Hunter}, {Hunter}, {Hurst}, {Husemann},
  {Hustak}, {Ilinca Ignat}, {Illingworth}, {Irish}, {Jackson}, {Jahromi},
  {Jakobsen}, {James}, {James}, {Januszewski}, {Jenkins}, {Jirdeh}, {Johnson},
  {Johnson}, {Jones}, {Jones}, {Jones}, {Jones}, {Jordan}, {Jordan}, {Jurczyk},
  {Jurling}, {Kaleida}, {Kalmanson}, {Kammerer}, {Kang}, {Kao}, {Karakla},
  {Kavanagh}, {Kelly}, {Kendrew}, {Kennedy}, {Kenny}, {Keski-kuha}, {Keyes},
  {Kidwell}, {Kinzel}, {Kirk}, {Kirkpatrick}, {Kirshenblat}, {Klaassen},
  {Knapp}, {Knight}, {Knollenberg}, {Koehler}, {Koekemoer}, {Kovacs}, {Kulp},
  {Kumari}, {Kyprianou}, {LaMassa}, {Labador}, {Labiano Ortega}, {Lagage},
  {Lajoie}, {Lallo}, {Lam}, {Lamb}, {Lambros}, {Lampenfield}, {Langston},
  {Larson}, {Law}, {Lawrence}, {Lee}, {Leisenring}, {Lepo}, {Leveille},
  {Levenson}, {Levine}, {Levy}, {Lewis}, {Lewis}, {Libralato}, {Lightsey},
  {Link}, {Liu}, {Lo}, {Lockwood}, {Logue}, {Long}, {Long}, {Loomis},
  {Lopez-Caniego}, {Alvarez}, {Love-Pruitt}, {Lucy}, {Luetzgendorf}, {Maghami},
  {Maiolino}, {Major}, {Malla}, {Malumuth}, {Manjavacas}, {Mannfolk},
  {Marrione}, {Marston}, {Martel}, {Maschmann}, {Masci}, {Masciarelli},
  {Maszkiewicz}, {Mather}, {McKenzie}, {McLean}, {McMaster}, {Melbourne},
  {Mel{\'e}ndez}, {Menzel}, {Merz}, {Meyett}, {Meza}, {Miskey}, {Misselt},
  {Moller}, {Morrison}, {Morse}, {Moseley}, {Mosier}, {Mountain}, {Mueckay},
  {Mueller}, {Mullally}, {Murphy}, {Murray}, {Murray}, {Mustelier},
  {Muzerolle}, {Mycroft}, {Myers}, {Myrick}, {Nanavati}, {Nance}, {Nayak},
  {Naylor}, {Nelan}, {Nickson}, {Nielson}, {Nieto-Santisteban}, {Nikolov},
  {Noriega-Crespo}, {O'Shaughnessy}, {O'Sullivan}, {Ochs}, {Ogle}, {Oleszczuk},
  {Olmsted}, {Osborne}, {Ottens}, {Owens}, {Pacifici}, {Pagan}, {Page}, {Park},
  {Parrish}, {Patapis}, {Paul}, {Pauly}, {Pavlovsky}, {Pedder}, {Peek},
  {Pena-Guerrero}, {Pennanen}, {Perez}, {Perna}, {Perriello}, {Phillips},
  {Pietraszkiewicz}, {Pinaud}, {Pirzkal}, {Pitman}, {Piwowar}, {Platais},
  {Player}, {Plesha}, {Pollizi}, {Polster}, {Pontoppidan}, {Porterfield},
  {Proffitt}, {Pueyo}, {Pulliam}, {Quirt}, {Quispe Neira}, {Ramos Alarcon},
  {Ramsay}, {Rapp}, {Rapp}, {Rauscher}, {Ravindranath}, {Rawle}, {Regan},
  {Reichard}, {Reis}, {Ressler}, {Rest}, {Reynolds}, {Rhue}, {Richon},
  {Rickman}, {Ridgaway}, {Ritchie}, {Rix}, {Robberto}, {Robinson}, {Robinson},
  {Robinson}, {Rock}, {Rodriguez}, {Rodriguez Del Pino}, {Roellig}, {Rohrbach},
  {Roman}, {Romelfanger}, {Rose}, {Roteliuk}, {Roth}, {Rothwell}, {Rowlands},
  {Roy}, {Royer}, {Royle}, {Rui}, {Rumler}, {Runnels}, {Russ}, {Rustamkulov},
  {Ryden}, {Ryer}, {Sabata}, {Sabatke}, {Sabbi}, {Samuelson}, {Sapp},
  {Sappington}, {Sargent}, {Sauer}, {Scheithauer}, {Schlawin}, {Schlitz},
  {Schmitz}, {Schneider}, {Schreiber}, {Schulze}, {Schwab}, {Scott}, {Sembach},
  {Shanahan}, {Shaughnessy}, {Shaw}, {Shawger}, {Shay}, {Sheehan}, {Shen},
  {Sherman}, {Shiao}, {Shih}, {Shivaei}, {Sienkiewicz}, {Sing}, {Sirianni},
  {Sivaramakrishnan}, {Skipper}, {Sloan}, {Slocum}, {Slowinski}, {Smith},
  {Smith}, {Smith}, {Smith}, {Snyder}, {Soh}, {Sohn}, {Soto}, {Spencer},
  {Stallcup}, {Stansberry}, {Starr}, {Starr}, {Stewart}, {Stiavelli},
  {Straughn}, {Strickland}, {Stys}, {Summers}, {Sun}, {Sunnquist}, {Swade},
  {Swam}, {Swaters}, {Swoish}, {Taylor}, {Taylor}, {Te Plate}, {Tea}, {Teague},
  {Telfer}, {Temim}, {Thatte}, {Thompson}, {Thompson}, {Thomson}, {Tikkanen},
  {Tippet}, {Todd}, {Toolan}, {Tran}, {Trejo}, {Truong}, {Tsukamoto},
  {Tustain}, {Tyra}, {Ubeda}, {Underwood}, {Uzzo}, {Van Campen}, {Vandal},
  {Vandenbussche}, {Vila}, {Volk}, {Wahlgren}, {Waldman}, {Walker}, {Wander},
  {Warfield}, {Warner}, {Wasiak}, {Watkins}, {Weaver}, {Weilert}, {Weiser},
  {Weiss}, {Weissman}, {Welty}, {West}, {Wheate}, {Wheatley}, {Wheeler},
  {White}, {Whiteaker}, {Whitehouse}, {Whiteleather}, {Whitman}, {Williams},
  {Willmer}, {Willoughby}, {Wilson}, {Wirth}, {Wislowski}, {Wolf}, {Wolfe},
  {Wolff}, {Workman}, {Wright}, {Wu}, {Wu}, {Wymer}, {Yates}, {Yeager},
  {Yeates}, {Yerger}, {Yoon}, {Young}, {Yu}, {Zak}, {Zeidler}, {Zhou},
  {Zielinski}, {Zincke}, \& {Zonak}}]{RigbyJWST2022}
{Rigby}, J., {Perrin}, M., {McElwain}, M., {et~al.} 2022, arXiv e-prints,
  arXiv:2207.05632, \dodoi{10.48550/arXiv.2207.05632}

\bibitem[{{Rubin} {et~al.}(2014){Rubin}, {Prochaska}, {Koo}, {Phillips},
  {Martin}, \& {Winstrom}}]{Rubin2014}
{Rubin}, K. H.~R., {Prochaska}, J.~X., {Koo}, D.~C., {et~al.} 2014, \apj, 794,
  156, \dodoi{10.1088/0004-637X/794/2/156}

\bibitem[{{Rubin} {et~al.}(2018){Rubin}, {O'Meara}, {Cooksey}, {Matuszewski},
  {Rizzi}, {Doppmann}, {Kwok}, {Martin}, {Moore}, {Morrissey}, \&
  {Neill}}]{Rubin2018}
{Rubin}, K. H.~R., {O'Meara}, J.~M., {Cooksey}, K.~L., {et~al.} 2018, \apj,
  859, 146, \dodoi{10.3847/1538-4357/aaaeb7}

\bibitem[{{Rudie} {et~al.}(2019){Rudie}, {Steidel}, {Pettini}, {Trainor},
  {Strom}, {Hummels}, {Reddy}, \& {Shapley}}]{Rudie2019}
{Rudie}, G.~C., {Steidel}, C.~C., {Pettini}, M., {et~al.} 2019, \apj, 885, 61,
  \dodoi{10.3847/1538-4357/ab4255}

\bibitem[{{Rudie} {et~al.}(2012){Rudie}, {Steidel}, {Trainor}, {Rakic},
  {Bogosavljevi{\'c}}, {Pettini}, {Reddy}, {Shapley}, {Erb}, \&
  {Law}}]{Rudie2012}
{Rudie}, G.~C., {Steidel}, C.~C., {Trainor}, R.~F., {et~al.} 2012, \apj, 750,
  67, \dodoi{10.1088/0004-637X/750/1/67}

\bibitem[{{Simcoe} {et~al.}(2012){Simcoe}, {Sullivan}, {Cooksey}, {Kao},
  {Matejek}, \& {Burgasser}}]{Simcoe2012}
{Simcoe}, R.~A., {Sullivan}, P.~W., {Cooksey}, K.~L., {et~al.} 2012, \nat, 492,
  79, \dodoi{10.1038/nature11612}

\bibitem[{{Stocke} {et~al.}(2013){Stocke}, {Keeney}, {Danforth}, {Shull},
  {Froning}, {Green}, {Penton}, \& {Savage}}]{Stocke2013}
{Stocke}, J.~T., {Keeney}, B.~A., {Danforth}, C.~W., {et~al.} 2013, \apj, 763,
  148, \dodoi{10.1088/0004-637X/763/2/148}

\bibitem[{{Tumlinson} {et~al.}(2017){Tumlinson}, {Peeples}, \&
  {Werk}}]{Tumlinson2017}
{Tumlinson}, J., {Peeples}, M.~S., \& {Werk}, J.~K. 2017, \araa, 55, 389,
  \dodoi{10.1146/annurev-astro-091916-055240}

\bibitem[{{Tumlinson} {et~al.}(2011){Tumlinson}, {Thom}, {Werk}, {Prochaska},
  {Tripp}, {Weinberg}, {Peeples}, {O'Meara}, {Oppenheimer}, {Meiring}, {Katz},
  {Dav{\'e}}, {Ford}, \& {Sembach}}]{Tumlinson2011}
{Tumlinson}, J., {Thom}, C., {Werk}, J.~K., {et~al.} 2011, Science, 334, 948,
  \dodoi{10.1126/science.1209840}

\bibitem[{{Tumlinson} {et~al.}(2013){Tumlinson}, {Thom}, {Werk}, {Prochaska},
  {Tripp}, {Katz}, {Dav{\'e}}, {Oppenheimer}, {Meiring}, {Ford}, {O'Meara},
  {Peeples}, {Sembach}, \& {Weinberg}}]{Tumlinson2013}
---. 2013, \apj, 777, 59, \dodoi{10.1088/0004-637X/777/1/59}

\bibitem[{{Wang} {et~al.}(2023){Wang}, {Yang}, {Hennawi}, {Fan}, {Sun},
  {Champagne}, {Costa}, {Habouzit}, {Endsley}, {Li}, {Lin}, {Meyer},
  {Schindler}, {Wu}, {Ba{\~n}ados}, {Barth}, {Bhowmick}, {Bieri}, {Blecha},
  {Bosman}, {Cai}, {Colina}, {Connor}, {Davies}, {Decarli}, {De Rosa}, {Drake},
  {Egami}, {Eilers}, {Evans}, {Farina}, {Haiman}, {Jiang}, {Jin}, {Jun},
  {Kakiichi}, {Khusanova}, {Kulkarni}, {Li}, {Liu}, {Loiacono}, {Lupi},
  {Mazzucchelli}, {Onoue}, {Pudoka}, {Rojas-Ruiz}, {Shen}, {Strauss}, {Tee},
  {Trakhtenbrot}, {Trebitsch}, {Venemans}, {Volonteri}, {Walter}, {Xie}, {Yue},
  {Zhang}, {Zhang}, \& {Zou}}]{Wang2023}
{Wang}, F., {Yang}, J., {Hennawi}, J.~F., {et~al.} 2023, arXiv e-prints,
  arXiv:2304.09894, \dodoi{10.48550/arXiv.2304.09894}

\bibitem[{{Werk} {et~al.}(2013){Werk}, {Prochaska}, {Thom}, {Tumlinson},
  {Tripp}, {O'Meara}, \& {Peeples}}]{Werk2013}
{Werk}, J.~K., {Prochaska}, J.~X., {Thom}, C., {et~al.} 2013, \apjs, 204, 17,
  \dodoi{10.1088/0067-0049/204/2/17}

\bibitem[{{Werk} {et~al.}(2014){Werk}, {Prochaska}, {Tumlinson}, {Peeples},
  {Tripp}, {Fox}, {Lehner}, {Thom}, {O'Meara}, {Ford}, {Bordoloi}, {Katz},
  {Tejos}, {Oppenheimer}, {Dav{\'e}}, \& {Weinberg}}]{Werk2014}
{Werk}, J.~K., {Prochaska}, J.~X., {Tumlinson}, J., {et~al.} 2014, \apj, 792,
  8, \dodoi{10.1088/0004-637X/792/1/8}

\bibitem[{{Williams} {et~al.}(2018){Williams}, {Curtis-Lake}, {Hainline},
  {Chevallard}, {Robertson}, {Charlot}, {Endsley}, {Stark}, {Willmer},
  {Alberts}, {Amorin}, {Arribas}, {Baum}, {Bunker}, {Carniani}, {Crandall},
  {Egami}, {Eisenstein}, {Ferruit}, {Husemann}, {Maseda}, {Maiolino}, {Rawle},
  {Rieke}, {Smit}, {Tacchella}, \& {Willott}}]{Williams2018}
{Williams}, C.~C., {Curtis-Lake}, E., {Hainline}, K.~N., {et~al.} 2018, \apjs,
  236, 33, \dodoi{10.3847/1538-4365/aabcbb}

\bibitem[{{Wotta} {et~al.}(2019){Wotta}, {Lehner}, {Howk}, {O'Meara},
  {Oppenheimer}, \& {Cooksey}}]{Wotta2019}
{Wotta}, C.~B., {Lehner}, N., {Howk}, J.~C., {et~al.} 2019, \apj, 872, 81,
  \dodoi{10.3847/1538-4357/aafb74}

\bibitem[{{Wu} {et~al.}(2021){Wu}, {Cai}, {Neeleman}, {Finlator}, {Zhang},
  {Prochaska}, {Wang}, {Emonts}, {Fan}, {Keating}, {Wang}, {Yang}, {Hennawi},
  \& {Wang}}]{Wu2021}
{Wu}, Y., {Cai}, Z., {Neeleman}, M., {et~al.} 2021, Nature Astronomy, 5, 1110,
  \dodoi{10.1038/s41550-021-01471-4}

\bibitem[{{Yang} {et~al.}(2020){Yang}, {Wang}, {Fan}, {Hennawi}, {Davies},
  {Yue}, {Banados}, {Wu}, {Venemans}, {Barth}, {Bian}, {Boutsia}, {Decarli},
  {Farina}, {Green}, {Jiang}, {Li}, {Mazzucchelli}, \& {Walter}}]{Yang2020}
{Yang}, J., {Wang}, F., {Fan}, X., {et~al.} 2020, \apjl, 897, L14,
  \dodoi{10.3847/2041-8213/ab9c26}

\bibitem[{{Zhu} {et~al.}(2014){Zhu}, {M{\'e}nard}, {Bizyaev}, {Brewington},
  {Ebelke}, {Ho}, {Kinemuchi}, {Malanushenko}, {Malanushenko}, {Marchante},
  {More}, {Oravetz}, {Pan}, {Petitjean}, \& {Simmons}}]{Zhu2014}
{Zhu}, G., {M{\'e}nard}, B., {Bizyaev}, D., {et~al.} 2014, \mnras, 439, 3139,
  \dodoi{10.1093/mnras/stu186}

\end{thebibliography}
\bibliographystyle{aasjournal}

\appendix 

In this section we present the JWST false color stamps, NIRCam WFSS spectra and ground based absorption spectroscopy of the {\mgii} absorption systems at $z\sim 4.22$ (Figure \ref{fig:appendix1}) and $z\sim 4.5$ (Figure \ref{fig:appendix2}), respectively.

\begin{figure*}\setcounter{figure}{0} \renewcommand{\thefigure}{A.\arabic{figure}} 
\centering
\includegraphics[width=0.85\columnwidth]{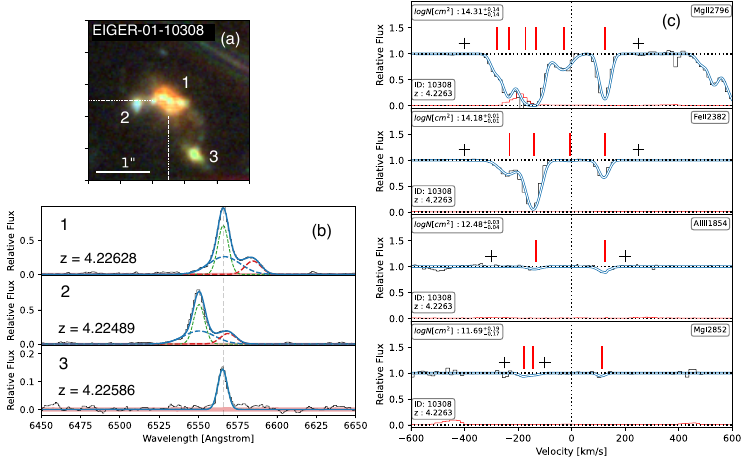}
\caption{Compilation of CGM absorption around merging galaxy system EIGER-01-10308 at $z\sim 4.22$. \textit{Panel (a)}: The false color JWST broad band image of the system is presented. Merging components are marked with a number. Tidal tail between component 1 and 3 can be clearly seen.  \textit{Panel (b)}: Extracted NIRCam WFSS 1D spectrum associated with each marked merging component is presented in individual subplots. H-$\alpha$ emission components are fitted with Gaussian profiles to measure the redshift of merging components. \textit{Panel (c)}: \mgii, \feii, \aliii, and \mgi\ absorption associated with this system are presented with their corresponding Voigt profile fits.
\label{fig:appendix1}}
\end{figure*}

\begin{figure*}\setcounter{figure}{1} \renewcommand{\thefigure}{A.\arabic{figure}} 
\centering
\includegraphics[width=0.85\columnwidth]{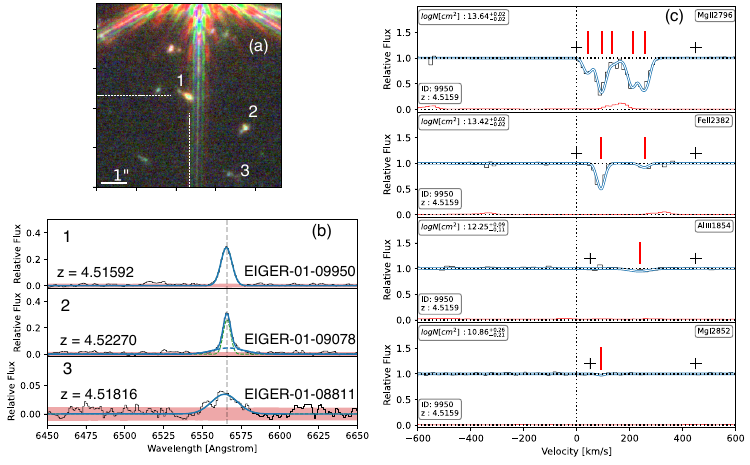}
\caption{Compilation of CGM absorption around three kinematically close galaxies at $z\sim 4.5$. \textit{Panel (a)}: The false color JWST broad band image of three galaxies. All galaxies are within 50 kpc of the quasar sightline and $<$ 370 {\kms} of each other.  \textit{Panel (b)}: 1D NIRCam WFSS spectra associated with each marked galaxies. H-$\alpha$ emission components are fitted with Gaussian profiles to measure the redshift of merging components. \textit{Panel (c)}: \mgii, \feii, \aliii, and \mgi\ absorption associated with this system are presented with their corresponding Voigt profile fits. 
\label{fig:appendix2}}
\end{figure*}




\end{document}